\documentclass[12pt, draftcls, onecolumn]{IEEEtran}
\usepackage{multicol}
\usepackage{etoolbox}
\makeatletter
\patchcmd{\@makecaption}
  {\scshape}
  {}
  {}
  {}
\makeatletter
\patchcmd{\@makecaption}
  {\\}
  {.\ }
  {}
  {}
\makeatother

\usepackage{amsfonts}
\usepackage{amsmath}
\usepackage{mathrsfs}
\usepackage{mathtools}
\usepackage{amsfonts}
\usepackage{amssymb}
\usepackage{graphicx}
\usepackage{epsfig}
\usepackage{psfrag}{}
\usepackage{array}
\usepackage{cases}
\usepackage{eufrak}
\usepackage{cite,graphicx,amssymb,color}
\usepackage{algorithmic}
\usepackage{algorithm}
\usepackage{setspace}
\usepackage{subfigure}
\usepackage{bm}
\usepackage{multirow}
\usepackage{threeparttable}
\usepackage{array}
\usepackage{makecell}
\usepackage{soul}
\usepackage{float}
\usepackage{booktabs}
\newfloat{figtab}{htb}{fgtb}
\makeatletter
  \newcommand\figcaption{\def\@captype{figure}\caption}
  \newcommand\tabcaption{\def\@captype{table}\caption}
\makeatother

\newtheorem{Thm}{Theorem}
\newtheorem{Lem}{Lemma}

\newtheorem{Exam}{Example}

\newtheorem{Prob}{Problem}
\newtheorem{Rem}{Remark}

\newtheorem{Proof}{Proof}

\IEEEoverridecommandlockouts
\begin{document}
\title{An Optimization Framework for General Rate Splitting for General Multicast}
\author{\IEEEauthorblockN{Lingzhi Zhao, Ying Cui, Sheng Yang, and Shlomo Shamai (Shitz)}\thanks{Lingzhi Zhao and Ying Cui are with the Department of Electronic Engineering, Shanghai Jiao Tong University, Shanghai, China. 
Sheng Yang is with the Laboratory of Signals and Systems, CentraleSup\'elec-CNRS-Paris-Saclay University, Gif-sur-Yvette, France. Shlomo Shamai (Shitz) is with the Technion–Israel Institute of Technology, Haifa, Israel.}}

\maketitle

\begin{abstract}
Immersive video, such as virtual reality (VR) and multi-view videos, is growing in popularity. Its wireless streaming is an instance of general multicast, extending conventional unicast and multicast, whose effective design is still open. This paper investigates general rate splitting for general multicast. Specifically, we consider a multi-carrier single-cell wireless network where a multi-antenna base station (BS) communicates to multiple single-antenna users via general multicast. We consider linear beamforming at the BS and joint decoding at each user in the slow fading and fast fading scenarios. In the slow fading scenario, we consider the maximization of the weighted sum average rate, which is a challenging nonconvex stochastic problem with numerous variables. To reduce computational complexity, we decouple the original nonconvex stochastic problem into multiple nonconvex deterministic problems, one for each system channel state. Then, we propose an iterative algorithm for each deterministic problem to obtain a \textcolor{black}{Karush-Kuhn-Tucker (KKT)} point using the concave-convex procedure (CCCP). In the fast fading scenario, we consider the maximization of the weighted sum ergodic rate. This problem is more challenging than the one for the slow fading scenario, as it is not separable. First, we propose a stochastic iterative algorithm to obtain a KKT point using stochastic successive convex approximation (SSCA) and the exact penalty method. Then, we propose two low-complexity iterative algorithms to obtain feasible points with promising performance for two cases of channel distributions using approximation and CCCP. The proposed optimization framework generalizes the existing ones for rate splitting for various types of services. 
Finally, we numerically show substantial gains of the proposed solutions over existing schemes in both scenarios and reveal the design insights of general rate splitting for general multicast.
\end{abstract}

\begin{IEEEkeywords}
General multicast, general rate splitting, linear beamforming, joint decoding, optimization, concave-convex procedure (CCCP), stochastic successive convex approximation (SSCA).
\end{IEEEkeywords}

\section{Introduction}
Conventional mobile Internet services include (traditional) video, audio, web browsing, social networking, software downloading, etc. These services can be supported by unicast, \textcolor{black}{unicast with a common message,} single-group multicast, and multi-group multicast. 
Immersive video, such as 360 video (projection of virtual reality (VR) spherical video onto a rectangle) and multi-view video (a key technique in free-viewpoint television, naked-eye 3D and VR), is growing in popularity. It is predicted that the VR market will reach 87.97 billion USD by 2025\cite{web}. When watching a 360 video, 
the tiles in a user's current FoV plus a safe margin are usually transmitted to the user in case of FoV change. On the other hand, when watching a multi-view video, 
a user's current view and adjacent views are usually transmitted to the user in case of a view switch. In wireless streaming of a popular immersive video 
to multiple users simultaneously, multiple messages (e.g., tiles for 360 video and views for multi-view video) are transmitted to each user, and one message may be intended for multiple users\cite{TMM20,TCOM20}, as illustrated in Fig. 1. This emerging service plays an important role in online gaming, self-driving, and cloud meeting, etc. but cannot perfectly adapt to the conventional transmission schemes mentioned above. This motivates us to consider general multicast (also referred to as general connection\cite{TON17,TON18} and general groupcast \cite{ISIT17}) where one message can be intended for any user. Clearly, general multicast includes conventional unicast, \textcolor{black}{unicast with a common message,} single-group multicast, and multi-group multicast as special cases.

\begin{figure}[t]
\begin{center}
 \subfigure[\small{360 video}]
 {\resizebox{4.5cm}{!}{\includegraphics{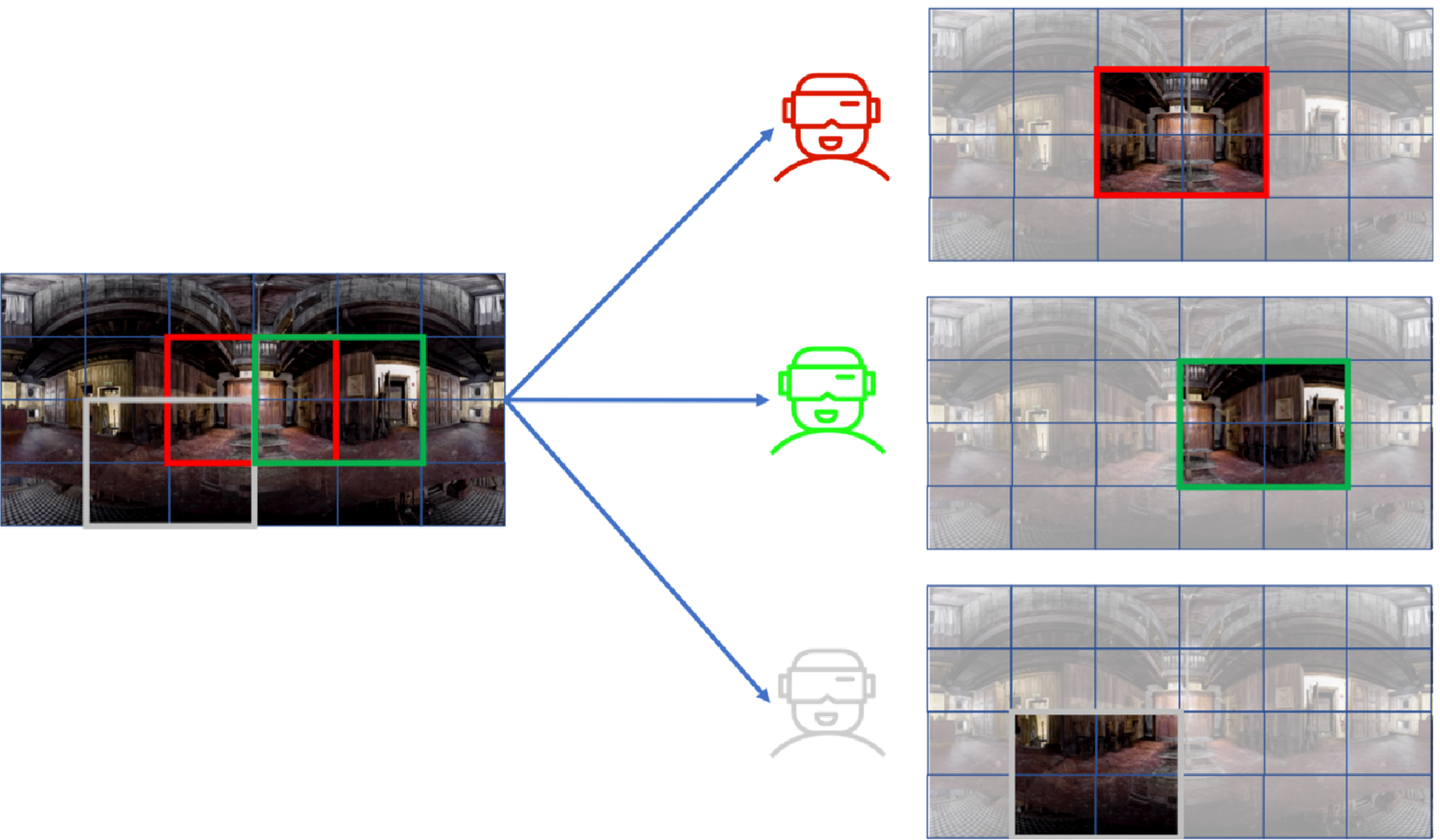}}}
 \subfigure[\small{Multi-view video}]
 {\resizebox{4.5cm}{!}{\includegraphics{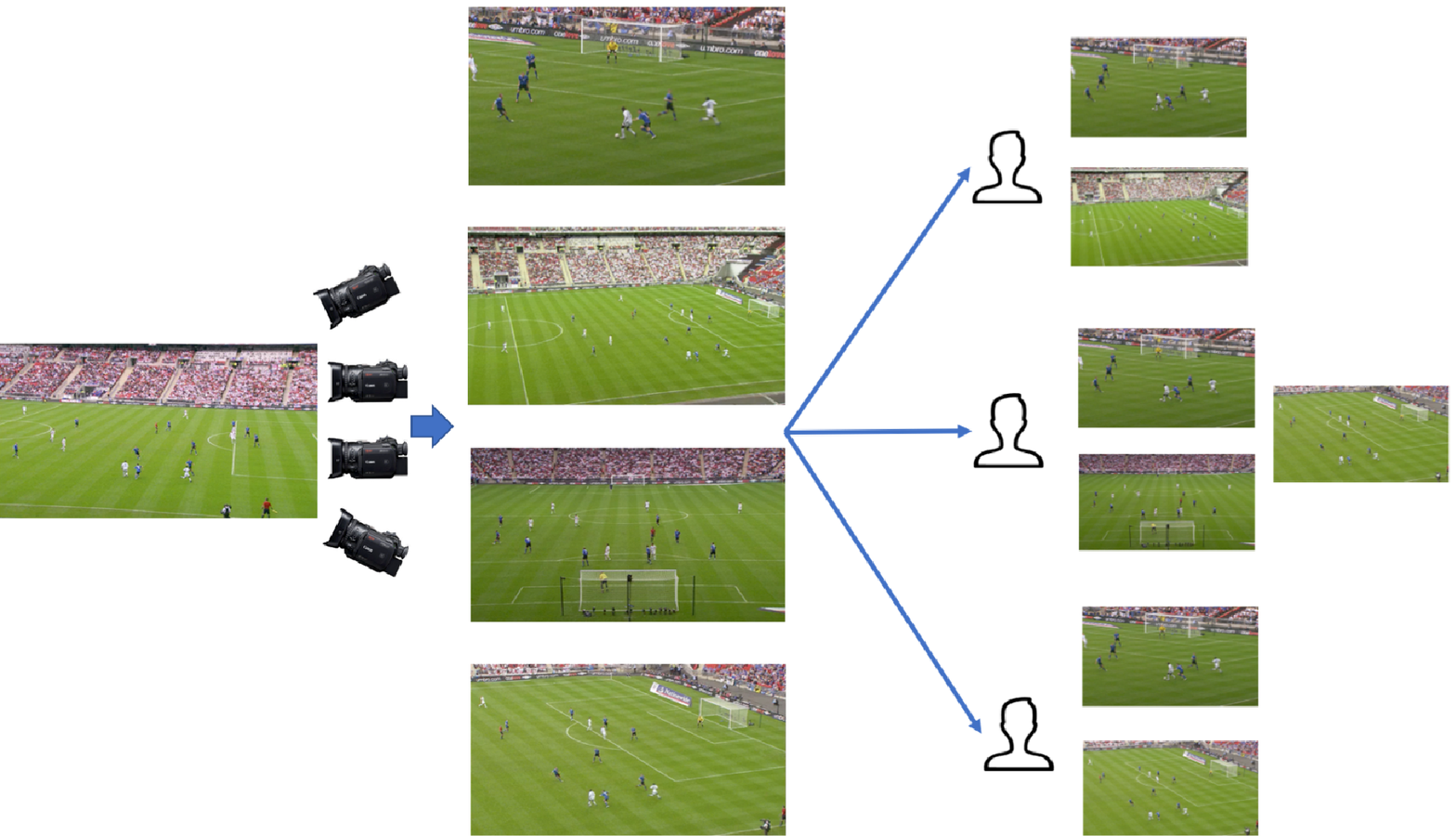}}}
 \end{center}
   \caption{\small{Applications of general multicast.}}
   \label{fig:applications}
\end{figure}

References\cite{TMM20,TCOM20,TWC21,CL20} are pioneer works for supporting wireless streaming of a 360 video\cite{TMM20,TWC21} and \textcolor{black}{wireless streaming of} a multi-view video\cite{TCOM20,CL20}, which are instances of general multicast. In \cite{TMM20,TCOM20,TWC21,CL20}, Orthogonal Multiple Access (OMA), such as Time Division Multiple Access (TDMA)\cite{TMM20,TCOM20} and Orthogonal Frequency Division Multiple Access (OFDMA)\cite{TWC21,CL20}, is adopted to convert general multicast to per resource block single-group multicast. 
\textcolor{black}{While} the OMA-based mechanisms are easy to implement, \textcolor{black}{spatial multiplexing gain is not exploited}. On the other hand, non-orthogonal transmission \textcolor{black}{mechanisms achieve higher} transmission efficiency but \textcolor{black}{are also more challenging due to interference}. Space Division Multiple Access (SDMA) and Non-Orthogonal Multiple Access (NOMA) are \textcolor{black}{two solutions.}
The cost to suppress interference in SDMA can be high when the channels for some users are spatially aligned, \textcolor{black}{while decoding interference in NOMA may not be possible when the interfering message rate is too high.} \textcolor{black}{Thus, SDMA and NOMA may have unsatisfactory performance.} Rate splitting\cite{arxiv22,JSAC21} is introduced to partially suppress interference and partially decode interference to circumvent these limitations.

Rate splitting is originally proposed to effectively support unicast services\cite{TIT1981}. Specifically, \cite{TIT1981} investigates the simplest form of rate splitting for unicast, \textcolor{black}{hereafter} called 1-layer rate splitting, \textcolor{black}{for the two-user interference channel}. The idea of 1-layer rate splitting is as follows. First, each individual message is split into one private part and one common part, respectively. Then, the common parts of all the messages are re-assembled into one common message that is multicasted to all the users, and the private parts are unicasted to the
corresponding receivers, respectively. In this way, part of the
interference can be removed since it is decodable \textcolor{black}{by design}. Later, \textcolor{black}{in \cite{TIT13_yang}, 1-layer rate splitting is applied to the multi-antenna broadcast channel (BC) and shown to provide a strict sum degree of freedom gain of a BC when only imperfect channel state information at the transmitter is available.} In \cite{arxiv21,TSP16,TWC20,TCOM16,TCOM21_Yang}, \textcolor{black}{the authors investigate the precoder optimization} of 1-layer rate splitting for unicast \textcolor{black}{for Gaussian multiple-input multiple-output channels}. In particular, the rates and beamforming vectors for common and private messages are optimized to maximize the sum rate\cite{arxiv21,TCOM21_Yang}, worst-case rate\cite{TSP16,TWC20} or ergodic sum rate\cite{TCOM16}. On the one hand, \cite{EJWC18,JSAC21} extend 1-layer rate splitting for unicast to general rate splitting \textcolor{black}{for unicast} and study the optimization of three-user unicast\cite{EJWC18} and multi-user unicast\cite{JSAC21}, \textcolor{black}{respectively, with linear precoding.} On the other hand, \cite{TCOM19,wcnc21,TWC17,TVT20_A,TVT20} generalize 1-layer rate splitting for unicast to 1-layer rate splitting for unicast \textcolor{black}{together} with a multicast message intended for all users \cite{TCOM19} and multi-group multicast\cite{wcnc21,TWC17,TVT20_A,TVT20}, \textcolor{black}{respectively,} and investigate \textcolor{black}{the optimizations} with linear beamforming. Note that most works\cite{arxiv21,TSP16,TWC20,TCOM21_Yang,EJWC18,JSAC21,TCOM19,TWC17,TVT20_A,TVT20} \textcolor{black}{focus on the slow fading scenario, whereas \cite{TCOM16},\textcolor{black}{\cite{wcnc21}} concentrates on the fast fading scenario.}

\textcolor{black}{Optimization-based random linear network coding designs for general multicast have been studied in \cite{TON17,TON18} for wired networks.
Besides,} general rate splitting for general multicast has been studied in \cite{ISIT17} for discrete memoryless broadcast channels. Here, we are interested in Gaussian fading channels and specifically the linear beamforming design from the optimization perspective.
In general rate splitting \textcolor{black}{for general multicast}, \textcolor{black}{each message intended for a user group is split into sub-messages, one for each subset of users containing the user group. Then, the sub-messages intended for the same group of users are re-assembled and multicasted to the group. In this way, each user group decodes the desired message and part of the message of any other user group.} \textcolor{black}{Note that general rate splitting for general multicast produces more sub-messages and enables more  flexible and effective interference reduction than brute-force application of general rate splitting for unicast\cite{EJWC18,JSAC21} to general multicast.\footnote{\textcolor{black}{One can brute-forcely apply general rate splitting\cite{EJWC18,JSAC21} for unicast to general multicast by treating a group of users who request the same message as a virtual user.}}} Besides, the optimizations of rate splitting for unicast and its slight generalization in \cite{TCOM19,wcnc21,TWC17,TVT20_A,TVT20} cannot apply to general multicast. Therefore, for general multicast in both slow fading and fast fading scenarios, the optimization of general rate splitting with linear beamforming remains an open problem.

This paper \textcolor{black}{intends to shed some light on the above issue}. Specifically, we consider a multi-carrier single-cell wireless network, where a multi-antenna base station (BS) communicates to multiple single-antenna users via general multicast. Our main contributions are summarized below.
\begin{itemize}
  \item 
  We present general rate splitting for general multicast and illustrate its connection with conventional unicast, \textcolor{black}{unicast with a common message,} single-group multicast, and multi-group multicast. We adopt linear beamforming at the BS and joint decoding at each user and characterize the corresponding rate regions in the slow fading and fast fading scenarios.
  \item In the slow fading scenario, we optimize the transmission beamforming vectors and rates of sub-message units to maximize the weighted sum average rate. Note that the proposed problem formulation can reduce to those for general rate splitting for unicast\cite{JSAC21}, \textcolor{black}{unicast with a common message\cite{TCOM19},} single-group multicast\cite{TSP06}, and 1-layer rate splitting for multi-group multicast\cite{TWC17}. This problem is a challenging nonconvex stochastic problem with a large number of variables. To reduce computational complexity, we decouple the original nonconvex stochastic problem into multiple nonconvex deterministic problems, one for each system channel state. Then, for each nonconvex deterministic problem, we propose an iterative algorithm to obtain a \textcolor{black}{Karush-Kuhn-Tucker (KKT)} point using the concave-convex procedure (CCCP). 
  \item In the fast fading scenario, we optimize the transmission beamforming vectors and rates of sub-message units to maximize the weighted sum ergodic rate and show that the problem formulation can reduce to the one for 1-layer rate splitting for unicast\cite{TCOM16}. This problem is more challenging than the one for the slow fading scenario, as it is not separable. First, we propose a stochastic iterative algorithm to obtain a KKT point using stochastic successive convex approximation (SSCA) and the exact penalty method. Then, we propose two low-complexity iterative algorithms to obtain feasible points with promising performance for two cases of channel distributions, i.e., spatially correlated channel and independent and identically distributed (i.i.d.) channel, using approximation and CCCP. It is noteworthy that the proposed optimization framework also provides general rate splitting designs for other service types in the fast fading scenario.
  \item We compare the complexities of the proposed solutions in the slow fading and fast fading scenarios. We also numerically demonstrate substantial gains of the proposed solutions over existing schemes in both scenarios and reveal the design insights of general rate splitting for general multicast.
  \end{itemize}

\begin{table}[t]
\caption{\small{KEY NOTAION}}
\begin{center}
\textcolor{black}{
\resizebox{9cm}{!}{
\begin{tabular}{|c|c|} 
\hline 
Notation & Description \\ \hline  
$K$ & number of users\\ \hline
$\mathcal{K}$ & set of $K$ user indices\\ \hline
$I$ & number of messages\\ \hline
$\mathcal{I}$ &  set of the indices of $I$ messages \\ \hline
$\mathcal{I}_{k}$ &  set of the indices of messages requested by user $k$  \\ \hline
$\mathcal{S}$ &  subset of $\mathcal{K}$  \\ \hline
$\mathcal{P}_{\mathcal{S}}$ & set of the indices of the messages requested by each user in $\mathcal{S}$ and not requested by any user in $\mathcal{K}\backslash\mathcal{S}$   \\ \hline
$\boldsymbol{\mathcal{G}}_{\mathcal{S}}$ & subsets of $\mathcal{K}$ that contain $\mathcal{S}$   \\ \hline
$R_{\mathcal{S}}$ & rate of the message unit $\mathcal{P}_{\mathcal{S}}$   \\ \hline
$R_{\mathcal{S}}(\mathbf{h})$ & rate of the message unit $\mathcal{P}_{\mathcal{S}}(\mathbf{h})$ under $\mathbf{h}$   \\ \hline
$M$ &   number of the antennas \\ \hline
$N$ & number of the subcarriers \\ \hline
$B$ & bandwidth of each subcarrier \\\hline
$\mathbf{w}_{\mathcal{G},n}(\mathbf{h})$ & beamforming vector for transmitting $\tilde{\mathcal{P}}_{\mathcal{G}}(\mathbf{h})$ on subcarrier $n$ under $\mathbf{h}$ \\\hline
$\mathbf{w}_{\mathcal{G},n}$ & constant beamforming vector for transmitting $\tilde{\mathcal{P}}_{\mathcal{G}}$ on subcarrier $n$ \\\hline
\end{tabular}
}} \label{table0}
\end{center}
\end{table}

\textbf{Notation:} We represent vectors by boldface lowercase letters (e.g., $\mathbf{x}$), matrices by boldface uppercase
letters (e.g., $\mathbf{X}$), scalar constants by non-boldface letters (e.g., $x$), sets by calligraphic letters
(e.g., $\mathcal{X}$ ), and sets of sets by boldface calligraphic letters (e.g., $\boldsymbol{\mathcal{X}}$). The notation $x_{i}$ represents the $i$-th element of vector $\mathbf{x}$. The symbol $(\cdot)^{H}$ denotes complex conjugate transpose operator. $\|\cdot\|_{2}$ denotes the Euclidean norm of a vector. $\Re\{\cdot\}$ denotes the real part of a complex number. $\mathbb{E}[\cdot]$ denotes the statistical expectation. $\mathbf{I}_{M\times M}$ denotes the $M\times M$ identity matrix. $\mathbb{C}$ denotes the set of complex numbers.

\section{System Model}\label{sec:system}
\textcolor{black}{In this section, we first introduce} general multicast \textcolor{black}{in a single-cell wireless network} and briefly illustrate its connection with \textcolor{black}{unicast, \textcolor{black}{unicast with a common message,} single-group multicast, and multi-group multicast.} Then, we present the physical layer model and general rate splitting with joint decoding for general multicast. \textcolor{black}{The key notation used in this paper is listed in Table \ref{table0}.}

\subsection{General Multicast}
We consider a \textcolor{black}{single-cell wireless network consisting of} one BS and $K$ users. Let $\mathcal{K} \triangleq \{1,\ldots,K\}$ denote the set of user indices. \textcolor{black}{The BS has $I$ independent messages.} Let $\mathcal{I}\triangleq \{1,\ldots, I\}$ denote the set of \textcolor{black}{the indices} of $I$ messages. We consider general multicast. Specifically, each user $k\in\mathcal{K}$ can request arbitrary $I_{k}$ messages in $\mathcal{I}$, denoted by $\mathcal{I}_{k}\subseteq\mathcal{I}$, \textcolor{black}{from the BS. We do not have any assumptions on $\mathcal{I}_{k},~k\in\mathcal{K}$ except that each message in $\mathcal{I}$ is requested by at least one user, i.e., $\cup_{k\in\mathcal{K}}\mathcal{I}_{k} = \mathcal{I}$. }

\begin{figure}[t]
\begin{center}
 \subfigure[\small{$K = 2$, $I = 6$, $\mathcal{I}_{1} = \{1,2,5,6\}$, $\mathcal{I}_{2} = \{2,3,6,7\}$.}]
 {\resizebox{8cm}{!}{\includegraphics{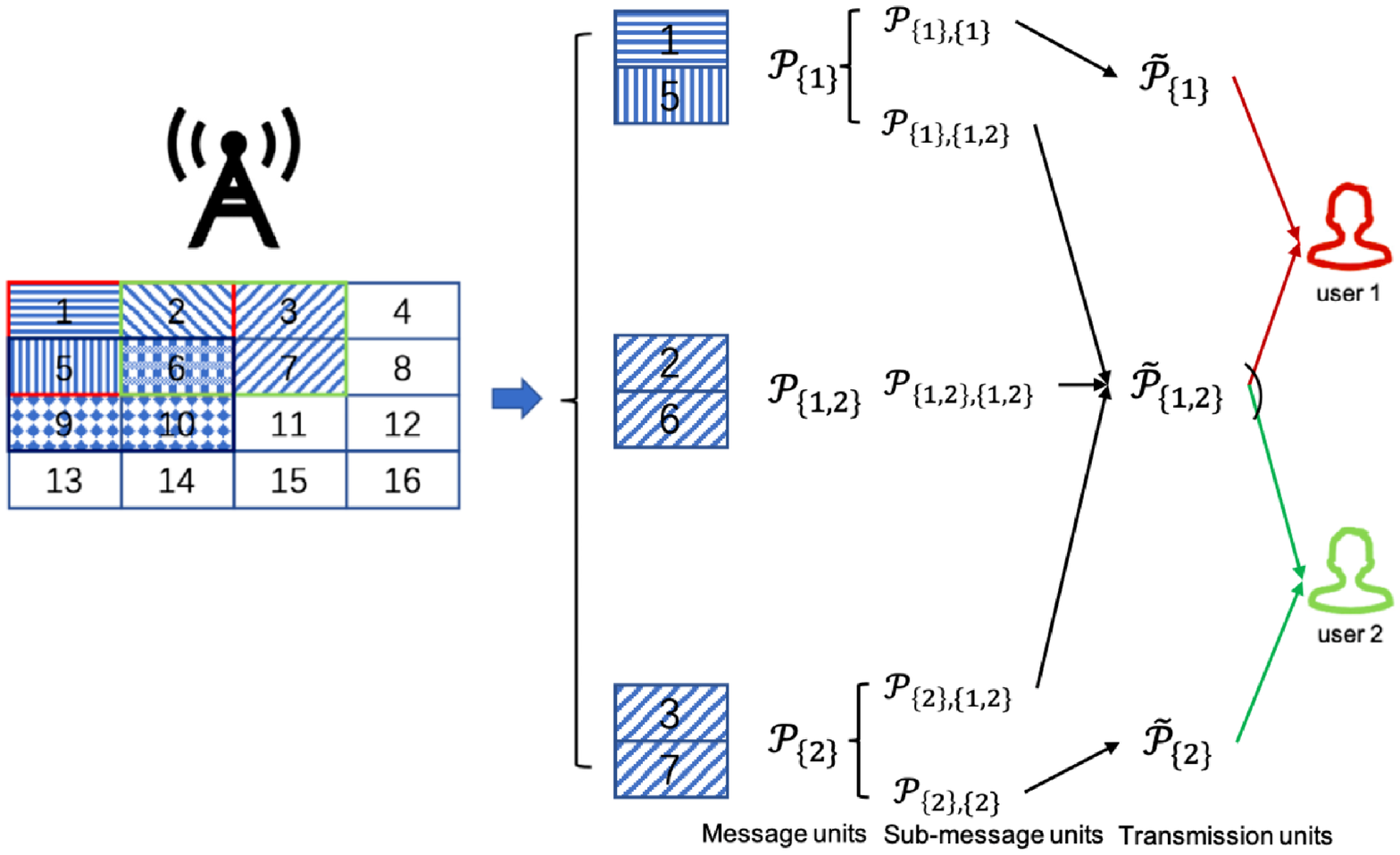}}}
 \subfigure[\small{$K = 3$, $I = 8$, $\mathcal{I}_{1} = \{1,2,5,6\}$, $\mathcal{I}_{2} = \{2,3,6,7\}$, $\mathcal{I}_{3} = \{5,6,9,10\}$.}]
 {\resizebox{8cm}{!}{\includegraphics{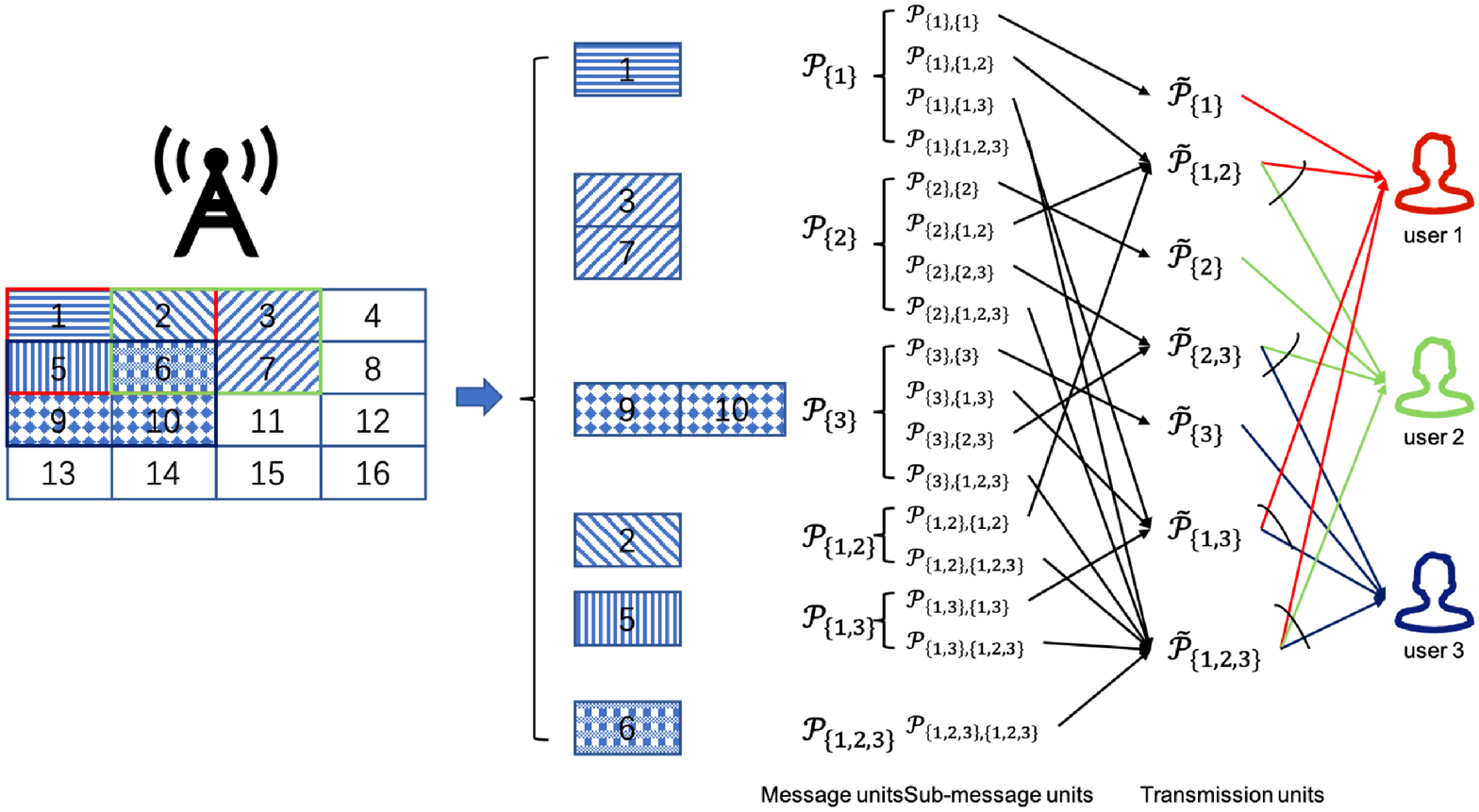}}}
 \end{center}
   \caption{\small{\textcolor{black}{Wireless streaming of a tiled 360 video to multiple users.} The 360 video is divided into $4\times 4$ tiles. The users have different FoVs which overlap to certain extent. }}
   \label{system_model}
\end{figure}

\textcolor{black}{To facilitate serving the $K$ users, we partition the message set $\mathcal{I}$ according to the requests from the $K$ users.} For all $\mathcal{S}\subseteq \mathcal{K},\mathcal{S}\neq \emptyset$, let 
\begin{align}
\mathcal{P}_{\mathcal{S}} \triangleq \left(\bigcap_{k\in\mathcal{S}}\mathcal{I}_{k}\right)\bigcap\left(\mathcal{I} - \bigcup_{k\in\mathcal{K}\backslash\mathcal{S}}\mathcal{I}_{k}\right)\label{eq:partition}
\end{align}
denote the set of \textcolor{black}{the indices of messages} requested by each user in $\mathcal{S}$ and not requested by any user in $\mathcal{K}\backslash\mathcal{S}$ \cite{TMM20}. Define
\begin{align}
\boldsymbol{\mathcal{P}}\triangleq \{\mathcal{P}_{\mathcal{S}}|\mathcal{P}_{\mathcal{S}}\neq \emptyset, \mathcal{S}\subseteq\mathcal{K},\mathcal{S}\neq \emptyset \},\nonumber\\
\boldsymbol{\mathcal{S}} \triangleq \{\mathcal{S}|\mathcal{P}_{\mathcal{S}}\neq \emptyset, \mathcal{S}\subseteq\mathcal{K},\mathcal{S}\neq \emptyset\}.\nonumber
\end{align}
Thus, $\boldsymbol{\mathcal{P}}$ forms a partition of $\mathcal{I}$ and $\boldsymbol{\mathcal{S}}$ specifies the user groups corresponding to the partition. We refer to each element in $\boldsymbol{\mathcal{P}}$ as a message unit.\footnote{\textcolor{black}{$\boldsymbol{\mathcal{P}}$ and $\boldsymbol{\mathcal{S}}$ are assumed to be given in \cite{ISIT17}}.} \textcolor{black}{We can see that different message units in $\boldsymbol{\mathcal{P}}$ are requested by different user groups in $\boldsymbol{\mathcal{S}}$.}
\begin{Exam}[Illustration of $\boldsymbol{\mathcal{P}}$ and $\boldsymbol{\mathcal{S}}$ for Two-User Case]
\textcolor{black}{As illustrated in Fig. \ref{system_model} (a), we consider $K = 2$, $I = 6$, $\mathcal{I}_{1} = \{1,2,5,6\}$, $\mathcal{I}_{2} = \{2,3,6,7\}$. Then, we have $\mathcal{P}_{\{1\}} = \{1,5\}$, $\mathcal{P}_{\{2\}} = \{3,7\}$, $\mathcal{P}_{\{1,2\}} = \{2,6\}$, $\boldsymbol{\mathcal{P}} = \{\mathcal{P}_{\{1\}},\mathcal{P}_{\{2\}},\mathcal{P}_{\{1,2\}}\}$, and $\boldsymbol{\mathcal{S}} = \{\{1\},\{2\},\{1,2\}\}$. There are 3 message units that are requested by 3 groups of users, respectively. For example, message unit $\mathcal{P}_{\{1\}}$ is requested only by user 1 and message unit $\mathcal{P}_{\{1,2\}}$ is requested by user 1 and user 2.\footnote{\textcolor{black}{This general multicast scenario coincides with unicast with a common message.}}}
\end{Exam}

\begin{Exam}[Illustration of $\boldsymbol{\mathcal{P}}$ and $\boldsymbol{\mathcal{S}}$ \textcolor{black}{for Three-User Case}]
As illustrated in \textcolor{black}{Fig. \ref{system_model} (b)}, we consider $K = 3$, $I = 8$, $\mathcal{I}_{1} = \{1,2,5,6\}$, $\mathcal{I}_{2} = \{2,3,6,7\}$, $\mathcal{I}_{3} = \{5,6,9,10\}$. Then, we have $\mathcal{P}_{\{1\}} = \{1\}$, $\mathcal{P}_{\{2\}} = \{3,7\}$, $\mathcal{P}_{\{3\}} = \{9,10\}$, $\mathcal{P}_{\{1,2\}} = \{2\}$, $\mathcal{P}_{\{1,3\}} = \{5\}$, $\mathcal{P}_{\{1,2,3\}} = \{6\}$, $\boldsymbol{\mathcal{P}} = \{\mathcal{P}_{\{1\}},\mathcal{P}_{\{2\}},\mathcal{P}_{\{3\}},\mathcal{P}_{\{1,2\}},\mathcal{P}_{\{1,3\}},\mathcal{P}_{\{1,2,3\}}\}$, and $\boldsymbol{\mathcal{S}} = \{\{1\},\{2\},\{3\},\{1,2\},\{1,3\},\{1,2,3\}\}$. There are 6 message units that are requested by 6 groups of users, respectively. For example, message unit $\mathcal{P}_{\{1\}}$ is requested only by user 1, message unit $\mathcal{P}_{\{1,2\}}$ is requested by user 1 and user 2, \textcolor{black}{and} message unit $\mathcal{P}_{\{1,2,3\}}$ is requested by user 1, user 2, and user 3.
\end{Exam}

\begin{Rem}[Connection with Unicast and Multicast]
\textcolor{black}{The considered general multicast includes conventional unicast, \textcolor{black}{unicast with a common message,} single-group multicast, and multi-group multicast as special cases. \textcolor{black}{(\romannumeral1) When $I = K, I_{k} = 1,k\in\mathcal{K}$, and $\mathcal{I}_{k}\not= \mathcal{I}_{k'}, k,k'\in\mathcal{K}, k\not=k'$, general multicast reduces to unicast\textcolor{black}{\cite{arxiv21,TSP16,TWC20,TCOM16,TCOM21_Yang,EJWC18,JSAC21}}.} \textcolor{black}{In this case, $\boldsymbol{\mathcal{P}} = \{\{1\},\{2\},\ldots,\{K\}\}$ and $\boldsymbol{\mathcal{S}} = \{\{1\},\{2\},\ldots, \{K\}\}$.} \textcolor{black}{(\romannumeral2) When $I = K + 1, I_{k} = 2,k\in\mathcal{K}$, $\mathcal{I}_{k}\not= \mathcal{I}_{k'}, k,k'\in\mathcal{K}, k\not=k',$ and $|\cap_{k\in\mathcal{K}}\mathcal{I}_{k}| = 1$, general multicast reduces to unicast with a common message\cite{TCOM19}.} \textcolor{black}{In this case, $\boldsymbol{\mathcal{P}} = \{\{1\},\{2\},\ldots,\{K\},\mathcal{K}\}$ and $\boldsymbol{\mathcal{S}} = \{\{1\},\{2\},\ldots, \{K\},\mathcal{K}\}$.} \textcolor{black}{(\romannumeral3) When $I = 1$, implying $I_{k} = 1,k\in\mathcal{K}$, and $\mathcal{I}_{k} = \mathcal{I}_{k'},k,k'\in\mathcal{K},k\not=k'$, general multicast becomes single-group multicast\textcolor{black}{\cite{TSP06}}. \textcolor{black}{In this case, $\boldsymbol{\mathcal{P}} = \{\{1\}\}$ and $\boldsymbol{\mathcal{S}} = \{\mathcal{K}\}$}}. (\romannumeral4) When $1< I < K$ and $I_{k} = 1, k\in\mathcal{K}$, general multicast reduces to multi-group ($I$-group) multicast\textcolor{black}{\cite{wcnc21,TWC17,TVT20_A,TVT20}}. In this case, $\boldsymbol{\mathcal{P}} = \{\{1\},\{2\},\ldots,\{I\}\}$ and $\boldsymbol{\mathcal{S}} = \left\{\{k\in\mathcal{K}|\mathcal{I}_{k} = \{1\}\},\{k\in\mathcal{K}|\mathcal{I}_{k} = \{2\}\},\ldots, \{k\in\mathcal{K}|\mathcal{I}_{k} = \{I\}\}\right\}$}. \textcolor{black}{The general multicast considered in this paper, \textcolor{black}{general connection in \cite{TON17,TON18}}, and general groupcast considered in \cite{ISIT17} mean the same.}
\end{Rem}

\subsection{General Rate Splitting}\label{sec:rs_model}

We consider rate splitting in the most general form for general multicast to serve the $K$ users\cite{ISIT17}. \textcolor{black}{It allows each user group to decode not only the desired message unit $\mathcal{P}_{\mathcal{S}}$ but also part of the message unit of any other user group, $\mathcal{P}_{\mathcal{S}'}$ for all $\mathcal{S}' \not= \mathcal{S}, \mathcal{S}'\in\boldsymbol{\mathcal{S}}$, to flexibly reduce the interference level.} For all $\mathcal{S}\in\boldsymbol{\mathcal{S}}$, let $\boldsymbol{\mathcal{G}}_{\mathcal{S}} \triangleq \{\mathcal{X}|\mathcal{S}\subseteq\mathcal{X}\subseteq \mathcal{K}\}$.
Namely, $\boldsymbol{\mathcal{G}}_{\mathcal{S}}$ collects all $2^{K - |\mathcal{S}|}$ subsets of $\mathcal{K}$ that contain $\mathcal{S}$. Define $\boldsymbol{\mathcal{G}} \triangleq \bigcup_{\mathcal{S}\in\boldsymbol{\mathcal{S}}} \boldsymbol{\mathcal{G}}_{\mathcal{S}}$. \textcolor{black}{Obviously, $\boldsymbol{\mathcal{S}} \subseteq \boldsymbol{\mathcal{G}}$}.
First, we split each message unit $\mathcal{P}_{\mathcal{S}}$ into \textcolor{black}{$2^{K - |\mathcal{S}|}$} sub-message units.
Accordingly, the rate of the message unit $\mathcal{P}_{\mathcal{S}}$, denoted by $R_{\mathcal{S}}$, is split into the rates of the $2^{K - |\mathcal{S}|}$ sub-message units, 
denoted by $R_{\mathcal{S},\mathcal{G}},\mathcal{G}\in\boldsymbol{\mathcal{G}}_{\mathcal{S}}$\footnote{When $\mathcal{S} = \mathcal{K}$, $\boldsymbol{\mathcal{G}}_{S} =\{\mathcal{S}\}$ and the message unit $\mathcal{P}_{\mathcal{S}}$ will not be split. For ease of exposition, we let $R_{\mathcal{S}} = R_{\mathcal{S},\mathcal{S}}$ for $\mathcal{S} = \mathcal{K}$.} i.e.,
\begin{align}
R_{\mathcal{S}} = \sum\nolimits_{\mathcal{G}\in\boldsymbol{\mathcal{G}}_{\mathcal{S}}}\label{eq:rate_split}
R_{\mathcal{S},\mathcal{G}},~\mathcal{S}\in\boldsymbol{\mathcal{S}}.
\end{align}
\textcolor{black}{Let $\boldsymbol{\mathcal{S}}_{\mathcal{G}} \triangleq \{\mathcal{S}\in\boldsymbol{\mathcal{S}} | \mathcal{S}\subseteq\mathcal{G} \} $}. Then, for all $\mathcal{G}\in\boldsymbol{\mathcal{G}}$, we re-assemble the sub-message units 
to form a transmission unit $\widetilde{\mathcal{P}}_{\mathcal{G}}$ with rate:
\begin{align}
&\widetilde{R}_{\mathcal{G}} = \sum\nolimits_{\mathcal{S}\in\boldsymbol{\mathcal{S}}_{\mathcal{G}}}
R_{\mathcal{S},\mathcal{G}},
~\mathcal{G}\in\boldsymbol{\mathcal{G}}.\label{eq:rate_combine}
\end{align}
\textcolor{black}{That is, we first split $|\boldsymbol{\mathcal{S}}|$ message units, $\mathcal{P}_{\mathcal{S}},\mathcal{S}\in\boldsymbol{\mathcal{S}}$, into $\sum_{\mathcal{S}\in\boldsymbol{\mathcal{S}}}2^{K-|\mathcal{S}|}$ sub-message units
and then we re-assemble these sub-message units to form $|\boldsymbol{\mathcal{G}}|$ transmission units, $\widetilde{\mathcal{P}}_{\mathcal{G}},\mathcal{G}\in\boldsymbol{\mathcal{G}}$.}
\begin{Exam}[Illustration of $\boldsymbol{\mathcal{G}}$ and General Rate Splitting for Two-User Case]
\textcolor{black}{For Example 1, we have $\mathcal{G}_{\{1\}} = \{\{1\}, \{1,2\}\}$, $\mathcal{G}_{\{2\}} = \{\{2\}, \{1,2\}\}$, $\boldsymbol{\mathcal{G}} = \{\{1\}, \{2\}, \{1,2\}\}$. As shown in Fig. \ref{system_model} (a), 
we first split 3 message units into 5 sub-message units and then re-assemble the 5 sub-message units to form 3 transmission units.}
\end{Exam}

\begin{Exam}[Illustration of $\boldsymbol{\mathcal{G}}$ and General Rate Splitting for \textcolor{black}{Three-User Case}]
For Example 2, we have $\mathcal{G}_{\{1\}} = \{\{1\}, \{1,2\}, \{1,3\}, \{1,2,3\}\}$, $\mathcal{G}_{\{2\}} = \{\{2\}, \{1,2\}, \{2,3\}, \{1,2,3\}\}$, $\mathcal{G}_{\{3\}} = \{\{3\}, \{1,3\}, \{2,3\}, \{1,2,3\}\}$, $\mathcal{G}_{\{1,2\}} = \{\{1,2\}, \{1,2,3\}\}$, $\mathcal{G}_{\{1,3\}} = \{\{1,3\}, \{1,2,3\}\}$, $\boldsymbol{\mathcal{G}} = \{\{1\}, \{2\}, \{3\}, \{1,2\}, \{1,3\}, \{2,3\}, \{1,2,3\}\}$. As shown in \textcolor{black}{Fig. \ref{system_model} (b)}, 
we first split 6 message units into 17 sub-message units and then re-assemble the 17 sub-message units to form 7 transmission units.
\end{Exam}

\begin{Rem}[Connection with Rate Splitting for Unicast and Multicast]
(\romannumeral1) When general multicast \textcolor{black}{degrades to} unicast, the proposed general rate splitting reduces to the general rate splitting for unicast proposed in our previous work \cite{JSAC21}, which extends \textcolor{black}{the} one-layer rate splitting for unicast \cite{TCOM16}. \textcolor{black}{(\romannumeral2) When general multicast degrades to unicast with a common message, the proposed general rate splitting reduces to the one-layer rate splitting for unicast with a common message\cite{TCOM19}.} (\romannumeral3) When general multicast \textcolor{black}{degrades to} single-group multicast, the proposed general rate splitting reduces to the conventional single-group multicast transmission\textcolor{black}{\cite{TSP06}}. \textcolor{black}{(\romannumeral4) When general multicast degrades to multi-group multicast, the proposed general rate splitting reduces to the one-layer rate splitting for multi-group multicast \cite{TWC17}.}
\end{Rem}

\subsection{Physical Layer Model}
\textcolor{black}{The BS is equipped with $M$ antennas, and each user has one antenna.} We consider a multi-carrier system. Let $N$ and $\mathcal{N} \triangleq\{1,2, \ldots, N\}$ denote the number of subcarriers and the set of subcarrier indices, respectively. The bandwidth of each subcarrier is $B$ (in Hz). We consider a discrete-time system, i.e., time is divided into fixed-length slots. We adopt the block fading model, i.e., for each user and subcarrier, the channel remains constant within each slot and \textcolor{black}{is independent and identically distributed (i.i.d.) over slots.} Let $\boldsymbol{\vartheta}_{k,n} \in\boldsymbol{\mathcal{H}}$ denote the $M$-dimensional random channel vector for user $k$ and subcarrier $n$, where \textcolor{black}{$\boldsymbol{\mathcal{H}}\subseteq \mathbb{C}^{M}$} denotes the $M$-dimensional channel state space. Let $\boldsymbol{\vartheta}\triangleq (\boldsymbol{\vartheta}_{k,n})_{k\in\mathcal{K},n\in\mathcal{N}}\in\boldsymbol{\mathcal{H}}^{KN}$ denote the random system channel state, where $\boldsymbol{\mathcal{H}}^{KN}$ represents the system channel state space. Besides, let $\mathbf{h}\triangleq (\mathbf{h}_{k,n})_{k\in\mathcal{K},n\in\mathcal{N}}$ denote a realization of $\boldsymbol{\vartheta}$ in one slot, where $\mathbf{h}_{k,n}$ is a realization of $\boldsymbol{\vartheta}_{k,n}$. Assume that user $k\in\mathcal{K}$ knows his channel state $\mathbf{h}_{k}\triangleq (\mathbf{h}_{k,n})_{n\in\mathcal{N}}$ in each slot. \textcolor{black}{We consider two channel models which are presented below}.
\subsubsection{Slow Fading}
We assume that the system channel state is known to the BS in each slot and adopt transmission rate adaptation over slots. We consider channel coding within each slot and across the $N$ subcarriers. \textcolor{black}{For the sake of exposition,} we consider an arbitrary slot with system channel state $\mathbf{h}$ and \textcolor{black}{any user group $\mathcal{S}\in\boldsymbol{\mathcal{S}}$}. Let $\mathcal{P}_{\mathcal{S}}(\mathbf{h})$ denote the component of $\mathcal{P}_{\mathcal{S}}$ to be transmitted to user group $\mathcal{S}$ in the slot, \textcolor{black}{also referred to as message unit,} and $R_{\mathcal{S}}(\mathbf{h})$ denote the corresponding \textcolor{black}{(transmission)} rate of $\mathcal{P}_{\mathcal{S}}(\mathbf{h})$. Then, the average rate of message unit $\mathcal{P}_{\mathcal{S}}$ is given by $\mathbb{E}[R_{\mathcal{S}}(\boldsymbol{\vartheta})]$. We apply the general rate splitting scheme presented in Section~\ref{sec:rs_model} to transmit $\mathcal{P}_{\mathcal{S}}(\mathbf{h}),\mathcal{S}\in\boldsymbol{\mathcal{S}}$ to all user groups \textcolor{black}{in $\boldsymbol{\mathcal{S}}$} in the slot.
Specifically, for all $\mathcal{S}\in\boldsymbol{\mathcal{S}}$, message unit $\mathcal{P}_{\mathcal{S}}(\mathbf{h})$ is split into $2^{K - |\mathcal{S}|}$ sub-message units. 
The rate $R_{\mathcal{S}}(\mathbf{h})$ of $\mathcal{P}_{\mathcal{S}}(\mathbf{h})$ and the rates of the sub-message unit, denoted by $R_{\mathcal{S},\mathcal{G}}(\mathbf{h}),\mathcal{G}\in\boldsymbol{\mathcal{G}}_{\mathcal{S}}$, satisfy:
\begin{align}
&R_{\mathcal{S}}(\mathbf{h}) = \sum\nolimits_{\mathcal{G}\in\boldsymbol{\mathcal{G}}_{\mathcal{S}}}\label{eq:rate_split_perslot}
R_{\mathcal{S},\mathcal{G}}(\mathbf{h}),~\mathcal{S}\in\boldsymbol{\mathcal{S}},~\mathbf{h}\in\boldsymbol{\mathcal{H}}^{KN}.
\end{align}
Then, for all $\mathcal{G}\in\boldsymbol{\mathcal{G}}$, we re-assemble sub-message units 
to form transmission unit \textcolor{black}{$\widetilde{\mathcal{P}}_{\mathcal{G}}(\mathbf{h}) $} with rate:
\begin{align}
&\widetilde{R}_{\mathcal{G}}(\mathbf{h}) = \sum\nolimits_{\mathcal{S}\in\boldsymbol{\mathcal{S}}_{\mathcal{G}}}
R_{\mathcal{S},\mathcal{G}}(\mathbf{h}),
~\mathcal{G}\in\boldsymbol{\mathcal{G}},~\mathbf{h}\in\boldsymbol{\mathcal{H}}^{KN}.\label{eq:rate_combine_perslot}
\end{align}

For all $\mathcal{G}\in\boldsymbol{\mathcal{G}}$, transmission unit $\widetilde{\mathcal{P}}_{\mathcal{G}}(\mathbf{h})$ is encoded (channel coding) into codewords that span over the $N$ subcarriers. Let \textcolor{black}{$s_{\mathcal{G},n}(\mathbf{h}) \in \mathbb{C}$} denote a symbol for $\widetilde{\mathcal{P}}_{\mathcal{G}}(\mathbf{h})$ that is transmitted on the $n$-th subcarrier. For all $n\in\mathcal{N}$, let $\mathbf{s}_{n}(\mathbf{h}) \triangleq (s_{\mathcal{G},n}(\mathbf{h}))_{\mathcal{G}\in\boldsymbol{\mathcal{G}}}$, and assume that $\mathbb{E}[\mathbf{s}_{n}(\mathbf{h})\mathbf{s}_{n}^{H}(\mathbf{h})] = \mathbf{I}$. \textcolor{black}{We consider linear beamforming.} \textcolor{black}{For all $n\in\mathcal{N}$,} let $\mathbf{w}_{\mathcal{G},n}(\mathbf{h})\in\mathbb{C}^{M\times 1}$ denote the beamforming vector for \textcolor{black}{transmitting} $\widetilde{\mathcal{P}}_{\mathcal{G}}(\mathbf{h})$ on subcarrier $n$ \textcolor{black}{in the slot}. \textcolor{black}{Using superposition coding,} the transmitted signal on subcarrier $n$, \textcolor{black}{denoted by $\mathbf{x}_{n}(\mathbf{h})\in\mathbb{C}^{M\times 1}$,} is given by:
\begin{align}
\mathbf{x}_{n}(\mathbf{h})=\sum\nolimits_{\mathcal{G}\in\boldsymbol{\mathcal{G}}}
\mathbf{w}_{\mathcal{G},n}(\mathbf{h})s_{\mathcal{G},n}(\mathbf{h}),~n\in\mathcal{N},~\mathbf{h}\in\boldsymbol{\mathcal{H}}^{KN}.\label{eq:signal_slow_fading}
\end{align}
The transmission power on subcarrier $n\in\mathcal{N}$ is given by $\sum_{\mathcal{G} \in \boldsymbol{\mathcal{G}}}\| \mathbf{w}_{\mathcal{G},n}(\mathbf{h}) \|^2_{2}$, and the total transmission power \textcolor{black}{is given by} $\sum_{n \in \mathcal{N}}\sum_{\mathcal{G} \in \boldsymbol{\mathcal{G}}}\| \mathbf{w}_{\mathcal{G},n}(\mathbf{h}) \|^2_{2}$. \textcolor{black}{In the slow fading scenario, the total transmission power constraint is given by:}
\begin{align}
 \sum\nolimits_{n \in \mathcal{N}}\sum\nolimits_{\mathcal{G} \in \boldsymbol{\mathcal{G}}}\| \mathbf{w}_{\mathcal{G},n}(\mathbf{h}) \|^2_{2} \leq P,~\mathbf{h}\in\boldsymbol{\mathcal{H}}^{KN}. \label{eq:power_perslot}
\end{align}
Here, $P$ denotes the transmission power budget. \textcolor{black}{Define $\boldsymbol{\mathcal{G}}^{(k)} \triangleq \{\mathcal{G} \in \boldsymbol{\mathcal{G}} | k\in\mathcal{G} \}, k\in\mathcal{K}$.}
Then, the received signal at user $k\in\mathcal{K}$ on subcarrier $n\in\mathcal{N}$, \textcolor{black}{denoted by $y_{k,n}(\mathbf{h})\in\mathbb{C}$}, is given by:
\begin{align}
y_{k,n}(\mathbf{h})= & \mathbf{h}^{H}_{k,n}\mathbf{x}_{n}(\mathbf{h}) + z_{k,n} =  \mathbf{h}_{k,n}^H \sum\nolimits_{\mathcal{G}\in\boldsymbol{\mathcal{G}}^{(k)}}
\mathbf{w}_{\mathcal{G},n}(\mathbf{h})s_{\mathcal{G},n}(\mathbf{h})\nonumber\\
&+\mathbf{h}_{k,n}^H\sum\nolimits_{\mathcal{G}'\in\boldsymbol{\mathcal{G}}\backslash\boldsymbol{\mathcal{G}}^{(k)}} \mathbf{w}_{\mathcal{G}',n}(\mathbf{h})s_{\mathcal{G}',n}(\mathbf{h}) +z_{k,n},~k \in \mathcal{K},~n\in\mathcal{N},\mathbf{h}\in\boldsymbol{\mathcal{H}}^{KN},\label{eq:receive_signal}
\end{align}
where \textcolor{black}{the last equality is due to \eqref{eq:signal_slow_fading}, and }$z_{k,n}\sim C\mathcal{N}(0,\sigma^2)$ is the additive white gaussian noise (AWGN). \textcolor{black}{In \eqref{eq:receive_signal}, the first term represents the desired signal, and the second represents the interference.} \textcolor{black}{It is noteworthy that} the main idea of rate splitting is to make the \textcolor{black}{undesired} messages partially decodable in order to reduce interference \cite{JSAC21}.
To exploit the full potential of the general rate splitting for general multicast, we consider joint decoding \textcolor{black}{at each user}.\footnote{We can easily extend it to successive decoding as in \cite{JSAC21}.} That is, each user $k\in\mathcal{K}$ jointly decodes the desired transmission units $\widetilde{\mathcal{P}}_{\mathcal{G}}(\mathbf{h}),\mathcal{G}\in\boldsymbol{\mathcal{G}}^{(k)}$. Thus, in the slow fading scenario, the achievable rate region of the transmission units in the slot with system channel state $\mathbf{h}$ is described by the following constraints:
\begin{align}
\sum\limits_{\mathcal{G}\in\boldsymbol{\mathcal{X}}}\widetilde{R}_{\mathcal{G}}(\mathbf{h}) \leq B\sum\limits_{n\in\mathcal{N}}\log_{2}\left(1+\frac{\sum\nolimits_{\mathcal{G}\in\boldsymbol{\mathcal{X}}}|\mathbf{h}^{H}_{k,n}\mathbf{w}_{\mathcal{G},n}(\mathbf{h})|^{2}}{\sigma^2 +\sum\nolimits_{\mathcal{G}'\in\boldsymbol{\mathcal{G}}\backslash\boldsymbol{\mathcal{G}}^{(k)}}|\mathbf{h}^{H}_{k,n}\mathbf{w}_{\mathcal{G}^{'},n}(\mathbf{h})|^{2} }\right),~\boldsymbol{\mathcal{X}}\subseteq\boldsymbol{\mathcal{G}}^{(k)},k \in \mathcal{K},\mathbf{h}\in\boldsymbol{\mathcal{H}}^{KN},\label{eq:rateconstraints_perslot}
\end{align}
\textcolor{black}{where $\widetilde{R}_{\mathcal{G}}(\mathbf{h})$ is given by \eqref{eq:rate_combine_perslot}.} 

\textcolor{black}{
\begin{Exam}[Illustration of Physical Layer Design for Two-user Case in Slow Fading]For Example~1, the rates of message units $\mathcal{P}_{\{1\}}(\mathbf{h})$, $\mathcal{P}_{\{2\}}(\mathbf{h})$, and $\mathcal{P}_{\{1,2\}}(\mathbf{h})$ are $R_{\{1\}}(\mathbf{h})$, $R_{\{2\}}(\mathbf{h})$, and $R_{\{1,2\}}(\mathbf{h})$, respectively. 
The rates of transmission units $\tilde{\mathcal{P}}_{\{1\}}(\mathbf{h})$, $\tilde{\mathcal{P}}_{\{2\}}(\mathbf{h})$, and $\tilde{\mathcal{P}}_{\{1,2\}}(\mathbf{h})$ are $\tilde{R}_{\{1\}}(\mathbf{h}) = R_{\{1\},\{1\}}(\mathbf{h})$, $\tilde{R}_{\{2\}}(\mathbf{h}) = R_{\{2\},\{2\}}(\mathbf{h})$, and $\tilde{R}_{\{1,2\}}(\mathbf{h}) = R_{\{1,2\}}(\mathbf{h}) + R_{\{1\},\{1,2\}}(\mathbf{h}) + R_{\{2\},\{1,2\}}(\mathbf{h})$, respectively. 
The beamforming vectors for transmitting $\tilde{\mathcal{P}}_{\{1\}}(\mathbf{h})$, $\tilde{\mathcal{P}}_{\{2\}}(\mathbf{h})$, and $\tilde{\mathcal{P}}_{\{1,2\}}(\mathbf{h})$ on subcarrier $n\in\mathcal{N}$ are $\mathbf{w}_{\{1\},n}(\mathbf{h})$, $\mathbf{w}_{\{2\},n}(\mathbf{h})$, and $\mathbf{w}_{\{1,2\},n}(\mathbf{h})$,  respectively.
\end{Exam}
\begin{Exam}[Illustration of Physical Layer Design for Three-user Case in Slow Fading]
For Example~2, the rates of message units $\mathcal{P}_{\{1\}}(\mathbf{h})$, $\mathcal{P}_{\{2\}}(\mathbf{h})$, $\mathcal{P}_{\{3\}}(\mathbf{h})$, $\mathcal{P}_{\{1,2\}}(\mathbf{h})$, $\mathcal{P}_{\{1,3\}}(\mathbf{h})$, and $\mathcal{P}_{\{1,2,3\}}(\mathbf{h})$ are $R_{\{1\}}(\mathbf{h})$, $R_{\{2\}}(\mathbf{h})$, $R_{\{3\}}(\mathbf{h})$, $R_{\{1,2\}}(\mathbf{h})$, $R_{\{1,3\}}(\mathbf{h})$, and $R_{\{1,2,3\}}(\mathbf{h})$, respectively. 
The rates of transmission units $\tilde{\mathcal{P}}_{\{1\}}(\mathbf{h})$, $\tilde{\mathcal{P}}_{\{2\}}(\mathbf{h})$, $\tilde{\mathcal{P}}_{\{3\}}(\mathbf{h})$, $\tilde{\mathcal{P}}_{\{1,2\}}(\mathbf{h})$, $\tilde{\mathcal{P}}_{\{1,3\}}(\mathbf{h})$, $\tilde{\mathcal{P}}_{\{2,3\}}(\mathbf{h})$, and $\tilde{\mathcal{P}}_{\{1,2,3\}}(\mathbf{h})$ are $\tilde{R}_{\{1\}}(\mathbf{h}) = R_{\{1\},\{1\}}(\mathbf{h})$, $\tilde{R}_{\{2\}}(\mathbf{h}) = R_{\{2\},\{2\}}(\mathbf{h})$, $\tilde{R}_{\{3\}}(\mathbf{h}) = R_{\{3\},\{3\}}(\mathbf{h})$, $\tilde{R}_{\{1,2\}}(\mathbf{h}) = R_{\{1,2\},\{1,2\}}(\mathbf{h}) + R_{\{1\},\{1,2\}}(\mathbf{h}) + R_{\{2\},\{1,2\}}(\mathbf{h})$, $\tilde{R}_{\{1,3\}}(\mathbf{h}) = R_{\{1,3\},\{1,3\}}(\mathbf{h}) + R_{\{1\},\{1,3\}}(\mathbf{h}) + R_{\{3\},\{1,3\}}(\mathbf{h})$, $\tilde{R}_{\{2,3\}}(\mathbf{h}) = R_{\{2\},\{2,3\}}(\mathbf{h}) + R_{\{3\},\{2,3\}}(\mathbf{h})$, and $\tilde{R}_{\{1,2,3\}}(\mathbf{h}) = R_{\{1,2,3\}}(\mathbf{h}) + R_{\{1,3\},\{1,2,3\}}(\mathbf{h}) + R_{\{1,2\},\{1,2,3\}}(\mathbf{h}) + R_{\{1\},\{1,2,3\}}(\mathbf{h}) + R_{\{2\},\{1,2,3\}}(\mathbf{h}) + R_{\{3\},\{1,2,3\}}(\mathbf{h})$, respectively. 
The beamforming vectors for transmitting $\tilde{\mathcal{P}}_{\{1\}}(\mathbf{h})$, $\tilde{\mathcal{P}}_{\{2\}}(\mathbf{h})$, $\tilde{\mathcal{P}}_{\{3\}}(\mathbf{h})$, $\tilde{\mathcal{P}}_{\{1,2\}}(\mathbf{h})$, $\tilde{\mathcal{P}}_{\{1,3\}}(\mathbf{h})$, $\tilde{\mathcal{P}}_{\{2,3\}}(\mathbf{h})$, and $\tilde{\mathcal{P}}_{\{1,2,3\}}(\mathbf{h})$ on subcarrier $n\in\mathcal{N}$ are $\mathbf{w}_{\{1\},n}(\mathbf{h})$, $\mathbf{w}_{\{2\},n}(\mathbf{h})$, $\mathbf{w}_{\{3\},n}(\mathbf{h})$, $\mathbf{w}_{\{1,2\},n}(\mathbf{h})$, $\mathbf{w}_{\{1,3\},n}(\mathbf{h})$, $\mathbf{w}_{\{2,3\},n}(\mathbf{h})$, and $\mathbf{w}_{\{1,2,3\},n}(\mathbf{h})$,  respectively.
\end{Exam}
}

\subsubsection{Fast Fading}We assume that the system channel state is unknown to the BS in each slot, but its distribution is known to the BS. We consider channel coding over many slots and across the $N$ subcarriers. 
We apply the general rate splitting scheme presented in Section~\ref{sec:rs_model} to transmit $\mathcal{P}_{\mathcal{S}},\mathcal{S}\in\boldsymbol{\mathcal{S}}$ to all user groups in $\boldsymbol{\mathcal{S}}$ over many slots. For all $\mathcal{G}\in\boldsymbol{\mathcal{G}}$, transmission unit $\widetilde{\mathcal{P}}_{\mathcal{G}}$ is encoded into codewords that span over the $N$ subcarriers and many slots. Let \textcolor{black}{$s_{\mathcal{G},n}\in\mathbb{C}$} denote a symbol for $\widetilde{\mathcal{P}}_{\mathcal{G}}$ that is transmitted on the $n$-th subcarrier. Let $\mathbf{w}_{\mathcal{G},n}\in\mathbb{C}^{M\times 1}$ denote the constant beamforming vector for \textcolor{black}{transmitting} $\widetilde{\mathcal{P}}_{\mathcal{G}}$ on subcarrier $n$ over these slots. \textcolor{black}{Similarly, using superposition coding,} the transmitted signal on subcarrier $n$ is given by:
\begin{align}
\mathbf{x}_{n}=\sum\nolimits_{\mathcal{G}\in\boldsymbol{\mathcal{G}}}
\mathbf{w}_{\mathcal{G},n}s_{\mathcal{G},n},~n\in\mathcal{N}.\label{eq:signal_fast_fading}
\end{align}
The total transmission power \textcolor{black}{is given by $\sum_{n \in \mathcal{N}}\sum_{\mathcal{G} \in \boldsymbol{\mathcal{G}}}\| \mathbf{w}_{\mathcal{G},n} \|^2_{2}$. In the fast fading scenario,} the total transmission power constraint is given by:
\begin{align}
 \sum\nolimits_{n \in \mathcal{N}}\sum\nolimits_{\mathcal{G} \in \boldsymbol{\mathcal{G}}}\| \mathbf{w}_{\mathcal{G},n} \|^2_{2} \leq P. \label{eq:power}
\end{align}
Then, in a slot with system channel state $\mathbf{h}$, the received signal at user $k$ on subcarrier $n$, \textcolor{black}{denoted by $y_{k,n}\in\mathbb{C}$}, is given by:
\begin{align}
y_{k,n}& = \mathbf{h}^{H}_{k,n}\mathbf{x}_{n} + z_{k,n} \nonumber\\
&=  \mathbf{h}_{k,n}^H \sum\nolimits_{\mathcal{G}\in\boldsymbol{\mathcal{G}}^{(k)}}
\mathbf{w}_{\mathcal{G},n}s_{\mathcal{G},n}+
\mathbf{h}_{k,n}^H\sum\nolimits_{\mathcal{G}'\in\boldsymbol{\mathcal{G}}\backslash\boldsymbol{\mathcal{G}}^{(k)}} \mathbf{w}_{\mathcal{G}',n}s_{\mathcal{G}',n} +z_{k,n},~k \in \mathcal{K},~n\in\mathcal{N},\nonumber
\end{align}
\textcolor{black}{where the last equality is due to \eqref{eq:signal_fast_fading}}. Similarly, we consider joint decoding \textcolor{black}{at each user}. The achievable \textcolor{black}{ergodic} rate region of the transmission units over these slots is given by:
\begin{align}
\sum\limits_{\mathcal{G}\in\boldsymbol{\mathcal{X}}}\widetilde{R}_{\mathcal{G}} \leq B\sum\limits_{n\in\mathcal{N}}\mathbb{E}\left[\log_{2}\left(1+\frac{\sum\nolimits_{\mathcal{G}\in\boldsymbol{\mathcal{X}}}|\boldsymbol{\vartheta}^{H}_{k,n}\mathbf{w}_{\mathcal{G},n}|^{2}}{\sigma^2 +\sum\nolimits_{\mathcal{G}'\in\boldsymbol{\mathcal{G}}\backslash\boldsymbol{\mathcal{G}}^{(k)}}|\boldsymbol{\vartheta}^{H}_{k,n}\mathbf{w}_{\mathcal{G}^{'},n}|^{2} }\right)\right],~\boldsymbol{\mathcal{X}}\subseteq\boldsymbol{\mathcal{G}}^{(k)},k \in \mathcal{K},\label{eq:rateconstraints_overslots}
\end{align}
\textcolor{black}{where $\widetilde{R}_{\mathcal{G}}$ is given by \eqref{eq:rate_combine}.}

\textcolor{black}{
\begin{Exam}[Illustration of Physical Layer Design for Two-user Case in Fast Fading]
For Example~1, the rates of message units $\mathcal{P}_{\{1\}}$, $\mathcal{P}_{\{2\}}$, and $\mathcal{P}_{\{1,2\}}$ are $R_{\{1\}}$, $R_{\{2\}}$, and $R_{\{1,2\}}$, respectively. 
The rates of transmission units $\tilde{\mathcal{P}}_{\{1\}}$, $\tilde{\mathcal{P}}_{\{2\}}$, and $\tilde{\mathcal{P}}_{\{1,2\}}$ are $\tilde{R}_{\{1\}} = R_{\{1\},\{1\}}$, $\tilde{R}_{\{2\}} = R_{\{2\},\{2\}}$, and $\tilde{R}_{\{1,2\}} = R_{\{1,2\}} + R_{\{1\},\{1,2\}} + R_{\{2\},\{1,2\}}$, respectively. 
The beamforming vectors for transmitting $\tilde{\mathcal{P}}_{\{1\}}$, $\tilde{\mathcal{P}}_{\{2\}}$, and $\tilde{\mathcal{P}}_{\{1,2\}}$ on subcarrier $n\in\mathcal{N}$ are $\mathbf{w}_{\{1\},n}$, $\mathbf{w}_{\{2\},n}$, and $\mathbf{w}_{\{1,2\},n}$, respectively.
\end{Exam}
\begin{Exam}[Illustration of Physical Layer Design for Three-user Case in Fast Fading]
For Example~2, the rates of message units $\mathcal{P}_{\{1\}}$, $\mathcal{P}_{\{2\}}$, $\mathcal{P}_{\{3\}}$, $\mathcal{P}_{\{1,2\}}$, $\mathcal{P}_{\{1,3\}}$, and $\mathcal{P}_{\{1,2,3\}}$ are $R_{\{1\}}$, $R_{\{2\}}$, $R_{\{3\}}$, $R_{\{1,2\}}$, $R_{\{1,3\}}$, and $R_{\{1,2,3\}}$, respectively. 
The rates of transmission units $\tilde{\mathcal{P}}_{\{1\}}$, $\tilde{\mathcal{P}}_{\{2\}}$, $\tilde{\mathcal{P}}_{\{3\}}$, $\tilde{\mathcal{P}}_{\{1,2\}}$, $\tilde{\mathcal{P}}_{\{1,3\}}$, $\tilde{\mathcal{P}}_{\{2,3\}}$, and $\tilde{\mathcal{P}}_{\{1,2,3\}}$ are $\tilde{R}_{\{1\}} = R_{\{1\},\{1\}}$, $\tilde{R}_{\{2\}} = R_{\{2\},\{2\}}$, $\tilde{R}_{\{3\}} = R_{\{3\},\{3\}}$, $\tilde{R}_{\{1,2\}} = R_{\{1,2\},\{1,2\}} + R_{\{1\},\{1,2\}} + R_{\{2\},\{1,2\}}$, $\tilde{R}_{\{1,3\}} = R_{\{1,3\},\{1,3\}} + R_{\{1\},\{1,3\}} + R_{\{3\},\{1,3\}}$, $\tilde{R}_{\{2,3\}} = R_{\{2\},\{2,3\}} + R_{\{3\},\{2,3\}}$, and $\tilde{R}_{\{1,2,3\}} = R_{\{1,2,3\}} + R_{\{1,3\},\{1,2,3\}} + R_{\{1,2\},\{1,2,3\}} + R_{\{1\},\{1,2,3\}} + R_{\{2\},\{1,2,3\}} + R_{\{3\},\{1,2,3\}}$, respectively. 
The beamforming vectors for transmitting $\tilde{\mathcal{P}}_{\{1\}}$, $\tilde{\mathcal{P}}_{\{2\}}$, $\tilde{\mathcal{P}}_{\{3\}}$, $\tilde{\mathcal{P}}_{\{1,2\}}$, $\tilde{\mathcal{P}}_{\{1,3\}}$, $\tilde{\mathcal{P}}_{\{2,3\}}$, and $\tilde{\mathcal{P}}_{\{1,2,3\}}$ on subcarrier $n\in\mathcal{N}$ are $\mathbf{w}_{\{1\},n}$, $\mathbf{w}_{\{2\},n}$, $\mathbf{w}_{\{3\},n}$, $\mathbf{w}_{\{1,2\},n}$, $\mathbf{w}_{\{1,3\},n}$, $\mathbf{w}_{\{2,3\},n}$, and $\mathbf{w}_{\{1,2,3\},n}$, respectively.
\end{Exam}
}

\begin{figure}[t]
\begin{center}
 {\resizebox{10cm}{!}{\includegraphics{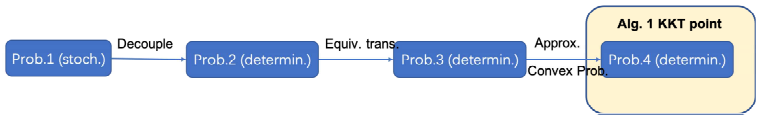}}}
\end{center}
   \caption{\small{Illustration of the proposed solution framework for the slow fading scenario. ``determin.'' and ``stoch.'' are short for ``deterministic'' and ``stochastic'', respectively. ``trans.'', ``Equiv.'', ``Approx.'', and ``Prob.'' are short for ``transformation'', ``Equivalent'', ``Approximation'', and ``Problem'', respectively.}}
   \label{fig:frame_slow}
\end{figure}

\section{Average Rate Maximization In Slow Fading Scenario}\label{sec:slowfading}
This section considers the slow fading scenario and maximizes the weighted sum average rate under the \textcolor{black}{achievable} rate constraints and total transmission power constraints. 
The proposed solution framework for the slow fading scenario is illustrated in Fig.~\ref{fig:frame_slow}.
\subsection{Optimization Problem Formulation}
We would like to optimize the transmission beamforming vectors $\mathbf{w}(\mathbf{h}) \triangleq (\mathbf{w}_{\mathcal{G},n}(\mathbf{h}))_{\mathcal{G}\in\boldsymbol{\mathcal{G}},n\in\mathcal{N}},\mathbf{h}\in\boldsymbol{\mathcal{H}}^{KN}$ and rates of the sub-message units $\mathbf{R}(\mathbf{h}) \triangleq (R_{\mathcal{S},\mathcal{G}}(\mathbf{h}))_{\mathcal{S}\in\boldsymbol{\mathcal{S}}, \mathcal{G}\in\boldsymbol{\mathcal{G}}},\mathbf{h}\in\boldsymbol{\mathcal{H}}^{KN}$ to maximize the \textcolor{black}{weighted sum average rate},\footnote{\textcolor{black}{The proposed problem formulation and solution method can be readily extended to maximize the sum average rate and worst-case average rate as in \cite{JSAC21}.}} $\sum\nolimits_{\mathcal{S}\in\boldsymbol{\mathcal{S}}} \alpha_{\mathcal{S}} \mathbb{E}\left[R_{\mathcal{S}}(\boldsymbol{\vartheta})\right]$, \textcolor{black}{where the coefficient $\alpha_{\mathcal{S}} \geq 0$ denotes the weight for message unit $\mathcal{P}_{\mathcal{S}}$, subject to the total transmission power constraints in \eqref{eq:power_perslot} and the achievable rate constraints in \eqref{eq:rateconstraints_perslot}.}
Therefore, we formulate the following optimization problem.

\begin{Prob}[Weighted Sum Average Rate Maximization]\label{prob:UMwT_LMsg_1slot}
\begin{align}
U^{(\text{slow})\star} \triangleq \max_{(\mathbf{w}(\mathbf{h}))_{\mathbf{h}\in\mathcal{H}^{KN}},(\mathbf{R}(\mathbf{h}))_{\mathbf{h}\in\mathcal{H}^{KN}}\succeq 0}\quad & 
\sum\nolimits_{\mathcal{S}\in\boldsymbol{\mathcal{S}}} \alpha_{\mathcal{S}} \mathbb{E}\left[R_{\mathcal{S}}(\boldsymbol{\vartheta})\right]\nonumber\\
\mathrm{s.t.}\quad &\eqref{eq:power_perslot},~\eqref{eq:rateconstraints_perslot}, \notag
\end{align}
\textcolor{black}{where $R_{\mathcal{S}}(\cdot)$ is given by \eqref{eq:rate_split_perslot}.}
\end{Prob}

As the objective function and constraints of Problem~\ref{prob:UMwT_LMsg_1slot} are seperable with $\mathbf{h}\in\mathcal{H}^{KN}$, Problem~\ref{prob:UMwT_LMsg_1slot} is seperable and can be readily solved by solving the following problems for all $\mathbf{h}\in\mathcal{H}^{KN}$ in parallel.


\begin{Prob}[Weighted Sum Rate Maximization for $\mathbf{h}$]\label{prob:UMwT_LMsg_1slot_decouple}
\begin{align}
&\max_{\mathbf{w}(\mathbf{h}),\mathbf{R}(\mathbf{h})\succeq 0}\quad  
\sum\limits_{\mathcal{S}\in\boldsymbol{\mathcal{S}}} \alpha_{\mathcal{S}} R_{\mathcal{S}}(\mathbf{h})\nonumber\\
&\mathrm{s.t.}\quad \sum_{n \in \mathcal{N}}\sum_{\mathcal{G} \in \boldsymbol{\mathcal{G}}}\| \mathbf{w}_{\mathcal{G},n}(\mathbf{h}) \|^2_{2} \leq P, \label{eq:power_perslot_decouple}\\
& \sum\limits_{\mathcal{G}\in\boldsymbol{\mathcal{X}}}\widetilde{R}_{\mathcal{G}}(\mathbf{h}) \leq B\sum\limits_{n\in\mathcal{N}}\log_{2}\left(1+\frac{\sum\nolimits_{\mathcal{G}\in\boldsymbol{\mathcal{X}}}|\mathbf{h}^{H}_{k,n}\mathbf{w}_{\mathcal{G},n}(\mathbf{h})|^{2}}{\sigma^2 +\sum\nolimits_{\mathcal{G}'\in\boldsymbol{\mathcal{G}}\backslash\boldsymbol{\mathcal{G}}^{(k)}}|\mathbf{h}^{H}_{k,n}\mathbf{w}_{\mathcal{G}^{'},n}(\mathbf{h})|^{2} }\right),~\boldsymbol{\mathcal{X}}\subseteq\boldsymbol{\mathcal{G}}^{(k)},k \in \mathcal{K}.\label{eq:rateconstraints_perslot_decouple}
\end{align}
\end{Prob}

\begin{Rem}[Connection with Rate Splitting for Unicast and Multicast]
(\romannumeral1) When general multicast degrades to unicast, Problem \ref{prob:UMwT_LMsg_1slot_decouple} reduces to the weighted sum rate maximization problem \textcolor{black}{for general rate splitting for unicast} in \cite{JSAC21}. \textcolor{black}{(\romannumeral2) When general multicast degrades to unicast with a common message, Problem~\ref{prob:UMwT_LMsg_1slot_decouple} can be viewed as a generalization of the weighted sum rate maximization for unicast with a common message in \cite{TCOM19}}. (\romannumeral3) When general multicast degrades to single-group multicast, Problem~\ref{prob:UMwT_LMsg_1slot_decouple} reduces to the rate maximization problem for single-group multicast in \cite{TSP06}. \textcolor{black}{(\romannumeral4) When general multicast degrades to multi-group multicast, Problem~\ref{prob:UMwT_LMsg_1slot_decouple} \textcolor{black}{can be viewed as a generalization of} the weighted sum rate maximization for multi-group multicast in \cite{TWC17}. }
\end{Rem}

Therefore, we can focus on solving Problem~\ref{prob:UMwT_LMsg_1slot_decouple} \textcolor{black}{for all $\mathbf{h}$.} Note that the objective function is linear, the constraint in \eqref{eq:power_perslot_decouple} is convex, and the constraints in \eqref{eq:rateconstraints_perslot_decouple} are nonconvex. Thus, Problem~\ref{prob:UMwT_LMsg_1slot_decouple} is nonconvex.\footnote{\textcolor{black}{There are generally no effective methods for solving a nonconvex problem optimally. The goal of solving a nonconvex problem is usually to design an iterative algorithm to obtain a stationary point or a KKT point (which satisfies necessary conditions for optimality if strong
duality holds).}}

\subsection{KKT Point}\label{subsec_slow_solution}
In this subsection, we propose an iterative algorithm to obtain a KKT point of Problem~\ref{prob:UMwT_LMsg_1slot_decouple} using CCCP. 
First, we transform Problem~\ref{prob:UMwT_LMsg_1slot_decouple} into the following equivalent problem by introducing \textcolor{black}{auxiliary variables $\mathbf{e}(\mathbf{h}) \triangleq \left(e_{k,n,\boldsymbol{\mathcal{X}}}(\mathbf{h})\right)
_{\boldsymbol{\mathcal{X}}\subseteq\boldsymbol{\mathcal{G}}^{(k)},k\in\mathcal{K},n\in\mathcal{N}}$ and
$\mathbf{u}(\mathbf{h}) \triangleq \left(u_{k,n,\boldsymbol{\mathcal{X}}}(\mathbf{h})\right)_{\boldsymbol{\mathcal{X}}\subseteq\boldsymbol{\mathcal{G}}^{(k)},k\in\mathcal{K},n\in\mathcal{N}}$ and extra constraints:}
\begin{align}
&\sum_{\mathcal{G}\in\boldsymbol{\mathcal{X}}}\sum_{\mathcal{S}\in\boldsymbol{\mathcal{S}}_{\mathcal{G}}}
R_{\mathcal{S},\mathcal{G}}(\mathbf{h}) = \sum_{n\in\mathcal{N}} e_{k,n,\boldsymbol{\mathcal{X}}}(\mathbf{h}),~\boldsymbol{\mathcal{X}}\subseteq\boldsymbol{\mathcal{G}}^{(k)},k\in\mathcal{K},\label{eq:DC_R<e_perslot}\\
& 2^{\frac{e_{k,n,\boldsymbol{\mathcal{X}}}(\mathbf{h})}{B}} \leq  u_{k,n,\boldsymbol{\mathcal{X}}}(\mathbf{h}),~\boldsymbol{\mathcal{X}}\subseteq\boldsymbol{\mathcal{G}}^{(k)},k\in\mathcal{K},n\in\mathcal{N}, \label{eq:DC_e<u_perslot}
\end{align}
\begin{align}
& \sum\limits_{\mathcal{G}'\in\boldsymbol{\mathcal{G}}\backslash\boldsymbol{\mathcal{G}}^{(k)}}|\mathbf{h}^{H}_{k,n}\mathbf{w}_{\mathcal{G}^{'},n}(\mathbf{h})|^{2}
+\sigma^2 - \frac{\sum\limits_{\mathcal{G}\in\boldsymbol{\mathcal{X}}}|
\mathbf{h}^{H}_{k,n}\mathbf{w}_{\mathcal{G},n}(\mathbf{h})|^{2}+
\sum\limits_{\mathcal{G}'\in\boldsymbol{\mathcal{G}}\backslash\boldsymbol{\mathcal{G}}^{(k)}}|\mathbf{h}^{H}_{k,n}\mathbf{w}_{\mathcal{G}^{'},n}(\mathbf{h})|^{2}
+\sigma^2}{u_{k,n,\boldsymbol{\mathcal{X}}}(\mathbf{h})} \leq 0,\nonumber\\
&~~~~~~~~~~~~~~~~~~~~~~~~~~~~~~~~~~~~~~~~~~~~~~~~~~~~~~~~~~\boldsymbol{\mathcal{X}}\subseteq\boldsymbol{\mathcal{G}}^{(k)},k\in\mathcal{K},n\in\mathcal{N}.\label{eq:equivalentDCfunction_perslot}
\end{align}

\begin{Prob}[Equivalent DC Problem of Problem \ref{prob:UMwT_LMsg_1slot_decouple}]\label{prob:UMwT_LMsg_1slot_decouple_DC}
\begin{align}
\max_{\mathbf{w}(\mathbf{h}),\mathbf{R}(\mathbf{h})\succeq 0,\mathbf{e}(\mathbf{h}),\mathbf{u}(\mathbf{h})}\quad &\sum\limits_{\mathcal{S}\in\boldsymbol{\mathcal{S}}} \alpha_{\mathcal{S}} \sum_{\mathcal{G}\in\boldsymbol{\mathcal{G}}_{\mathcal{S}}}R_{\mathcal{S},\mathcal{G}}(\mathbf{h})  \nonumber\\
\mathrm{s.t.}\quad &\eqref{eq:power_perslot_decouple},~\eqref{eq:DC_R<e_perslot},~\eqref{eq:DC_e<u_perslot},~\eqref{eq:equivalentDCfunction_perslot}.\nonumber
\end{align}
Let $(\mathbf{w}^{\star}(\mathbf{h}), \mathbf{R}^{\star}(\mathbf{h}), \mathbf{e}^{\star}(\mathbf{h}), \mathbf{u}^{\star}(\mathbf{h}))$ denote an optimal solution of Problem~\ref{prob:UMwT_LMsg_1slot_decouple_DC}.
\end{Prob}

\begin{Lem}[Equivalence Between Problem~\ref{prob:UMwT_LMsg_1slot_decouple} and Problem~\ref{prob:UMwT_LMsg_1slot_decouple_DC}]\label{lemma_DC_transform}
$(\mathbf{w}^{\star}(\mathbf{h}), \mathbf{R}^{\star}(\mathbf{h}), \mathbf{e}^{\star}(\mathbf{h}), \mathbf{u}^{\star}(\mathbf{h}))$ satisfies
\textcolor{black}{$2^{\frac{e^{\star}_{k,n,\boldsymbol{\mathcal{X}}}(\mathbf{h})}{B}} = u^{\star}_{k,n,\boldsymbol{\mathcal{X}}}(\mathbf{h}),~\boldsymbol{\mathcal{X}}\subseteq\boldsymbol{\mathcal{G}}^{(k)},k\in\mathcal{K},n\in\mathcal{N}.$}
Furthermore, Problem~\ref{prob:UMwT_LMsg_1slot_decouple} and Problem~\ref{prob:UMwT_LMsg_1slot_decouple_DC} are equivalent.
\end{Lem}
\textcolor{black}{
\begin{Proof}
First, by introducing auxiliary variables $\mathbf{e}(\mathbf{h})$ and $\mathbf{u}(\mathbf{h})$ and extra constraints:
\begin{align}
2^{\frac{e_{k,n,\boldsymbol{\mathcal{X}}}(\mathbf{h})}{B}} =  u_{k,n,\boldsymbol{\mathcal{X}}}(\mathbf{h}),~\boldsymbol{\mathcal{X}}\subseteq\boldsymbol{\mathcal{G}}^{(k)},k\in\mathcal{K},n\in\mathcal{N}, \label{eq:DC_e<u_perslot_proof}
\end{align}
we can equivalently transform Problem~\ref{prob:UMwT_LMsg_1slot_decouple} into the following problem:
\begin{align}
\max_{\mathbf{w}(\mathbf{h}),\mathbf{R}(\mathbf{h})\succeq 0,\mathbf{e}(\mathbf{h}),\mathbf{u}(\mathbf{h})}\quad &\sum\limits_{\mathcal{S}\in\boldsymbol{\mathcal{S}}} \alpha_{\mathcal{S}} \sum_{\mathcal{G}\in\boldsymbol{\mathcal{G}}_{\mathcal{S}}}R_{\mathcal{S},\mathcal{G}}(\mathbf{h})  \nonumber\\
\mathrm{s.t.}\quad &\eqref{eq:power_perslot_decouple},~\eqref{eq:DC_R<e_perslot},~\eqref{eq:equivalentDCfunction_perslot},~\eqref{eq:DC_e<u_perslot_proof}\nonumber
\end{align}
Let $(\mathbf{w}^{\ddag}(\mathbf{h}),\mathbf{R}^{\ddag}(\mathbf{h}),\mathbf{e}^{\ddag}(\mathbf{h}),\mathbf{u}^{\ddag}(\mathbf{h}))$ denote an optimal solution.
It is obvious that $(\mathbf{w}^{\ddag}(\mathbf{h}),\mathbf{R}^{\ddag}(\mathbf{h}))$ is an optimal solution of Problem \ref{prob:UMwT_LMsg_1slot_decouple}. Next, we transform the above problem to Problem~\ref{prob:UMwT_LMsg_1slot_decouple_DC} by relaxing the constrains in \eqref{eq:DC_e<u_perslot_proof} to the constraints in \eqref{eq:DC_e<u_perslot}. By contradiction and the monotonicity of the objective function with respect to (w.r.t.) $\mathbf{R}$ in Problem~\ref{prob:UMwT_LMsg_1slot_decouple_DC}, we can show that the constraints in \eqref{eq:DC_e<u_perslot} are active at the optimal solution. Thus, $(\mathbf{w}^{\ddag}(\mathbf{h}),\mathbf{R}^{\ddag}(\mathbf{h}),\mathbf{e}^{\ddag}(\mathbf{h}),\mathbf{u}^{\ddag}(\mathbf{h}))$ is an optimal solution of Problem~\ref{prob:UMwT_LMsg_1slot_decouple_DC}. Therefore, we can show Lemma~\ref{lemma_DC_transform}.$\hfill\blacksquare$
\end{Proof}
}

Based on Lemma~\ref{lemma_DC_transform}, solving Problem~\ref{prob:UMwT_LMsg_1slot_decouple} is equivalent to solving Problem~\ref{prob:UMwT_LMsg_1slot_decouple_DC}. 
\textcolor{black}{Problem~\ref{prob:UMwT_LMsg_1slot_decouple_DC} is a difference of convex functions (DC) programming (one type of nonconvex problems), and a KKT point can be obtained by CCCP\cite{JSAC21}.\footnote{CCCP can exploit the partial convexity and usually converges faster to a KKT point than conventional gradient methods.} The main idea of CCCP is to solve a sequence of successively refined approximate convex problems, each of which is obtained by linearizing the concave part and preserving the remaining convex part in the DC problem.}
\textcolor{black}{Specifically, at the $i$-th iteration, the approximate convex problem of Problem~\ref{prob:UMwT_LMsg_1slot_decouple_DC} is given as follows. Let $(\mathbf{w}^{(i)}(\mathbf{h}),\mathbf{R}^{(i)}(\mathbf{h}), \mathbf{e}^{(i)}(\mathbf{h}),\mathbf{u}^{(i)}(\mathbf{h}))$ denote an optimal solution of the following problem.}

\begin{Prob}[Approximation of Problem~\ref{prob:UMwT_LMsg_1slot_decouple_DC} at Iteration $i$]\label{prob:UMwT_LMsg_1slot_decouple_DC_approxi}
\begin{align}
&\max_{\mathbf{w}(\mathbf{h}),\mathbf{R}(\mathbf{h})\succeq 0,\mathbf{e}(\mathbf{h}),\mathbf{u}(\mathbf{h})}\quad \sum\limits_{\mathcal{S}\in\boldsymbol{\mathcal{S}}} \alpha_{\mathcal{S}} \sum_{\mathcal{G}\in\boldsymbol{\mathcal{G}}_{\mathcal{S}}}R_{\mathcal{S},\mathcal{G}}(\mathbf{h})  \nonumber\\
&\mathrm{s.t.}\quad~\eqref{eq:power_perslot_decouple},~\eqref{eq:DC_R<e_perslot},~\eqref{eq:DC_e<u_perslot},\nonumber\\
&L_{k,n,\boldsymbol{\mathcal{X}}}(\mathbf{w}_{n}(\mathbf{h}), u_{k,n,\boldsymbol{\mathcal{X}}}(\mathbf{h});\mathbf{w}^{(i-1)}_{n}(\mathbf{h}), u^{(i-1)}_{k,n,\boldsymbol{\mathcal{X}}}(\mathbf{h})) \leq 0,~\boldsymbol{\mathcal{X}}\subseteq\boldsymbol{\mathcal{G}}^{(k)},k\in\mathcal{K},n\in\mathcal{N},\label{eq:DC_approximation_1slot}
\end{align}
where $\mathbf{w}^{(i-1)}_{n}(\mathbf{h}) \triangleq (\mathbf{w}^{(i-1)}_{\mathcal{G},n}(\mathbf{h}))_{\mathcal{G}\in\boldsymbol{\mathcal{X}} \cup (\boldsymbol{\mathcal{G}}\backslash\boldsymbol{\mathcal{G}}^{(k)})}$, $\mathbf{w}_{n}(\mathbf{h}) \triangleq (\mathbf{w}_{\mathcal{G},n}(\mathbf{h}))_{\mathcal{G}\in\boldsymbol{\mathcal{X}} \cup (\boldsymbol{\mathcal{G}}\backslash\boldsymbol{\mathcal{G}}^{(k)})}$, and
\begin{align}
 &L_{k,n,\boldsymbol{\mathcal{X}}}(\mathbf{w}_{n}(\mathbf{h}), u_{k,n,\boldsymbol{\mathcal{X}}}(\mathbf{h});\mathbf{w}^{(i-1)}_{n}(\mathbf{h}), u^{(i-1)}_{k,n,\boldsymbol{\mathcal{X}}}(\mathbf{h})) \triangleq \sum\limits_{\mathcal{G}'\in\boldsymbol{\mathcal{G}}\backslash\boldsymbol{\mathcal{G}}^{(k)}}|\mathbf{h}^{H}_{k,n}\mathbf{w}_{\mathcal{G}^{'},n}(\mathbf{h})|^{2}
+\sigma^2 \nonumber\\
&+ \frac{\left(\sum\limits_{\mathcal{G}\in\boldsymbol{\mathcal{X}}}
|\mathbf{h}^{H}_{k,n}\mathbf{w}^{(i-1)}_{\mathcal{G},n}(\mathbf{h})|^{2}+
\sum\limits_{\mathcal{G}'\in\boldsymbol{\mathcal{G}}\backslash\boldsymbol{\mathcal{G}}^{(k)}}|\mathbf{h}^{H}_{k,n}\mathbf{w}^{(i-1)}_{\mathcal{G}^{'},n}(\mathbf{h})|^{2}
+\sigma^2
\right)u_{k,n,\boldsymbol{\mathcal{X}}}(\mathbf{h})}{\left(u^{(i-1)}_{k,n,\boldsymbol{\mathcal{X}}}(\mathbf{h})\right)^2} \nonumber\\
&- \frac{2\Re\left\{\sum\limits_{\mathcal{G}\in\boldsymbol{\mathcal{X}}}
\mathbf{w}_{\mathcal{G},n}^{(i-1)H}(\mathbf{h})\mathbf{h}_{k,n}\mathbf{h}^{H}_{k,n}
\mathbf{w}_{\mathcal{G},n}(\mathbf{h})+\sum\limits_{\mathcal{G}'\in\boldsymbol{\mathcal{G}}\backslash\boldsymbol{\mathcal{G}}^{(k)}}
\mathbf{w}_{\mathcal{G}^{'},n}^{(i-1)H}(\mathbf{h})\mathbf{h}_{k,n}\mathbf{h}^{H}_{k,n}
\mathbf{w}_{\mathcal{G}^{'},n}(\mathbf{h})\right\}
+2\sigma^2}{u^{(i-1)}_{k,n,\boldsymbol{\mathcal{X}}}(\mathbf{h})},\nonumber\\
&~~~~~~~~~~~~~~~~~~~~~~~~~~~~~~~~~~~~~~~~~~~~~~~~~~~~~~\boldsymbol{\mathcal{X}}\subseteq\boldsymbol{\mathcal{G}}^{(k)},k\in\mathcal{K},n\in\mathcal{N}.\label{eq:slow_iter_define}
\end{align}
\end{Prob}

Problem \ref{prob:UMwT_LMsg_1slot_decouple_DC_approxi} is convex and can be solved efficiently using standard convex optimization methods\cite{boyd2004convex}. Problem~\ref{prob:UMwT_LMsg_1slot_decouple_DC_approxi} has $MN|\boldsymbol{\mathcal{G}}| + \sum_{S\in\boldsymbol{S}}2^{K-|\mathcal{S}|} + 2N\sum_{k\in\mathcal{K}} (2^{|\boldsymbol{\mathcal{G}}^{(k)}|} -1 )$ variables and $1 + (2N + 1)\sum_{k\in\mathcal{K}} (2^{|\boldsymbol{\mathcal{G}}^{(k)}|} -1 )$ constraints. Thus, when an interior point method is applied, the computational complexity for solving Problem \ref{prob:UMwT_LMsg_1slot_decouple_DC_approxi} is $\mathcal{O}(K^{3.5}2^{1.75 \times 2^{K}})$ as $K \rightarrow \infty$\cite{boyd2004convex}. The details of CCCP for obtaining a KKT point of Problem~\ref{prob:UMwT_LMsg_1slot_decouple_DC} are summarized in Algorithm~\ref{alg:CCCP_perslot}.\footnote{\textcolor{black}{In practice, we can run each of Algorithms 1-4 multiple times with randomly selected feasible initial points to obtain multiple KKT points and choose the KKT point with the best objective value.}} \textcolor{black}{As the number of iterations of Algorithm~\ref{alg:CCCP_perslot} does not scale with the problem size\cite{MOR21}, the computational complexity for Algorithm~\ref{alg:CCCP_perslot} is the same as that for solving Problem \ref{prob:UMwT_LMsg_1slot_decouple_DC_approxi}, i.e., $\mathcal{O}(K^{3.5}2^{1.75 \times 2^{K}})$ as $K \rightarrow \infty$.}

\begin{Thm}[Convergence of Algorithm~\ref{alg:CCCP_perslot}]\label{claim_DC_perslot}
As $i \rightarrow \infty$, $\left(\mathbf{w}^{(i)}(\mathbf{h}),\mathbf{R}^{(i)}(\mathbf{h}), \mathbf{e}^{(i)}(\mathbf{h}), \mathbf{u}^{(i)}(\mathbf{h})\right)$ obtained by Algorithm~\ref{alg:CCCP_perslot} converges to a KKT point of Problem \ref{prob:UMwT_LMsg_1slot_decouple_DC} \cite{TSP17}.
\end{Thm}

\begin{Proof}
\textcolor{black}{The constraints in \eqref{eq:power_perslot_decouple}, \eqref{eq:DC_R<e_perslot}, \eqref{eq:DC_e<u_perslot} are convex, and the constraint function in \eqref{eq:equivalentDCfunction_perslot} can be regarded as a difference between two convex functions, i.e., $\sum\nolimits_{\mathcal{G}'\in\boldsymbol{\mathcal{G}}\backslash\boldsymbol{\mathcal{G}}^{(k)}}|\mathbf{h}^{H}_{k,n}\mathbf{w}_{\mathcal{G}^{'},n}(\mathbf{h})|^{2} +\sigma^2$ and $\frac{\sum\nolimits_{\mathcal{G}\in\boldsymbol{\mathcal{X}}}|
\mathbf{h}^{H}_{k,n}\mathbf{w}_{\mathcal{G},n}(\mathbf{h})|^{2}+
\sum\nolimits_{\mathcal{G}'\in\boldsymbol{\mathcal{G}}\backslash\boldsymbol{\mathcal{G}}^{(k)}}|\mathbf{h}^{H}_{k,n}\mathbf{w}_{\mathcal{G}^{'},n}(\mathbf{h})|^{2}
+\sigma^2}{u_{k,n,\boldsymbol{\mathcal{X}}}(\mathbf{h})}$. Therefore, Problem~\ref{prob:UMwT_LMsg_1slot_decouple_DC} is a DC programming. 
Linearizing $\frac{\sum\nolimits_{\mathcal{G}\in\boldsymbol{\mathcal{X}}}|\mathbf{h}^{H}_{k,n}\mathbf{w}_{\mathcal{G},n}(\mathbf{h})|^{2}+\sum\nolimits_{\mathcal{G}'\in\boldsymbol{\mathcal{G}}\backslash\boldsymbol{\mathcal{G}}^{(k)}}|\mathbf{h}^{H}_{k,n}\mathbf{w}_{\mathcal{G}^{'},n}(\mathbf{h})|^{2}+\sigma^2}{u_{k,n,\boldsymbol{\mathcal{X}}}(\mathbf{h})}$ at $(\mathbf{w}^{(i-1)}_{n}(\mathbf{h}), u^{(i-1)}_{k,n,\boldsymbol{\mathcal{X}}}(\mathbf{h}))$ and preserving $\sum\nolimits_{\mathcal{G}'\in\boldsymbol{\mathcal{G}}\backslash\boldsymbol{\mathcal{G}}^{(k)}}|\mathbf{h}^{H}_{k,n}\mathbf{w}_{\mathcal{G}^{'},n}(\mathbf{h})|^{2} +\sigma^2$ give $L_{k,n,\boldsymbol{\mathcal{X}}}(\mathbf{w}_{n}(\mathbf{h}), u_{k,n,\boldsymbol{\mathcal{X}}}(\mathbf{h});\mathbf{w}^{(i-1)}_{n}(\mathbf{h}), u^{(i-1)}_{k,n,\boldsymbol{\mathcal{X}}}(\mathbf{h}))$ in \eqref{eq:slow_iter_define}. Thus, Algorithm~1 implements CCCP.
It has been validated in \cite{TSP17} that solving DC programming through CCCP always returns a KKT point. Therefore, we can show Theorem~\ref{claim_DC_perslot}.$\hfill\blacksquare$}
\end{Proof}

\textcolor{black}{This paper focuses mainly on exploiting the full potential of general rate splitting for general multicast. The computational complexity of Algorithm~1 can be formidable with a large $K$. We can use successive decoding to reduce the computational complexity to $\mathcal{O}(K^{1.5}|\boldsymbol{\mathcal{S}}|^{1.5}2^{2K})$ as in \cite{JSAC21}.\footnote{\textcolor{black}{Note that $|\boldsymbol{\mathcal{S}}|$ may not scale with $K$, and how $|\boldsymbol{\mathcal{S}}|$ scales with $K$ relies on the user requests. In the case of $|\boldsymbol{\mathcal{S}}| = \mathcal{O}(1)$, the reduced computational complexity is $\mathcal{O}(K^{1.5}2^{2K})$, as $K \rightarrow \infty$.}} We can also apply rate splitting with a reduced number of layers together with successive decoding to further reduce the computational complexity to $\mathcal{O}(K^{1.5}|\boldsymbol{\mathcal{S}}|^{1.5}(|\boldsymbol{\mathcal{G}}_{\text{lb}}|^{2} + |\boldsymbol{\mathcal{G}}_{\text{lb}}||\boldsymbol{\mathcal{S}}|K + |\boldsymbol{\mathcal{S}}|^{2}K^{2}))$ as in\cite{JSAC21}, for some $\boldsymbol{\mathcal{G}}_{\text{lb}}$ satisfying $\boldsymbol{\mathcal{S}}\subseteq \boldsymbol{\mathcal{G}}_{\text{lb}} \subseteq \boldsymbol{\mathcal{G}}.$ Note that $|\boldsymbol{\mathcal{G}}_{\text{lb}}|$ represents the reduced number of layers and satisfies $|\boldsymbol{\mathcal{S}}| \leq |\boldsymbol{\mathcal{G}}_{\text{lb}}| \leq 2^{K} - 1$.\footnote{In the case of $|\boldsymbol{\mathcal{S}}| = \mathcal{O}(1)$ and $|\boldsymbol{\mathcal{G}}_{\text{lb}}| = \mathcal{O}(1)$, the reduced computational complexity is $\mathcal{O}(N^{3.5}K^{3.5})$ as $K \rightarrow \infty$.} Low-complexity optimization methods are beyond the scope of this paper.}

\begin{algorithm}[t] \caption{Obtaining a KKT Point of Problem~\ref{prob:UMwT_LMsg_1slot_decouple_DC}}
\small{\begin{algorithmic}[1]
\STATE Initialization: Choose any feasible point of Problem~\ref{prob:UMwT_LMsg_1slot_decouple_DC} $(\mathbf{w}^{(0)}(\mathbf{h}),\mathbf{R}^{(0)}(\mathbf{h}), \mathbf{e}^{(0)}(\mathbf{h}),\mathbf{u}^{(0)}(\mathbf{h}))$ and set $i=0$.
\STATE \textbf{repeat}
\STATE Obtain an optimal solution $(\mathbf{w}^{(i)}(\mathbf{h}),\mathbf{R}^{(i)}(\mathbf{h}), \mathbf{e}^{(i)}(\mathbf{h}),\mathbf{u}^{(i)}(\mathbf{h}))$ of Problem~\ref{prob:UMwT_LMsg_1slot_decouple_DC_approxi} with an interior point method. 
\STATE Set $i=i+1$.
\STATE \textbf{until} the convergence criterion $\|(\mathbf{w}^{(i)}(\mathbf{h}),\mathbf{R}^{(i)}(\mathbf{h}), \mathbf{e}^{(i)}(\mathbf{h}),\mathbf{u}^{(i)}(\mathbf{h})) - (\mathbf{w}^{(i-1)}(\mathbf{h}),\mathbf{R}^{(i-1)}(\mathbf{h}), \mathbf{e}^{(i-1)}(\mathbf{h}),\mathbf{u}^{(i-1)}(\mathbf{h}))\|_{2} \leq \epsilon$ is met.
\end{algorithmic}}\normalsize\label{alg:CCCP_perslot}
\end{algorithm}

\begin{figure}[t]
\begin{center}
 {\resizebox{10cm}{!}{\includegraphics{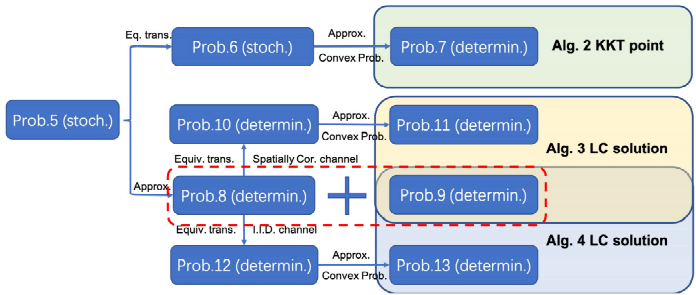}}}
\end{center}
   \caption{\small{Illustration of the proposed solution framework for the fast fading scenario. \textcolor{black}{``Cor.'' and ``LC'' are short for ``Correlated'' and ``Low Complexity'', respectively. The interpretations of the remaining short forms can be found in Fig~\ref{fig:frame_slow}}.}}
   \label{fig:frame_fast}
\end{figure}

\section{Ergodic Rate Maximization In Fast Fading Scenario}\label{sec:fast_fading}
This section considers the fast fading scenario and maximizes the weighted sum ergodic rate under the \textcolor{black}{total transmission power constraint} and \textcolor{black}{achievable ergodic rate constraints}. 
\textcolor{black}{The proposed solution framework for the fast fading scenario is illustrated in Fig.~\ref{fig:frame_fast}.} \textcolor{black}{It is noteworthy that in the fast fading scenario when general multicast degrades to unicast, the proposed solutions based on general rate splitting outperform the one based on 1-layer rate splitting in \cite{TCOM16} due to the fact that our solution can exploit the full potential of rate splitting and obtain a KKT point.}

\subsection{Optimization Problem Formulation}
We would like to optimize the transmission beamforming vectors $\mathbf{w} \triangleq (\mathbf{w}_{\mathcal{G},n})_{\mathcal{G}\in\boldsymbol{\mathcal{G}},n\in\mathcal{N}}$ and rates of sub-message units $\mathbf{R} \triangleq (R_{\mathcal{S},\mathcal{G}})_{\mathcal{S}\in\boldsymbol{\mathcal{S}}, \mathcal{G}\in\boldsymbol{\mathcal{G}}}$ to maximize the \textcolor{black}{weighted sum ergodic rate}, $\sum\nolimits_{\mathcal{S}\in\boldsymbol{\mathcal{S}}} \alpha_{\mathcal{S}} R_{\mathcal{S}}$, subject to the total transmission power constraint in \eqref{eq:power} and the achievable ergodic rate constraints in \eqref{eq:rateconstraints_overslots}. 
Therefore, we formulate the following problem.

\begin{Prob}[Weighted Sum \textcolor{black}{Ergodic} Rate Maximization]\label{prob:UMwT_LMsg_slots}
\begin{align}
U^{(\text{fast})\star} \triangleq \max_{\mathbf{w},\mathbf{R}\succeq 0}\quad & 
\sum\limits_{\mathcal{S}\in\boldsymbol{\mathcal{S}}} \alpha_{\mathcal{S}} R_{\mathcal{S}} \nonumber\\
\mathrm{s.t.}\quad &\eqref{eq:power},~\eqref{eq:rateconstraints_overslots}, \notag
\end{align}
\textcolor{black}{where $R_{\mathcal{S}}$ is given by \eqref{eq:rate_split}.}
\end{Prob}
\begin{Rem}[Connection with Rate Splitting for Unicast and Multicast]
\textcolor{black}{When general multicast degrades to unicast and general rate splitting reduces to 1-layer rate splitting, Problem~\ref{prob:UMwT_LMsg_slots} can reduce to the weighted sum ergodic rate maximization problem for unicast in \cite{TCOM16}. Besides, Problem~\ref{prob:UMwT_LMsg_slots} can be regards as a generalization the weighted sum ergodic rate maximization problem for other service types.}
\end{Rem}

Note that the objective function is linear, the constraint in \eqref{eq:power} is convex, and the constraints in \eqref{eq:rateconstraints_overslots} involve expectations and are nonconvex. \textcolor{black}{Thus, Problem \ref{prob:UMwT_LMsg_slots} is a more challenging nonconvex stochastic problem than the one in the slow fading scenario as it is not separable and has stochastic constraints.}

\subsection{KKT Point}
In this subsection, we propose a stochastic iterative algorithm to obtain a KKT point of Problem~\ref{prob:UMwT_LMsg_slots} using SSCA together with \textcolor{black}{the exact penalty method}\cite{WCL20}. 
First, we equivalently transform Problem~\ref{prob:UMwT_LMsg_slots} to a stochastic optimization problem whose objective function is the weighted sum of the original objective function and the penalty for violating the nonconvex constraints with expectations in \eqref{eq:rateconstraints_overslots}.

\begin{Prob}[Equivalent Problem of Problem \ref{prob:UMwT_LMsg_slots}]\label{prob:UMwT_LMsg_slots_eq}
\begin{align}
\max_{\mathbf{w},\mathbf{R}\succeq 0,\mathbf{s}\succeq 0}\quad & 
\sum\limits_{\mathcal{S}\in\boldsymbol{\mathcal{S}}} \alpha_{\mathcal{S}} \sum_{\mathcal{G}\in\boldsymbol{\mathcal{G}}_{\mathcal{S}}}R_{\mathcal{S},\mathcal{G}} - \rho \sum_{k\in\mathcal{K}}\sum_{\boldsymbol{\mathcal{X}}\subseteq\boldsymbol{\mathcal{G}}^{(k)}}s_{\boldsymbol{\mathcal{X}},k}\nonumber\\
\mathrm{s.t.}\quad &\eqref{eq:power}, \notag\\
&\sum\nolimits_{\mathcal{G}\in\boldsymbol{\mathcal{X}}}\widetilde{R}_{\mathcal{G}} - B\sum\nolimits_{n\in\mathcal{N}}\mathbb{E}\left[ \log_{2}\left(1+\frac{\sum\nolimits_{\mathcal{G}\in\boldsymbol{\mathcal{X}}}|\boldsymbol{\vartheta}^{H}_{k,n}\mathbf{w}_{\mathcal{G},n}|^{2}}{\sigma^2 +\sum\nolimits_{\mathcal{G}'\in\boldsymbol{\mathcal{G}}\backslash\boldsymbol{\mathcal{G}}^{(k)}}|\boldsymbol{\vartheta}^{H}_{k,n}\mathbf{w}_{\mathcal{G}^{'},n}|^{2} }\right)\right] \leq s_{\boldsymbol{\mathcal{X}},k},\nonumber\\
&~~~~~~~~~~~~~~~~~~~~~~~~~~~~~~~~~~~~~~~~~~~~~~~~~~~~~~~~~~~~~~~~~~~~\boldsymbol{\mathcal{X}}\subseteq\boldsymbol{\mathcal{G}}^{(k)},k \in \mathcal{K},\label{eq:penalty}
\end{align}
where $\mathbf{s} \triangleq (s_{\boldsymbol{\mathcal{X}},k})_{\boldsymbol{\mathcal{X}}\subseteq\boldsymbol{\mathcal{G}}^{(k)},k \in \mathcal{K}}$ are slack variables, and $\rho > 0$ is a penalty parameter that trades off the original objective function and the slack penalty term.
\end{Prob}

\textcolor{black}{Next, at iteration $i$, we choose:}
\begin{align}
\overline{F}^{(i)}_{\boldsymbol{\mathcal{X}},k,n}(\mathbf{w}) = (1 - \omega^{(i)} ) \overline{F}^{(i-1)}_{\boldsymbol{\mathcal{X}},k,n}(\mathbf{w}) + \omega^{(i)}\hat{f}_{\boldsymbol{\mathcal{X}},k,n} (\mathbf{w}; \mathbf{w}^{(i-1)}; \mathbf{h}^{(i)}),~\boldsymbol{\mathcal{X}}\subseteq\boldsymbol{\mathcal{G}}^{(k)},~k \in \mathcal{K},~n\in\mathcal{N},\nonumber
\end{align}
with $\overline{F}^{(0)}_{\boldsymbol{\mathcal{X}},k,n}(\mathbf{w}) = 0$ as a convex surrogate function of $\mathbb{E}\left[\log_{2}\left(1+\frac{\sum\nolimits_{\mathcal{G}\in\boldsymbol{\mathcal{X}}}|\boldsymbol{\vartheta}^{H}_{k,n}\mathbf{w}_{\mathcal{G},n}|^{2}}{\sigma^2 +\sum\nolimits_{\mathcal{G}'\in\boldsymbol{\mathcal{G}}\backslash\boldsymbol{\mathcal{G}}^{(k)}}|\boldsymbol{\vartheta}^{H}_{k,n}\mathbf{w}_{\mathcal{G}^{'},n}|^{2} }\right)\right]$ \textcolor{black}{in \eqref{eq:penalty}}, where $\mathbf{h}^{(i)}$ is the realization of the random system channel state at iteration $i$, $\omega^{(i)}$ is a stepsize satisfying
$\omega^{(i)} > 0,~ \lim_{t\rightarrow\infty}\omega^{(i)} = 0, \sum_{i=1}^{\infty}\omega^{(i)} = \infty$,
and
\begin{align}
&\hat{f}_{\boldsymbol{\mathcal{X}},k,n} (\mathbf{w}; \mathbf{w}^{(i-1)},; \mathbf{h}^{(i)}) \triangleq   \log_{2}\left(1+\frac{\sum\nolimits_{\mathcal{G}\in\boldsymbol{\mathcal{X}}}|\mathbf{h}^{(i)H}_{k,n}\mathbf{w}^{(i-1)}_{\mathcal{G},n}|^{2}}{\sigma^2 +\sum\nolimits_{\mathcal{G}'\in\boldsymbol{\mathcal{G}}\backslash\boldsymbol{\mathcal{G}}^{(k)}}|\mathbf{h}^{(i)H}_{k,n}\mathbf{w}^{(i-1)}_{\mathcal{G}^{'},n}|^{2} }\right)\nonumber\\
&+ \frac{2\Re\left\{\sum\nolimits_{\mathcal{G}'\in\boldsymbol{\mathcal{G}}\backslash\boldsymbol{\mathcal{G}}^{(k)}}\mathbf{w}^{H(i-1)}_{\mathcal{G}^{'},n}\mathbf{h}^{(i)}_{k,n}\mathbf{h}^{(i)H}_{k,n}(\mathbf{w}_{\mathcal{G}^{'},n} - \mathbf{w}^{(i-1)}_{\mathcal{G}^{'},n})+ \sum\nolimits_{\mathcal{G}\in\boldsymbol{\mathcal{X}}}\mathbf{w}^{H(i-1)}_{\mathcal{G},n}\mathbf{h}^{(i)}_{k,n}\mathbf{h}^{(i)H}_{k,n}(\mathbf{w}_{\mathcal{G},n} - \mathbf{w}^{(i-1)}_{\mathcal{G},n})\right\}}{\ln2(\sigma^2 +\sum\nolimits_{\mathcal{G}'\in\boldsymbol{\mathcal{G}}\backslash\boldsymbol{\mathcal{G}}^{(k)}}|\mathbf{h}^{(i)H}_{k,n}\mathbf{w}^{(i-1)}_{\mathcal{G}^{'},n}|^{2} + \sum\nolimits_{\mathcal{G}\in\boldsymbol{\mathcal{X}}}|\mathbf{h}^{(i)H}_{k,n}\mathbf{w}^{(i-1)}_{\mathcal{G},n}|^{2})}\nonumber
\end{align}
\begin{align}
&- \frac{2\Re\left\{\sum\nolimits_{\mathcal{G}'\in\boldsymbol{\mathcal{G}}\backslash\boldsymbol{\mathcal{G}}^{(k)}}\mathbf{w}^{H(i-1)}_{\mathcal{G}^{'},n}\mathbf{h}^{(i)}_{k,n}\mathbf{h}^{(i)H}_{k,n}(\mathbf{w}_{\mathcal{G}^{'},n} - \mathbf{w}^{(i-1)}_{\mathcal{G}^{'},n})\right\}}{\ln2(\sigma^2 +\sum\nolimits_{\mathcal{G}'\in\boldsymbol{\mathcal{G}}\backslash\boldsymbol{\mathcal{G}}^{(k)}}|\mathbf{h}^{(i)H}_{k,n}\mathbf{w}^{(i-1)}_{\mathcal{G}^{'},n}|^{2} )} + \tau_{\boldsymbol{\mathcal{X}},k,n}\|\mathbf{w} - \mathbf{w}^{(i-1)}\|^{2},\nonumber\\
&~~~~~~~~~~~~~~~~~~~~~~~~~~~~~~~~~~~~~~~~~~~~~~~~~~~~~~~~~~~~~~~~~~~~\boldsymbol{\mathcal{X}}\subseteq\boldsymbol{\mathcal{G}}^{(k)},~k \in \mathcal{K},~n\in\mathcal{N}\label{eq:surrogate_function}
\end{align}
\textcolor{black}{is a convex approximation of $\mathbb{E}\left[\log_{2}\left(1+\frac{\sum\nolimits_{\mathcal{G}\in\boldsymbol{\mathcal{X}}}|\boldsymbol{\vartheta}^{H}_{k,n}\mathbf{w}_{\mathcal{G},n}|^{2}}{\sigma^2 +\sum\nolimits_{\mathcal{G}'\in\boldsymbol{\mathcal{G}}\backslash\boldsymbol{\mathcal{G}}^{(k)}}|\boldsymbol{\vartheta}^{H}_{k,n}\mathbf{w}_{\mathcal{G}^{'},n}|^{2} }\right)\right]$ around $\mathbf{w}^{(i-1)}$.
In \eqref{eq:surrogate_function}}, $\tau_{\boldsymbol{\mathcal{X}},k,n} >0$ can be any constant, and the term $\tau_{\boldsymbol{\mathcal{X}},k,n}\|\mathbf{w} - \mathbf{w}^{(i-1)}\|^{2}$ is used to ensure strong convexity. Then, the approximation convex problem at iteration $i$ is given as follows.

\begin{Prob}[Approximation of Problem \ref{prob:UMwT_LMsg_slots} at $i$-th Iteration]\label{prob:UMwT_LMsg_slots_eq_ssca}
\begin{align}
\max_{\mathbf{w},\mathbf{R}\succeq 0,\mathbf{s}\succeq 0}\quad & 
\sum\limits_{\mathcal{S}\in\boldsymbol{\mathcal{S}}} \alpha_{\mathcal{S}} \sum_{\mathcal{G}\in\boldsymbol{\mathcal{G}}_{\mathcal{S}}}R_{\mathcal{S},\mathcal{G}} + \rho \sum_{k\in\mathcal{K}}\sum_{\boldsymbol{\mathcal{X}}\subseteq\boldsymbol{\mathcal{G}}^{(k)}}s_{\boldsymbol{\mathcal{X}},k}\nonumber\\
\mathrm{s.t.}\quad &\eqref{eq:power}, \notag\\
&\sum\nolimits_{\mathcal{G}\in\boldsymbol{\mathcal{X}}}\widetilde{R}_{\mathcal{G}} - B\sum\nolimits_{n\in\mathcal{N}}\overline{F}^{(i)}_{\boldsymbol{\mathcal{X}},k,n}(\mathbf{w}) \leq s_{\boldsymbol{\mathcal{X}},k},~\boldsymbol{\mathcal{X}}\subseteq\boldsymbol{\mathcal{G}}^{(k)},k \in \mathcal{K}.\label{eq:penalty_iter}
\end{align}
Let $(\overline{\mathbf{w}}^{(i)},\mathbf{R}^{(i)},\mathbf{s}^{(i)})$ denote an optimal solution of Problem~\ref{prob:UMwT_LMsg_slots_eq_ssca}.
\end{Prob}


It is noted that Problem~\ref{prob:UMwT_LMsg_slots_eq_ssca} is a convex optimization problem which is always feasible \textcolor{black}{(as $s_{\boldsymbol{\mathcal{X}},k}$ can be arbitrarily large).} \textcolor{black}{We can obtain an optimal solution of Problem~\ref{prob:UMwT_LMsg_slots_eq_ssca}}, $(\overline{\mathbf{w}}^{(i)},\mathbf{R}^{(i)},\mathbf{s}^{(i)})$, with the convex optimization techniques \cite{boyd2004convex}. Given $\overline{\mathbf{w}}^{(i)}$, we then update $\mathbf{w}^{(i)}$ according to:
\begin{align}
\mathbf{w}^{(i)} = (1-\gamma^{(i)})\mathbf{w}^{(i-1)} + \gamma^{(i)}\overline{\mathbf{w}}^{(i)},\label{eq:w_update}
\end{align}
where $\gamma^{(i)}$ is a step size satisfying:
\begin{align}
&\gamma^{(i)} > 0, ~ \lim_{t\rightarrow\infty}\gamma^{(i)} = 0,~ \sum_{i=1}^{\infty}\gamma^{(i)} = \infty,~\sum_{t=1}^{\infty}\left(\gamma^{(i)}\right)^{2} < \infty,~\lim_{t\rightarrow\infty}\frac{\gamma^{(i)}}{\omega^{(i)}} = 0.\nonumber
\end{align}
The detailed procedure is summarized in Algorithm~\ref{alg:SSCA}. Note that Problem~\ref{prob:UMwT_LMsg_slots_eq_ssca} has $MN|\boldsymbol{\mathcal{G}}| + \sum_{S\in\boldsymbol{S}}2^{K-|\mathcal{S}|} + \sum_{k\in\mathcal{K}} (2^{|\boldsymbol{\mathcal{G}}^{(k)}|} -1 )$ variables and $1 + \sum_{k\in\mathcal{K}} (2^{|\boldsymbol{\mathcal{G}}^{(k)}|} -1 )$ constraints. Thus, when an interior point method is applied, the computational complexity for solving Problem \ref{prob:UMwT_LMsg_slots_eq_ssca} is $\mathcal{O}(K^{3.5}2^{1.75 \times 2^{K}})$ as $K \rightarrow \infty$ \cite{boyd2004convex}. \textcolor{black}{Note that the number of iterations of Algorithm~\ref{claim_SSCA}, $T$, is fixed, the computational complexity for Algorithm~\ref{claim_SSCA} is the same as that for solving Problem \ref{prob:UMwT_LMsg_slots_eq_ssca}, i.e., $\mathcal{O}(K^{3.5}2^{1.75 \times 2^{K}})$ as $K \rightarrow \infty$.} \textcolor{black}{As discussed in Section~\ref{subsec_slow_solution}, we can apply rate splitting with a reduced number of layers together with successive decoding to reduce the computational complexity to $\mathcal{O}(K^{1.5}|\boldsymbol{\mathcal{S}}|^{1.5}(|\boldsymbol{\mathcal{G}}_{\text{lb}}|^{2} + |\boldsymbol{\mathcal{G}}_{\text{lb}}||\boldsymbol{\mathcal{S}}|K + |\boldsymbol{\mathcal{S}}|^{2}K^{2}))$ as in\cite{JSAC21}.}

\begin{Thm}[Convergence of Algorithm \ref{alg:SSCA}]\label{claim_SSCA}
$\{(\mathbf{w}^{(i)},\mathbf{R}^{(i)}, \mathbf{s}^{(i)})\}$ generated by Algorithm~\ref{alg:SSCA} has a limit point, denote by $(\mathbf{w}^{\dag},\mathbf{R}^{\dag}, \mathbf{s}^{\dag})$, and the following statements hold. (\romannumeral1) $(\mathbf{w}^{\dag},\mathbf{R}^{\dag}, \mathbf{s}^{\dag})$ is a KKT point of Problem~\ref{prob:UMwT_LMsg_slots_eq}; (\romannumeral2) If $\mathbf{s}^{\dag} = \mathbf{0}$, then $(\mathbf{w}^{\dag},\mathbf{R}^{\dag})$ is a KKT point of Problem~\ref{prob:UMwT_LMsg_slots}.
\end{Thm}

\begin{Proof}
\textcolor{black}{Problem~\ref{prob:UMwT_LMsg_slots} and $\overline{F}^{(i)}_{\boldsymbol{\mathcal{X}},k,n}(\mathbf{w})$ satisfy Assumption~1 and Assumption~2 in \cite{WCL20}, respectively. Additionally, Algorithm~2 implements the stochastic iterative algorithm proposed in \cite{WCL20}. Therefore, by Theorem 2 of \cite{WCL20}, we can show \textcolor{black}{that} statement (\romannumeral1) and statement (\romannumeral2) hold.}$\hfill\blacksquare$
\end{Proof}

We can run Algorithm~2 multiple times, each with a random initial point $\mathbf{w}^{(0)}$ \textcolor{black}{and a sufficiently large $\rho > 0$}, until a KKT point $(\mathbf{w}^{\dag},\mathbf{R}^{\dag}, \mathbf{s}^{\dag})$ of Problem~\ref{prob:UMwT_LMsg_slots_eq} with $\mathbf{s}^{\dag} = \mathbf{0}$, i.e., a KKT point $(\mathbf{w}^{\dag},\mathbf{R}^{\dag})$ of Problem~\ref{prob:UMwT_LMsg_slots}, is obtained.


\begin{algorithm}[t] \caption{SSCA for Obtaining a KKT Point of Problem~\ref{prob:UMwT_LMsg_slots_eq}}
\small{\begin{algorithmic}[1]
\STATE Initialization: Choose any feasible point $\mathbf{w}^{(0)}$ of Problem \ref{prob:UMwT_LMsg_slots_eq_ssca} as the initial point and a sufficiently large $\rho > 0$.
\STATE \textbf{for} $i=0,1,\ldots,T$ \textbf{do}
\STATE \quad Obtain an optimal solution $(\overline{\mathbf{w}}^{(i)},\mathbf{R}^{(i)},\mathbf{s}^{(i)})$ of Problem~\ref{prob:UMwT_LMsg_slots_eq_ssca} with an interior point method, and update $\mathbf{w}^{(i)}$ according to \eqref{eq:w_update}.
\STATE \textbf{end for}
\STATE \textbf{Output}: $(\mathbf{w}^{(T)},\mathbf{R}^{(T)},\mathbf{s}^{(T)})$
\end{algorithmic}}\normalsize\label{alg:SSCA}
\end{algorithm}

\subsection{Low-Complexity Solutions}
\textcolor{black}{It is noteworthy that a stochastic iterative algorithm usually converges slowly and has low computation efficiency compared with its deterministic counterpart.} Thus, in this subsection, we propose two low-complexity iterative algorithms to obtain feasible points of Problem~\ref{prob:UMwT_LMsg_slots} with promising performance for two cases of channel distributions using approximation and CCCP. \textcolor{black}{By approximating $\mathbb{E}\left[ \log_{2}\left(1+\frac{\sum\nolimits_{\mathcal{G}\in\boldsymbol{\mathcal{X}}}|\boldsymbol{\vartheta}^{H}_{k,n}\mathbf{w}_{\mathcal{G},n}|^{2}}{\sigma^2 +\sum\nolimits_{\mathcal{G}'\in\boldsymbol{\mathcal{G}}\backslash\boldsymbol{\mathcal{G}}^{(k)}}|\boldsymbol{\vartheta}^{H}_{k,n}\mathbf{w}_{\mathcal{G}^{'},n}|^{2} }\right)\right]$ in \eqref{eq:rateconstraints_overslots} with\cite{IT}:
\begin{align}
&\log_{2}\left(1+\frac{\sum\nolimits_{\mathcal{G}\in\boldsymbol{\mathcal{X}}}\mathbb{E}\left[|\boldsymbol{\vartheta}^{H}_{k,n}\mathbf{w}_{\mathcal{G},n}|^{2}\right]}{\sigma^2 +\sum\nolimits_{\mathcal{G}'\in\boldsymbol{\mathcal{G}}\backslash\boldsymbol{\mathcal{G}}^{(k)}}\mathbb{E}\left[|\boldsymbol{\vartheta}^{H}_{k,n}\mathbf{w}_{\mathcal{G}^{'},n}|^{2}\right] }\right) = \log_{2}\left(1+\frac{\sum\nolimits_{\mathcal{G}\in\boldsymbol{\mathcal{X}}}\mathbf{w}_{\mathcal{G},n}^{H}\mathbf{Q}_{k,n}\mathbf{w}_{\mathcal{G},n}}{\sigma^2 +\sum\nolimits_{\mathcal{G}'\in\boldsymbol{\mathcal{G}}\backslash\boldsymbol{\mathcal{G}}^{(k)}}\mathbf{w}_{\mathcal{G}^{'},n}^{H}\mathbf{Q}_{k,n}\mathbf{w}_{\mathcal{G}^{'},n} }\right),\nonumber
\end{align}
where $\mathbf{Q}_{k,n} \triangleq \mathbb{E}\left[\boldsymbol{\vartheta}_{k,n}\boldsymbol{\vartheta}^{H}_{k,n}\right] \succeq \mathbf{0}$ represents the channel covariance matrix for user $k$ and subcarrier $n$, we approximate the stochastic nonconvex problem in Problem~\ref{prob:UMwT_LMsg_slots} with the following deterministic nonconvex problem.}

\begin{Prob}[Approximate Problem of Problem \ref{prob:UMwT_LMsg_slots} for Obtaining $\mathbf{w}^{\dag}$]\label{prob:UMwT_LMsg_slots_approxi}
\begin{align}
\max_{\mathbf{w},\mathbf{R}\succeq 0}&\quad  
\sum\limits_{\mathcal{S}\in\boldsymbol{\mathcal{S}}} \alpha_{\mathcal{S}} \sum_{\mathcal{G}\in\boldsymbol{\mathcal{G}}_{\mathcal{S}}}R_{\mathcal{S},\mathcal{G}} \nonumber\\
&\mathrm{s.t.}\quad \eqref{eq:power},\nonumber
\end{align}
\end{Prob}
\begin{align}
\sum\limits_{\mathcal{G}\in\boldsymbol{\mathcal{X}}}\widetilde{R}_{\mathcal{G}} \leq B\sum\limits_{n\in\mathcal{N}}\log_{2}\left(1+\frac{\sum\nolimits_{\mathcal{G}\in\boldsymbol{\mathcal{X}}}\mathbf{w}_{\mathcal{G},n}^{H}\mathbf{Q}_{k,n}\mathbf{w}_{\mathcal{G},n}}{\sigma^2 +\sum\nolimits_{\mathcal{G}'\in\boldsymbol{\mathcal{G}}\backslash\boldsymbol{\mathcal{G}}^{(k)}}\mathbf{w}_{\mathcal{G}^{'},n}^{H}\mathbf{Q}_{k,n}\mathbf{w}_{\mathcal{G}^{'},n} }\right),~\boldsymbol{\mathcal{X}}\subseteq\boldsymbol{\mathcal{G}}^{(k)},k \in \mathcal{K}.\label{eq:rateconstraints_overslots_approxi}
\end{align}

Let $\mathbf{w}^{\dag}$ denote the beamforming vectors corresponding to a KKT point of \textcolor{black}{Problem~\ref{prob:UMwT_LMsg_slots_approxi}} (which may not be feasible for Problem~\ref{prob:UMwT_LMsg_slots} because of the \textcolor{black}{adopted} approximation).
\textcolor{black}{Then, construct the following deterministic linear programming parametrized by $\mathbf{w}^{\dag}$.
\begin{Prob}[Approximate Problem of Problem~\ref{prob:UMwT_LMsg_slots} for Obtaining $\mathbf{R}^{\dag}$]\label{prob:UMwT_LMsg_slots_linear}
\begin{align}
&\max_{\mathbf{R}\succeq 0}\quad  
\sum\limits_{\mathcal{S}\in\boldsymbol{\mathcal{S}}} \alpha_{\mathcal{S}} \sum_{\mathcal{G}\in\boldsymbol{\mathcal{G}}_{\mathcal{S}}}R_{\mathcal{S},\mathcal{G}} \nonumber\\
&\mathrm{s.t.}\sum\limits_{\mathcal{G}\in\boldsymbol{\mathcal{X}}}\widetilde{R}_{\mathcal{G}} \leq B\sum\limits_{n\in\mathcal{N}}\mathbb{E}\left[\log_{2}\left(1+\frac{\sum\nolimits_{\mathcal{G}\in\boldsymbol{\mathcal{X}}}|\boldsymbol{\vartheta}^{H}_{k,n}\mathbf{w}^{\dag}_{\mathcal{G},n}|^{2}}{\sigma^2 +\sum\nolimits_{\mathcal{G}'\in\boldsymbol{\mathcal{G}}\backslash\boldsymbol{\mathcal{G}}^{(k)}}|\boldsymbol{\vartheta}^{H}_{k,n}\mathbf{w}^{\dag}_{\mathcal{G}^{'},n}|^{2} }\right)\right],~\boldsymbol{\mathcal{X}}\subseteq\boldsymbol{\mathcal{G}}^{(k)},k \in \mathcal{K}.\label{eq:rateconstraints_overslots_linear}
\end{align}
Let $\mathbf{R}^{\dag}$ denote an optimal solution of Problem~\ref{prob:UMwT_LMsg_slots_linear}.
\end{Prob}
\begin{Lem}[Feasibility of $(\mathbf{w}^{\dag},\mathbf{R}^{\dag})$ for Problem~\ref{prob:UMwT_LMsg_slots}]
$(\mathbf{w}^{\dag},\mathbf{R}^{\dag})$ is a feasible point of Problem~\ref{prob:UMwT_LMsg_slots}.
\end{Lem}
\begin{Proof}
$\mathbf{w}^{\dag}$ satisfies the constraint in \eqref{eq:power}. Besides, $(\mathbf{w}^{\dag},\mathbf{R}^{\dag})$ satisfies the constraints in \eqref{eq:rateconstraints_overslots}. Thus, $(\mathbf{w}^{\dag},\mathbf{R}^{\dag})$ is a feasible point of Problem~\ref{prob:UMwT_LMsg_slots}.$\hfill\blacksquare$
\end{Proof}}

\textcolor{black}{As the approximation error is usually small, $(\mathbf{w}^{\dag},\mathbf{R}^{\dag})$ can be viewed as an approximate solution of Problem~\ref{prob:UMwT_LMsg_slots}. Based on the numerical calculation of $\mathbb{E}\left[\log_{2}\left(1+\frac{\sum\nolimits_{\mathcal{G}\in\boldsymbol{\mathcal{X}}}|\boldsymbol{\vartheta}^{H}_{k,n}\mathbf{w}^{\dag}_{\mathcal{G},n}|^{2}}{\sigma^2 +\sum\nolimits_{\mathcal{G}'\in\boldsymbol{\mathcal{G}}\backslash\boldsymbol{\mathcal{G}}^{(k)}}|\boldsymbol{\vartheta}^{H}_{k,n}\mathbf{w}^{\dag}_{\mathcal{G}^{'},n}|^{2} }\right)\right],~\boldsymbol{\mathcal{X}}\subseteq\boldsymbol{\mathcal{G}}^{(k)},k \in \mathcal{K}$, we can easily solve Problem~\ref{prob:UMwT_LMsg_slots_linear} using standard optimization methods. 
Problem~\ref{prob:UMwT_LMsg_slots_linear} has $\sum_{S\in\boldsymbol{S}}2^{K-|\mathcal{S}|}$ variables and $\sum_{k\in\mathcal{K}}2^{|\boldsymbol{\mathcal{G}}^{(k)}|} -K$ constraints. Thus, the computational complexity for solving Problem~\ref{prob:UMwT_LMsg_slots_linear} using an interior point method is $\mathcal{O}(K^{1.5}2^{0.75\times 2^{K}})$. Moreover, it is expected that the deterministic nonconvex problem in Problem~\ref{prob:UMwT_LMsg_slots_approxi} can be more effectively solved than the stochastic nonconvex problem in Problem~\ref{prob:UMwT_LMsg_slots}. 
In what follows, we concentrate on} solving Problem \ref{prob:UMwT_LMsg_slots_approxi} in \textcolor{black}{two cases of channel distributions, namely spatially correlated channel ($\mathbf{Q}_{k,n}$ has non-zero non-diagonal elements, for all $k\in\mathcal{K},n\in\mathcal{N}$) and independent and identically distributed (i.i.d.) channel ($\mathbf{Q}_{k,n}$ is a diagonal matrix with identical diagonal elements, for all $k\in\mathcal{K},n\in\mathcal{N}$).}

\subsubsection{Spatially correlated channel}In this part, we consider the case of spatially correlated channel, i.e., \textcolor{black}{$\mathbf{Q}_{k,n}$ has non-zero non-diagonal elements, for all $k\in\mathcal{K},n\in\mathcal{N}.$}
First, \textcolor{black}{using the same method as in Section~\ref{sec:slowfading},} Problem~\ref{prob:UMwT_LMsg_slots_approxi} can be transformed into the following DC programming by introducing auxiliary variables and extra constraints.

\begin{Prob}[Equivalent DC Problem of Problem~\ref{prob:UMwT_LMsg_slots_approxi}]\label{prob:equivalentDC_slots_general}
\begin{align}
&\max_{\mathbf{w},\mathbf{R}\succeq 0,\mathbf{e},\mathbf{u}}\quad \sum\limits_{\mathcal{S}\in\boldsymbol{\mathcal{S}}} \alpha_{\mathcal{S}} \sum_{\mathcal{G}\in\boldsymbol{\mathcal{G}}_{\mathcal{S}}}R_{\mathcal{S},\mathcal{G}}  \nonumber\\
&\mathrm{s.t.}\quad \eqref{eq:power},\nonumber\\
&\sum_{\mathcal{G}\in\boldsymbol{\mathcal{X}}}\sum_{\mathcal{S}\in\boldsymbol{\mathcal{S}},\mathcal{S}\subseteq\mathcal{G}}
R_{\mathcal{S},\mathcal{G}} \leq \sum_{n\in\mathcal{N}} e_{k,n,\boldsymbol{\mathcal{X}}},~\boldsymbol{\mathcal{X}}\subseteq\boldsymbol{\mathcal{G}}^{(k)},k\in\mathcal{K},\label{eq:DC_R<e_slots}\\ 
& 2^{\frac{e_{k,n,\boldsymbol{\mathcal{X}}}}{B}} \leq  u_{k,n,\boldsymbol{\mathcal{X}}},~\boldsymbol{\mathcal{X}}\subseteq\boldsymbol{\mathcal{G}}^{(k)},k\in\mathcal{K},n\in\mathcal{N}, \label{eq:DC_e<u_slots}\\
& \sum\nolimits_{\mathcal{G}'\in\boldsymbol{\mathcal{G}}\backslash\boldsymbol{\mathcal{G}}^{(k)}}\mathbf{w}_{\mathcal{G}^{'},n}^{H}\mathbf{Q}_{k,n}\mathbf{w}_{\mathcal{G}^{'},n}
+\sigma^2 - \frac{\sum\nolimits_{\mathcal{G}\in\boldsymbol{\mathcal{X}}}\mathbf{w}_{\mathcal{G},n}^{H}\mathbf{Q}_{k,n}\mathbf{w}_{\mathcal{G},n}+
\sum\nolimits_{\mathcal{G}'\in\boldsymbol{\mathcal{G}}\backslash\boldsymbol{\mathcal{G}}^{(k)}}\mathbf{w}_{\mathcal{G}^{'},n}^{H}\mathbf{Q}_{k,n}\mathbf{w}_{\mathcal{G}^{'},n}
+\sigma^2}{u_{k,n,\boldsymbol{\mathcal{X}}}} \leq 0,\nonumber\\
&~~~~~~~~~~~~~~~~~~~~~~~~~~~~~~~~~~~~~~~~~~~~~~~~~~~~~~~~~~\boldsymbol{\mathcal{X}}\subseteq\boldsymbol{\mathcal{G}}^{(k)},k\in\mathcal{K},n\in\mathcal{N},\label{eq:equivalentDCfunction_general}
\end{align}
where $\mathbf{e} \triangleq \left(e_{k,n,\boldsymbol{\mathcal{X}}}\right)
_{\boldsymbol{\mathcal{X}}\subseteq\boldsymbol{\mathcal{G}}^{(k)},k\in\mathcal{K},n\in\mathcal{N}}$ and
$\mathbf{u}\triangleq \left(u_{k,n,\boldsymbol{\mathcal{X}}}\right)_{\boldsymbol{\mathcal{X}}\subseteq\boldsymbol{\mathcal{G}}^{(k)},k\in\mathcal{K},n\in\mathcal{N}}$ (with a slight abuse of notation). Let $(\mathbf{w}^{\star}, \mathbf{R}^{\star}, \mathbf{e}^{\star}, \mathbf{u}^{\star})$ denote an optimal solution of Problem~\ref{prob:equivalentDC_slots_general}.
\end{Prob}

\begin{Lem}[Equivalence Between Problem~\ref{prob:UMwT_LMsg_slots_approxi} and Problem~\ref{prob:equivalentDC_slots_general}]
$(\mathbf{w}^{\star}, \mathbf{R}^{\star}, \mathbf{e}^{\star}, \mathbf{u}^{\star})$ satisfies $2^{\frac{e^{\star}_{k,n,\boldsymbol{\mathcal{X}}}}{B}} = u^{\star}_{k,n,\boldsymbol{\mathcal{X}}},~\boldsymbol{\mathcal{X}}\subseteq\boldsymbol{\mathcal{G}}^{(k)},k\in\mathcal{K},n\in\mathcal{N}$.
Furthermore, Problem~\ref{prob:UMwT_LMsg_slots_approxi} and Problem~\ref{prob:equivalentDC_slots_general} are equivalent.
\end{Lem}

\begin{Proof}
The proof is similar to the one for Lemma~\ref{lemma_DC_transform} and is omitted here.$\hfill\blacksquare$
\end{Proof}

\textcolor{black}{Note that $\mathbf{e}$ and $\mathbf{u}$ are auxiliary variables, and \eqref{eq:DC_R<e_slots}, \eqref{eq:DC_e<u_slots}, and \eqref{eq:equivalentDCfunction_general} are extra constraints.} \textcolor{black}{Analogously, 
the approximation convex problem of Problem~\ref{prob:equivalentDC_slots_general} at iteration $i$ is given \textcolor{black}{below}. Let $(\mathbf{w}^{(i)},\mathbf{R}^{(i)}, \mathbf{e}^{(i)},\mathbf{u}^{(i)})$ denote an optimal solution of the following problem.}

\begin{Prob}[Approximation of Problem~\ref{prob:equivalentDC_slots_general} at Iteration $i$]\label{prob:equivalentDCiteration_general}
\begin{align}
&\max_{\mathbf{w},\mathbf{R}\succeq 0,\mathbf{e},\mathbf{u}}\quad \sum\limits_{\mathcal{S}\in\boldsymbol{\mathcal{S}}} \alpha_{\mathcal{S}} \sum_{\mathcal{G}\in\boldsymbol{\mathcal{G}}_{\mathcal{S}}}R_{\mathcal{S},\mathcal{G}}  \nonumber\\
&\mathrm{s.t.}\quad~\eqref{eq:power},~\eqref{eq:DC_R<e_slots},~\eqref{eq:DC_e<u_slots},\nonumber\\
&Q_{k,n,\boldsymbol{\mathcal{X}}}(\mathbf{w}_{n}, u_{k,n,\boldsymbol{\mathcal{X}}};\mathbf{w}^{(i-1)}_{n}, u^{(i-1)}_{k,n,\boldsymbol{\mathcal{X}}}) \leq 0,~\boldsymbol{\mathcal{X}}\subseteq\boldsymbol{\mathcal{G}}^{(k)},k\in\mathcal{K},n\in\mathcal{N},\label{eq:general_DC_iter}
\end{align}
where $\mathbf{w}_{n} \triangleq (\mathbf{w}_{\mathcal{G},n})_{\mathcal{G}\in\boldsymbol{\mathcal{X}} \cup (\boldsymbol{\mathcal{G}}\backslash\boldsymbol{\mathcal{G}}^{(k)})}$, $\mathbf{w}^{(i-1)}_{n} \triangleq (\mathbf{w}^{(i-1)}_{\mathcal{G},n})_{\mathcal{G}\in\boldsymbol{\mathcal{X}} \cup (\boldsymbol{\mathcal{G}}\backslash\boldsymbol{\mathcal{G}}^{(k)})}$, and 
\begin{align}
&Q_{k,n,\boldsymbol{\mathcal{X}}}(\mathbf{w}_{n}, u_{k,n,\boldsymbol{\mathcal{X}}};\mathbf{w}^{(i-1)}_{n}, u^{(i-1)}_{k,n,\boldsymbol{\mathcal{X}}})\triangleq \sum\limits_{\mathcal{G}'\in\boldsymbol{\mathcal{G}}\backslash\boldsymbol{\mathcal{G}}^{(k)}}\mathbf{w}_{\mathcal{G}^{'},n}^{H}\mathbf{Q}_{k,n}\mathbf{w}_{\mathcal{G}^{'},n}
+\sigma^2  \nonumber\\
&+ \frac{\left(\sum\nolimits_{\mathcal{G}\in\boldsymbol{\mathcal{X}}}\mathbf{w}_{\mathcal{G},n}^{(i-1)H}\mathbf{Q}_{k,n}\mathbf{w}^{(i-1)}_{\mathcal{G},n}+
\sum\nolimits_{\mathcal{G}'\in\boldsymbol{\mathcal{G}}\backslash\boldsymbol{\mathcal{G}}^{(k)}}\mathbf{w}_{\mathcal{G}^{'},n}^{(i-1)H}\mathbf{Q}_{k,n}\mathbf{w}^{(i-1)}_{\mathcal{G}^{'},n}
+\sigma^2
\right)u_{k,n,\boldsymbol{\mathcal{X}}}}{\left(u^{(i-1)}_{k,n,\boldsymbol{\mathcal{X}}}\right)^2} \nonumber
\end{align}
\begin{align}
&- \frac{2\Re\left\{\sum\nolimits_{\mathcal{G}\in\boldsymbol{\mathcal{X}}}\mathbf{w}_{\mathcal{G},n}^{(i-1)H}\mathbf{Q}_{k,n}\mathbf{w}_{\mathcal{G},n}+
\sum\nolimits_{\mathcal{G}'\in\boldsymbol{\mathcal{G}}\backslash\boldsymbol{\mathcal{G}}^{(k)}}\mathbf{w}_{\mathcal{G}^{'},n}^{(i-1)H}\mathbf{Q}_{k,n}\mathbf{w}_{\mathcal{G}^{'},n}
+\sigma^2\right\}
+2\sigma^2}{u^{(i-1)}_{k,n,\boldsymbol{\mathcal{X}}}},\nonumber\\
&~~~~~~~~~~~~~~~~~~~~~~~~~~~~~~~~~~~~~~~~~~~~~~~~~~~~~~~~~~~~~~~~~~\boldsymbol{\mathcal{X}}\subseteq\boldsymbol{\mathcal{G}}^{(k)},k\in\mathcal{K},n\in\mathcal{N}.\label{eq:fast_iter_define}
\end{align}
\end{Prob}

Similarly, Problem \ref{prob:equivalentDCiteration_general} is a convex problem and can be solved using standard convex optimization methods. 
Problem~\ref{prob:equivalentDCiteration_general} has $MN|\boldsymbol{\mathcal{G}}| + \sum_{S\in\boldsymbol{S}}2^{K-|\mathcal{S}|} + 2N\sum_{k\in\mathcal{K}} (2^{|\boldsymbol{\mathcal{G}}^{(k)}|} -1 )$ variables and $1 + (2N + 1)\sum_{k\in\mathcal{K}} (2^{|\boldsymbol{\mathcal{G}}^{(k)}|} -1 )$ constraints. Thus, when an interior point method is applied, the computational complexity for solving Problem \ref{prob:equivalentDCiteration_general} is $\mathcal{O}(K^{3.5}2^{1.75 \times 2^{K}})$ as $K\rightarrow \infty$ \cite{boyd2004convex}. \textcolor{black}{The details for obtaining a feasible point of Problem~\ref{prob:UMwT_LMsg_slots} are summarized in Algorithm~\ref{alg:CCCP_general}.} \textcolor{black}{Similarly, as the number of iterations of Algorithm~\ref{alg:CCCP_perslot} does not scale with the problem size\cite{MOR21}, we can conclude that the computational complexity of Steps 2-5 in Algorithm~\ref{alg:CCCP_general} is $\mathcal{O}(K^{3.5}2^{1.75 \times 2^{K}})$ as $K\rightarrow \infty$. Recall that the computational complexity for solving Problem~\ref{prob:UMwT_LMsg_slots_linear} is $\mathcal{O}(K^{1.5}2^{0.75\times 2^{K}})$. Thus, the computational complexity of Algorithm~\ref{alg:CCCP_general} is $\mathcal{O}(K^{3.5}2^{1.75 \times 2^{K}})$ as $K\rightarrow \infty$.} \textcolor{black}{Similarly, we can reduce the computational complexity to $\mathcal{O}(K^{1.5}|\boldsymbol{\mathcal{S}}|^{1.5}(|\boldsymbol{\mathcal{G}}_{\text{lb}}|^{2} + |\boldsymbol{\mathcal{G}}_{\text{lb}}||\boldsymbol{\mathcal{S}}|K + |\boldsymbol{\mathcal{S}}|^{2}K^{2}))$ as in\cite{JSAC21}.}
Analogously, we have the following convergence result.

\begin{Thm}[Convergence of Algorithm~\ref{alg:CCCP_general}]\label{claim_DC_general}
As $i \rightarrow \infty$, $(\mathbf{w}^{(i)},\mathbf{R}^{(i)}, \mathbf{e}^{(i)},\mathbf{u}^{(i)})$ obtained by Steps 1-5 of Algorithm~\ref{alg:CCCP_general} converges to a KKT point of Problem~\ref{prob:equivalentDC_slots_general} \cite{TSP17}.
\end{Thm}
\begin{Proof}
\textcolor{black}{The constraints in \eqref{eq:power}, \eqref{eq:DC_R<e_slots}, \eqref{eq:DC_e<u_slots} are convex, and the constraint function in \eqref{eq:equivalentDCfunction_general} can be regarded as a difference between two convex functions, i.e., $\sum\nolimits_{\mathcal{G}'\in\boldsymbol{\mathcal{G}}\backslash\boldsymbol{\mathcal{G}}^{(k)}}\mathbf{w}_{\mathcal{G}^{'},n}^{H}\mathbf{Q}_{k,n}\mathbf{w}_{\mathcal{G}^{'},n} +\sigma^2$ and $\frac{\sum\nolimits_{\mathcal{G}\in\boldsymbol{\mathcal{X}}}\mathbf{w}_{\mathcal{G},n}^{H}\mathbf{Q}_{k,n}\mathbf{w}_{\mathcal{G},n} + \sum\nolimits_{\mathcal{G}'\in\boldsymbol{\mathcal{G}}\backslash\boldsymbol{\mathcal{G}}^{(k)}}\mathbf{w}_{\mathcal{G}^{'},n}^{H}\mathbf{Q}_{k,n}\mathbf{w}_{\mathcal{G}^{'},n}
+\sigma^2}{u_{k,n,\boldsymbol{\mathcal{X}}}}$. Therefore, Problem~\ref{prob:equivalentDC_slots_general} is a DC programming. Similarly,
\textcolor{black}{linearizing $\frac{\sum\nolimits_{\mathcal{G}\in\boldsymbol{\mathcal{X}}}\mathbf{w}_{\mathcal{G},n}^{H}\mathbf{Q}_{k,n}\mathbf{w}_{\mathcal{G},n} + \sum\nolimits_{\mathcal{G}'\in\boldsymbol{\mathcal{G}}\backslash\boldsymbol{\mathcal{G}}^{(k)}}\mathbf{w}_{\mathcal{G}^{'},n}^{H}\mathbf{Q}_{k,n}\mathbf{w}_{\mathcal{G}^{'},n}
+\sigma^2}{u_{k,n,\boldsymbol{\mathcal{X}}}}$ at $(\mathbf{w}^{(i-1)}_{n}, u^{(i-1)}_{k,n,\boldsymbol{\mathcal{X}}})$ and preserving $\sum\nolimits_{\mathcal{G}'\in\boldsymbol{\mathcal{G}}\backslash\boldsymbol{\mathcal{G}}^{(k)}}\mathbf{w}_{\mathcal{G}^{'},n}^{H}\mathbf{Q}_{k,n}\mathbf{w}_{\mathcal{G}^{'},n} +\sigma^2$ give $Q_{k,n,\boldsymbol{\mathcal{X}}}(\mathbf{w}_{n}, u_{k,n,\boldsymbol{\mathcal{X}}};\mathbf{w}^{(i-1)}_{n}, u^{(i-1)}_{k,n,\boldsymbol{\mathcal{X}}})$ in \eqref{eq:fast_iter_define}. Thus, Steps
1-5 of Algorithm~\ref{alg:CCCP_general} implements CCCP.}
It has been validated in \cite{TSP17} that solving DC programming through CCCP always returns a KKT point. Therefore, we can show Theorem~\ref{claim_DC_general}.$\hfill\blacksquare$}
\end{Proof}

\begin{algorithm}[t] \caption{Obtaining a Feasible Point of Problem~\ref{prob:UMwT_LMsg_slots}}
\small{\begin{algorithmic}[1]
\STATE Initialization: Choose any feasible point $(\mathbf{w}^{(0)},\mathbf{R}^{(0)}, \mathbf{e}^{(0)},\mathbf{u}^{(0)})$, and set $i=0$.
\STATE \textbf{repeat}
\STATE Obtain an optimal solution $(\mathbf{w}^{(i)},\mathbf{R}^{(i)}, \mathbf{e}^{(i)},\mathbf{u}^{(i)})$ of Problem~\ref{prob:equivalentDCiteration_general} with an interior point method. 
\STATE Set $i=i+1$.
\STATE \textbf{until} the convergence criterion $\|(\mathbf{w}^{(i)},\mathbf{R}^{(i)}, \mathbf{e}^{(i)},\mathbf{u}^{(i)}) - (\mathbf{w}^{(i-1)},\mathbf{R}^{(i-1)}, \mathbf{e}^{(i-1)},\mathbf{u}^{(i-1)})\|_{2} \leq \epsilon$ is met.
\STATE Set $\mathbf{w}^{\dag} = \mathbf{w}^{(i)}$.
\STATE Calculate $\mathbb{E}\left[\log_{2}\left(1+\frac{\sum\nolimits_{\mathcal{G}\in\boldsymbol{\mathcal{X}}}|\boldsymbol{\vartheta}^{H}_{k,n}\mathbf{w}^{\dag}_{\mathcal{G},n}|^{2}}{\sigma^2 +\sum\nolimits_{\mathcal{G}'\in\boldsymbol{\mathcal{G}}\backslash\boldsymbol{\mathcal{G}}^{(k)}}|\boldsymbol{\vartheta}^{H}_{k,n}\mathbf{w}^{\dag}_{\mathcal{G}^{'},n}|^{2} }\right)\right], \boldsymbol{\mathcal{X}}\subseteq\boldsymbol{\mathcal{G}}^{(k)},k\in\mathcal{K}$ numerically and obtain an optimal solution $\mathbf{R}^{\dag}$ of Problem~\ref{prob:UMwT_LMsg_slots_linear} using standard optimization methods.
\end{algorithmic}}\normalsize\label{alg:CCCP_general}
\end{algorithm}

\subsubsection{I.I.D. channel}
In this part, we consider i.i.d. channel, i.e., \textcolor{black}{$\mathbf{Q}_{k,n}=\lambda\mathbf{I}_{M\times M}$} is a diagonal matrix with identical diagonal elements, \textcolor{black}{for all $k\in\mathcal{K},n\in\mathcal{N}$}. By changing variables, i.e., letting $t_{\mathcal{G},n} \triangleq \|\mathbf{w}_{\mathcal{G},n}\|^{2}_{2} \geq 0$, Problem \ref{prob:UMwT_LMsg_slots_approxi} can be equivalently converted into the following problem.

\begin{Prob}[Equivalent problem of Problem \ref{prob:UMwT_LMsg_slots_approxi} for i.i.d. channel]\label{prob:UMwT_LMsg_slots_approxi_special_iid}
\begin{align}
\max_{\mathbf{t}\succeq 0,\mathbf{R}\succeq 0}\quad & 
\sum\limits_{\mathcal{S}\in\boldsymbol{\mathcal{S}}} \alpha_{\mathcal{S}} \sum_{\mathcal{G}\in\boldsymbol{\mathcal{G}}_{\mathcal{S}}}R_{\mathcal{S},\mathcal{G}} \nonumber\\
\mathrm{s.t.}\quad &\sum_{n \in \mathcal{N}}\sum_{\mathcal{G} \in \boldsymbol{\mathcal{G}}}t_{\mathcal{G},n}  \leq P,\label{eq:t_power}\\
& \sum\limits_{\mathcal{G}\in\boldsymbol{\mathcal{X}}}\widetilde{R}_{\mathcal{G}} \leq B\sum\limits_{n\in\mathcal{N}}\log_{2}\left(1+\frac{\lambda\sum\nolimits_{\mathcal{G}\in\boldsymbol{\mathcal{X}}}t_{\mathcal{G},n}}{\sigma^2 +\lambda\sum\nolimits_{\mathcal{G}'\in\boldsymbol{\mathcal{G}}\backslash\boldsymbol{\mathcal{G}}^{(k)}}t_{\mathcal{G}^{'},n} }\right),~\boldsymbol{\mathcal{X}}\subseteq\boldsymbol{\mathcal{G}}^{(k)},k \in \mathcal{K},\label{eq:rateconstraints_overslots_approxi_iid}
\end{align}
where $\mathbf{t}\triangleq (t_{\mathcal{G},n})_{\mathcal{G}\in\boldsymbol{\mathcal{G}},n\in\mathcal{N}}$. Let $(\mathbf{t}^{\star},\mathbf{R}^{\star})$ denote an optimal solution of Problem~\ref{prob:UMwT_LMsg_slots_approxi_special_iid}.
\end{Prob}

\textcolor{black}{Note that Problem~\ref{prob:UMwT_LMsg_slots_approxi_special_iid} for i.i.d. channel is irrelevant to $M$, as we intend to optimize the time-invariant beamforming vectors according to channel statistics in the fast fading scenario.} Problem~\ref{prob:equivalentDC_slots_general} has $MN|\boldsymbol{\mathcal{G}}|$ complex optimization variables for beamforming (i.e., $\mathbf{w}$) whereas Problem~\ref{prob:UMwT_LMsg_slots_approxi_special_iid} has $N|\boldsymbol{\mathcal{G}}|$ real optimization variables for beamforming (i.e., $\mathbf{t}$). Thus, it is expected that Problem~\ref{prob:UMwT_LMsg_slots_approxi_special_iid} has lower computational complexity than Problem~\ref{prob:equivalentDC_slots_general}. Besides, we have the following result.

\begin{Lem}[Equivalence Between Problem~\ref{prob:UMwT_LMsg_slots_approxi} and Problem~\ref{prob:UMwT_LMsg_slots_approxi_special_iid}]
\textcolor{black}{Any $(\mathbf{w}^{\star},\mathbf{R}^{\star})$ with $\|\mathbf{w}^{\star}_{\mathcal{G},n}\|_{2}^{2} = t^{\star}_{\mathcal{G},n} $ is an optimal solution of Problem~\ref{prob:UMwT_LMsg_slots_approxi}.}
\end{Lem}
\begin{Proof}
When $\mathbf{Q}_{k,n}=\lambda\mathbf{I}$, we can transform $\log_{2}\left(1+\frac{\sum\nolimits_{\mathcal{G}\in\boldsymbol{\mathcal{X}}}\mathbf{w}_{\mathcal{G},n}^{H}\mathbf{Q}_{k,n}\mathbf{w}_{\mathcal{G},n}}{\sigma^2 +\sum\nolimits_{\mathcal{G}'\in\boldsymbol{\mathcal{G}}\backslash\boldsymbol{\mathcal{G}}^{(k)}}\mathbf{w}_{\mathcal{G}^{'},n}^{H}\mathbf{Q}_{k,n}\mathbf{w}_{\mathcal{G}^{'},n} }\right)$ to 

$\log_{2}\left(1+\frac{\lambda\sum\nolimits_{\mathcal{G}\in\boldsymbol{\mathcal{X}}}t_{\mathcal{G},n}}{\sigma^2 +\lambda\sum\nolimits_{\mathcal{G}'\in\boldsymbol{\mathcal{G}}\backslash\boldsymbol{\mathcal{G}}^{(k)}}t_{\mathcal{G}^{'},n} }\right)$ by letting $t_{\mathcal{G},n} \triangleq \|\mathbf{w}_{\mathcal{G},n}\|^{2}_{2} \geq 0$. Thus, the constraints in \eqref{eq:power} and \eqref{eq:rateconstraints_overslots_approxi} can be equivalently converted to the constraints in \eqref{eq:t_power} and \eqref{eq:rateconstraints_overslots_approxi_iid}, respectively. Thus, Problem~\ref{prob:UMwT_LMsg_slots_approxi_special_iid} is equal to Problem~\ref{prob:UMwT_LMsg_slots_approxi}.$\hfill\blacksquare$
\end{Proof}

\textcolor{black}{Problem \ref{prob:UMwT_LMsg_slots_approxi_special_iid} is a DC programming, and we can obtain a KKT point using CCCP. Specifically, the approximation convex problem at iteration $i$ is given \textcolor{black}{below}.}

\begin{Prob}[Approximation of Problem \ref{prob:UMwT_LMsg_slots_approxi_special_iid} at Iteration $i$]\label{prob:UMwT_LMsg_slots_approxi_special_iid_iter}
\begin{align}
\max_{\mathbf{t}\succeq 0,\mathbf{R}\succeq 0}&\quad  
\sum\limits_{\mathcal{S}\in\boldsymbol{\mathcal{S}}} \alpha_{\mathcal{S}} \sum_{\mathcal{G}\in\boldsymbol{\mathcal{G}}_{\mathcal{S}}}R_{\mathcal{S},\mathcal{G}} \nonumber\\
&\mathrm{s.t.}\quad \eqref{eq:t_power}, \notag\\
&\sum\limits_{\mathcal{G}\in\boldsymbol{\mathcal{X}}}\widetilde{R}_{\mathcal{G}} - B\sum\limits_{n\in\mathcal{N}}\log_{2}\left(\sigma^2 +\sum\nolimits_{\mathcal{G}'\in\boldsymbol{\mathcal{G}}\backslash\boldsymbol{\mathcal{G}}^{(k)}\cup\boldsymbol{\mathcal{X}}}t_{\mathcal{G}^{'},n} \right)  + B\sum\limits_{n\in\mathcal{N}}\log_{2}\left(\sigma^2 +\sum\nolimits_{\mathcal{G}'\in\boldsymbol{\mathcal{G}}\backslash\boldsymbol{\mathcal{G}}^{(k)}}t^{(i-1)}_{\mathcal{G}^{'},n} \right) \nonumber
\end{align}
\end{Prob}
\begin{align}
&~~~~~~~~~+ \frac{B}{\ln2}\sum\limits_{n\in\mathcal{N}}\frac{\sum\nolimits_{\mathcal{G}'\in\boldsymbol{\mathcal{G}}\backslash\boldsymbol{\mathcal{G}}^{(k)}}(t_{\mathcal{G}',n} - t^{(i-1)}_{\mathcal{G}^{'},n})}{(\sigma^2 +\sum\nolimits_{\mathcal{G}'\in\boldsymbol{\mathcal{G}}\backslash\boldsymbol{\mathcal{G}}^{(k)}}t^{(i-1)}_{\mathcal{G}^{'},n})} \leq 0,~\boldsymbol{\mathcal{X}}\subseteq\boldsymbol{\mathcal{G}}^{(k)},k \in \mathcal{K}.\label{eq:rateconstraints_overslots_special_iid_DC}
\end{align}
Let $(\mathbf{t}^{(i)}, \mathbf{R}^{(i)})$ denote an optimal solution of Problem \ref{prob:UMwT_LMsg_slots_approxi_special_iid_iter}, where $\mathbf{t}^{(i)}\triangleq (t_{\mathcal{G},n}^{(i)})_{\mathcal{G}\in\boldsymbol{\mathcal{X}}\cup(\boldsymbol{\mathcal{G}}\backslash\boldsymbol{\mathcal{G}}^{(k)}),n\in\mathcal{N}}$.

Problem \ref{prob:UMwT_LMsg_slots_approxi_special_iid_iter} is a convex problem and can be solved using standard convex optimization methods. Note that Problem~\ref{prob:UMwT_LMsg_slots_approxi_special_iid_iter} has $N|\boldsymbol{\mathcal{G}}| + \sum_{S\in\boldsymbol{S}}2^{K-|\mathcal{S}|}$ variables and $1 + \sum_{k\in\mathcal{K}} (2^{|\boldsymbol{\mathcal{G}}^{(k)}|} -1 )$ constraints. Thus, when an interior point method is applied, the computational complexity for solving Problem \ref{prob:UMwT_LMsg_slots_approxi_special_iid_iter} is $\mathcal{O}(K^{1.5}2^{0.75\times2^{K} + 4K})$ as $K\rightarrow \infty$ \cite{boyd2004convex}. The details for obtaining \textcolor{black}{a feasible point of Problem~\ref{prob:UMwT_LMsg_slots}} are summarized in Algorithm~\ref{alg:CCCP_special_iid}. \textcolor{black}{Similarly, the computational complexity of Steps 2-5 in Algorithm~\ref{alg:CCCP_special_iid} is $\mathcal{O}(K^{1.5}2^{0.75\times2^{K} + 4K})$ as $K\rightarrow \infty$. Noting that the computational complexity for solving Problem~\ref{prob:UMwT_LMsg_slots_linear} is $\mathcal{O}(K^{1.5}2^{0.75\times 2^{K}})$, the computational complexity of Algorithm~\ref{alg:CCCP_special_iid} is $\mathcal{O}(K^{1.5}2^{0.75\times2^{K} + 4K})$ as $K\rightarrow \infty$.} \textcolor{black}{Similarly, we can reduce the computational complexity to $\mathcal{O}(K^{1.5}|\boldsymbol{\mathcal{S}}|^{1.5}(|\boldsymbol{\mathcal{G}}_{\text{lb}}|^{2} + |\boldsymbol{\mathcal{G}}_{\text{lb}}||\boldsymbol{\mathcal{S}}| + |\boldsymbol{\mathcal{S}}|^{2}))$ as in\cite{JSAC21}.} \textcolor{black}{Analogously, we have the following convergence result.}

\begin{Thm}[Convergence of Algorithm \ref{alg:CCCP_special_iid}]\label{claim_DC_special_iid}
As $i \rightarrow \infty$, $(\mathbf{t}^{(i)},\mathbf{R}^{(i)})$ obtained by \textcolor{black}{Steps 1-5 of} Algorithm~\ref{alg:CCCP_special_iid} converges to a KKT point of Problem \ref{prob:UMwT_LMsg_slots_approxi_special_iid_iter} \cite{TSP17}.
\end{Thm}
\begin{Proof}
\textcolor{black}{The constraint in \eqref{eq:t_power} is linear and each constraint function of \eqref{eq:rateconstraints_overslots_approxi_iid}, $\sum\nolimits_{\mathcal{G}\in\boldsymbol{\mathcal{X}}}\widetilde{R}_{\mathcal{G}} - B\sum\nolimits_{n\in\mathcal{N}}\log_{2}\left(1+\frac{\lambda\sum\nolimits_{\mathcal{G}\in\boldsymbol{\mathcal{X}}}t_{\mathcal{G},n}}{\sigma^2 +\lambda\sum\nolimits_{\mathcal{G}'\in\boldsymbol{\mathcal{G}}\backslash\boldsymbol{\mathcal{G}}^{(k)}}t_{\mathcal{G}^{'},n} }\right)$ can be regarded as a difference between two convex functions, $\sum\nolimits_{\mathcal{G}\in\boldsymbol{\mathcal{X}}}\widetilde{R}_{\mathcal{G}} - B\sum\nolimits_{n\in\mathcal{N}}\log_{2}\left(\sigma^2 +\sum\nolimits_{\mathcal{G}'\in\boldsymbol{\mathcal{G}}\backslash\boldsymbol{\mathcal{G}}^{(k)}\cup\boldsymbol{\mathcal{X}}}t_{\mathcal{G}^{'},n} \right)$ and $-B\sum\nolimits_{n\in\mathcal{N}}\log_{2}\left(\sigma^2 +\sum\nolimits_{\mathcal{G}'\in\boldsymbol{\mathcal{G}}\backslash\boldsymbol{\mathcal{G}}^{(k)}}t_{\mathcal{G}^{'},n} \right)$. Thus, Problem \ref{prob:UMwT_LMsg_slots_approxi_special_iid} is a DC programming.
\textcolor{black}{Linearizing $-B\sum\nolimits_{n\in\mathcal{N}}\log_{2}\left(\sigma^2 +\sum\nolimits_{\mathcal{G}'\in\boldsymbol{\mathcal{G}}\backslash\boldsymbol{\mathcal{G}}^{(k)}}t_{\mathcal{G}^{'},n} \right)$ at $\mathbf{t}^{(i)}$ and preserving $\sum\nolimits_{\mathcal{G}\in\boldsymbol{\mathcal{X}}}\widetilde{R}_{\mathcal{G}} - B\sum\nolimits_{n\in\mathcal{N}}\log_{2}\left(\sigma^2 +\sum\nolimits_{\mathcal{G}'\in\boldsymbol{\mathcal{G}}\backslash\boldsymbol{\mathcal{G}}^{(k)}\cup\boldsymbol{\mathcal{X}}}t_{\mathcal{G}^{'},n} \right)$ give the constraints in \eqref{eq:rateconstraints_overslots_special_iid_DC}. Thus, Step 1-5 of Algorithm~4 implements CCCP.}
It has been validated in \cite{TSP17} that solving DC programming through CCCP always returns a KKT point. Therefore, we can show Theorem~4.$\hfill\blacksquare$}
\end{Proof}

\begin{algorithm}[t] \caption{Obtaining a Feasible Point of Problem~\ref{prob:UMwT_LMsg_slots}}
\small{\begin{algorithmic}[1]
\STATE Initialization: Choose any feasible point $(\mathbf{t}^{(0)},\mathbf{R}^{(0)})$, and set $i=0$.
\STATE \textbf{repeat}
\STATE Obtain an optimal solution $(\mathbf{t}^{(i)},\mathbf{R}^{(i)})$ of Problem~\ref{prob:UMwT_LMsg_slots_approxi_special_iid_iter} with an interior point method. 
\STATE Set $i=i+1$.
\STATE \textbf{until} the convergence criterion $\|(\mathbf{t}^{(i)},\mathbf{R}^{(i)}) - (\mathbf{t}^{(i-1)},\mathbf{R}^{(i-1)})\|_{2} \leq \epsilon$ is met.
\STATE Set $\mathbf{w}^{\dag} = \mathbf{w}^{(i)}$, where $\|\mathbf{w}^{(i)}_{\mathcal{G},n}\|^{2}_{2} = t^{(i)}_{\mathcal{G},n}.$
\STATE Calculate $\mathbb{E}\left[\log_{2}\left(1+\frac{\sum\nolimits_{\mathcal{G}\in\boldsymbol{\mathcal{X}}}|\boldsymbol{\vartheta}^{H}_{k,n}\mathbf{w}^{\dag}_{\mathcal{G},n}|^{2}}{\sigma^2 +\sum\nolimits_{\mathcal{G}'\in\boldsymbol{\mathcal{G}}\backslash\boldsymbol{\mathcal{G}}^{(k)}}|\boldsymbol{\vartheta}^{H}_{k,n}\mathbf{w}^{\dag}_{\mathcal{G}^{'},n}|^{2} }\right)\right],\boldsymbol{\mathcal{X}}\subseteq\boldsymbol{\mathcal{G}}^{(k)},k \in \mathcal{K}$ numerically and obtain an optimal solution $\mathbf{R}^{\dag}$ of Problem~\ref{prob:UMwT_LMsg_slots_linear} using standard optimization methods.
\end{algorithmic}}\normalsize\label{alg:CCCP_special_iid}
\end{algorithm}
\section{Comparisons Between Slow Fading Scenario and Fast Fading Scenario}
First, we compare the computational complexities for the weighted sum average rate maximization in the slow fading scenario and the weighted sum ergodic rate maximization in the fast fading scenario. Algorithm~1 (for obtaining a KKT point in the slow fading scenario based on CCCP) has higher computational complexity than Algorithm~2 (for obtaining a KKT point in the fast fading scenario based on SSCA). \textcolor{black}{However,} noting that SSCA converges more slowly than CCCP in general, Algorithm~2 has longer computational time to achieve satisfactory performance than Algorithm~1. Besides, Algorithm~3 (for obtaining a low complexity solution in the fast fading scenario with spatially correlated channel based on CCCP) has the same computational complexity as Algorithm~1. 
\textcolor{black}{Last,} Algorithm~4 (for obtaining a low complexity solution in the fast fading scenario with i.i.d. channel based on CCCP) has lower computational complexity than Algorithm~1. 

Next, we compare the practical implementation complexities for \textcolor{black}{the solutions} in slow fading scenario and the fast fading scenario. In the slow fading scenario, the BS must obtain the instantaneous channel condition and adjust the beamforming vectors at each slot according to the instantaneous \textcolor{black}{system channel state}. 
By contrast, in the fast fading scenario, the BS requires only the knowledge of channel statistics and adopts constant beamforming vectors, relying on channel statistics over slots. \textcolor{black}{Therefore}, the practical implementation complexities in the slow fading scenario are \textcolor{black}{obviously} higher than the those in the fast fading scenario.

Finally, we compare the optimal values of the weighted sum average rate maximization in the slow fading scenario and weighted sum ergodic rate maximization in the fast fading scenario. As the beamforming vectors in the slow fading scenario are adjusted at each slot according to the instantaneous system channel state, it is expected that the optimal value in the slow fading scenario is larger than that in the fast fading scenario. Although the gain in the optimal value cannot be shown analytical, it will be numerically verified in Section \ref{sec:simulation}.

\section{Numerical Results}\label{sec:simulation}

In this section, we numerically evaluate \textcolor{black}{the proposed solutions obtained by} Algorithm~\ref{alg:CCCP_perslot} for the slow fading scenario, namely Slow-Prop-RS and Algorithm~\ref{alg:SSCA}, Algorithm~\ref{alg:CCCP_general}, and Algorithm~\ref{alg:CCCP_special_iid} for the fast fading scenario, namely Fast-Prop-RS, Fast-Prop-COR-RS, and Fast-Prop-IID-RS, \textcolor{black}{respectively}. In the slow fading scenario, we consider three baseline schemes, namely Slow-1L-RS, \textcolor{black}{Slow-NOMA}, and Slow-OFDMA. Slow-1L-RS and \textcolor{black}{Slow-NOMA (both with joint decoding for fair comparison)} extend 1-layer rate splitting (where each message is split into one private part and only one common part) \cite{TSP16} and NOMA\textcolor{black}{\cite{TCOM07}}, \textcolor{black}{both} for unicast in one slot, to general multicast in the \textcolor{black}{considered slow fading scenario.} More specifically, Slow-1L-RS and \textcolor{black}{Slow-NOMA} implement Algorithm~\ref{alg:CCCP_perslot} to obtain KKT points of Problem~\ref{prob:UMwT_LMsg_1slot_decouple} with $\mathcal{G}_{\mathcal{S}} = \{\mathcal{S},\mathcal{K}\}, \mathcal{S}\in\boldsymbol{\mathcal{S}}$ and with $\mathcal{G}_{\mathcal{S}} = \{\mathcal{S}\}, \mathcal{S}\in\boldsymbol{\mathcal{S}}$, respectively.\footnote{\textcolor{black}{Slow-NOMA adopts NOMA with joint decoding, which outperforms the commonly adopted NOMA with successive decoding\cite{TCOM07}. Note that it is challenging to optimize the subsequent decoding order for NOMA in general multicast.}}
Slow-OFDMA adopts OFDMA and considers the maximum ratio transmission (MRT) on each subcarrier and optimizes the subcarrier and power allocation\cite{TWC21}. In the fast fading scenario, we consider one baseline scheme, namely Fast-1L-RS.\footnote{As existing works on NOMA and OFDMA mainly concentrate on one single slot, \textcolor{black}{we do not consider their extensions to the fast fading scenario, whose performance can be far from satisfactory.}} Similarly, Fast-1L-RS \textcolor{black}{(with joint decoding)} extends 1-layer rate splitting for unicast \textcolor{black}{in the fast fading scenario\cite{TCOM16} to general multicast in the fast fading scenario. In particular,} Fast-1L-RS implements Algorithm~\ref{alg:SSCA} to obtain a KKT point of Problem~\ref{prob:UMwT_LMsg_slots} with $\mathcal{G}_{\mathcal{S}} = \{\mathcal{S},\mathcal{K}\}, \mathcal{S}\in\boldsymbol{\mathcal{S}}$. \textcolor{black}{Note that for the proposed solutions, $\mathcal{G}_{\mathcal{S}} = \{\mathcal{X}|\mathcal{S}\subseteq\mathcal{X}\subseteq \mathcal{K}\}$.} All algorithms adopt the same stopping criterion that the change in the objective function between two consecutive iterations is smaller than 0.1.
In the slow fading scenario, we generate 100 random realizations of $\boldsymbol{\vartheta}$, solve the weighted sum rate maximization problem for each realization, and evaluate the weighted sum average rate of each scheme over the 100 random realizations. In the sequel, the weighted sum average rate in the slow fading scenario and the weighted sum ergodic rate in the fast fading scenario are \textcolor{black}{both} termed the weighted sum rate for ease of presentation.

In the simulation, we set $K = 3$, $I = 7$, $I_{1} = \{1,4,5,7\}$, $I_{2} = \{2,4,6,7\}$, and $I_{3} = \{3,5,6,7\}$.
As a result, we have $\mathcal{P}_{\{1\}} = \{1\}$, $\mathcal{P}_{\{2\}} = \{2\}$, $\mathcal{P}_{\{3\}} = \{3\}$, $\mathcal{P}_{\{1,2\}} = \{4\}$, $\mathcal{P}_{\{1,3\}} = \{5\}$, $\mathcal{P}_{\{2,3\}} = \{6\}$, and $\mathcal{P}_{\{1,2,3\}} = \{7\}$. Additionally, we set $\alpha_{\mathcal{S}} = 1/7,\mathcal{S}\in\boldsymbol{\mathcal{S}}$, $B$ = 30 kHz, $N$ = 72, and $\sigma^{2} = 10^{-9}$ W. We consider two cases of channel distributions, i.e., spatially
correlated channel with the correlation following the one-ring scattering model as in \cite{JSAC21} and i.i.d. channel with $\textcolor{black}{\boldsymbol{\vartheta}_{k,n}}\sim \mathcal{CN}(0,\textcolor{black}{\mathbf{I}_{M\times M})},k\in\mathcal{K},n\in\mathcal{N}$. Note that the channel covariance matrix in the case of a spatially correlated channel is normalized for a fair comparison. When applying the one-ring scattering model, let $G$ denote the number of user groups. We set the same angular spreads for the $G$ groups and the same azimuth angle for the users in each group as in\cite{JSAC21}. Note that $G$ is related to the channel correlation among users. Specifically, the correlation increases as $G$ decreases. \textcolor{black}{When $G = 1$, 
users belong to one group and have the same channel covariance matrix. When $G = 3$, 
users are in different groups and have different channel covariance matrices.}

\begin{figure*}[t]
\begin{center}
 \subfigure[\small{Weighted sum rate versus $M$.}]
 {\resizebox{4.5cm}{!}{\includegraphics{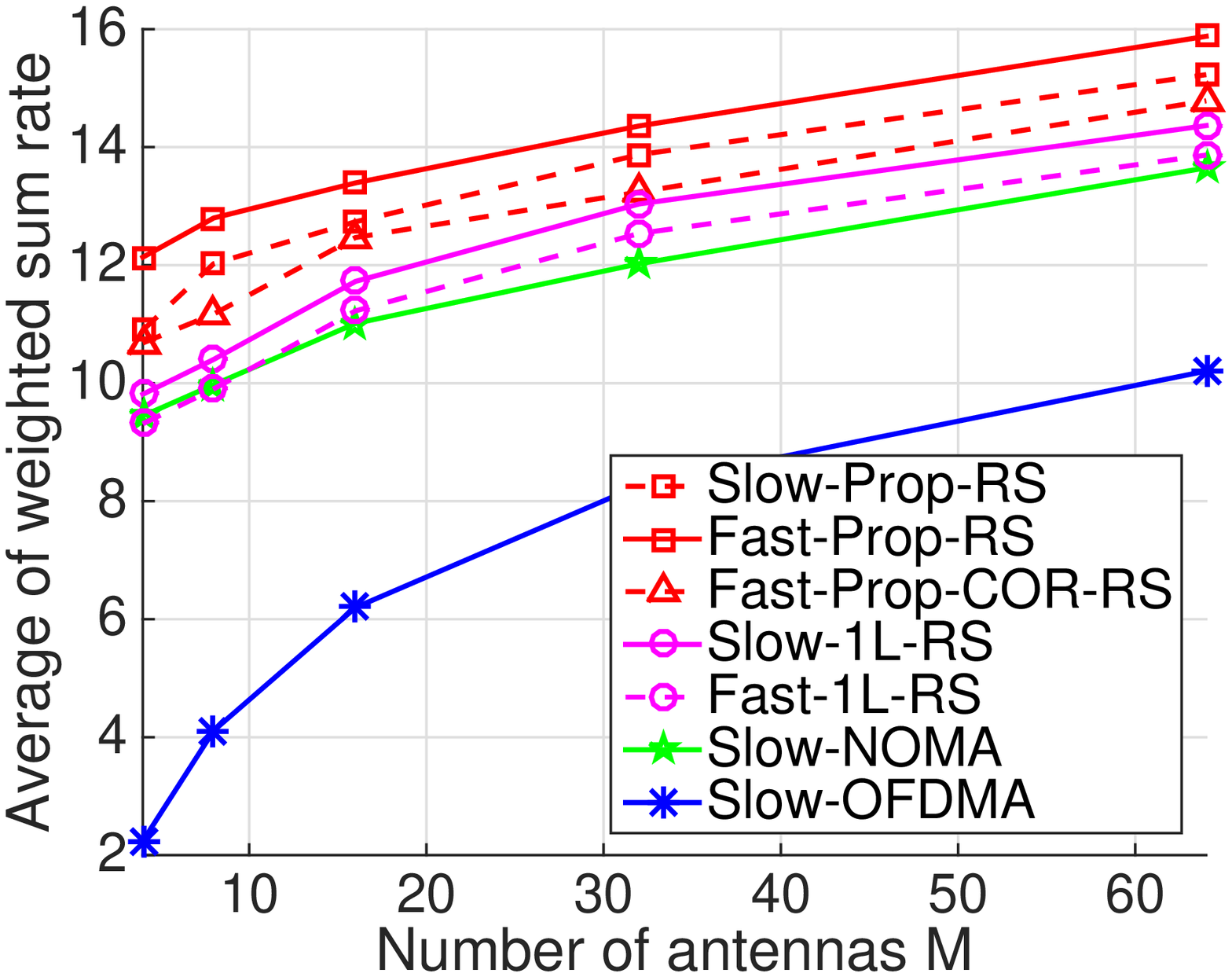}}}
 \subfigure[\small{Weighted sum rate versus $P$.}]
 {\resizebox{4.5cm}{!}{\includegraphics{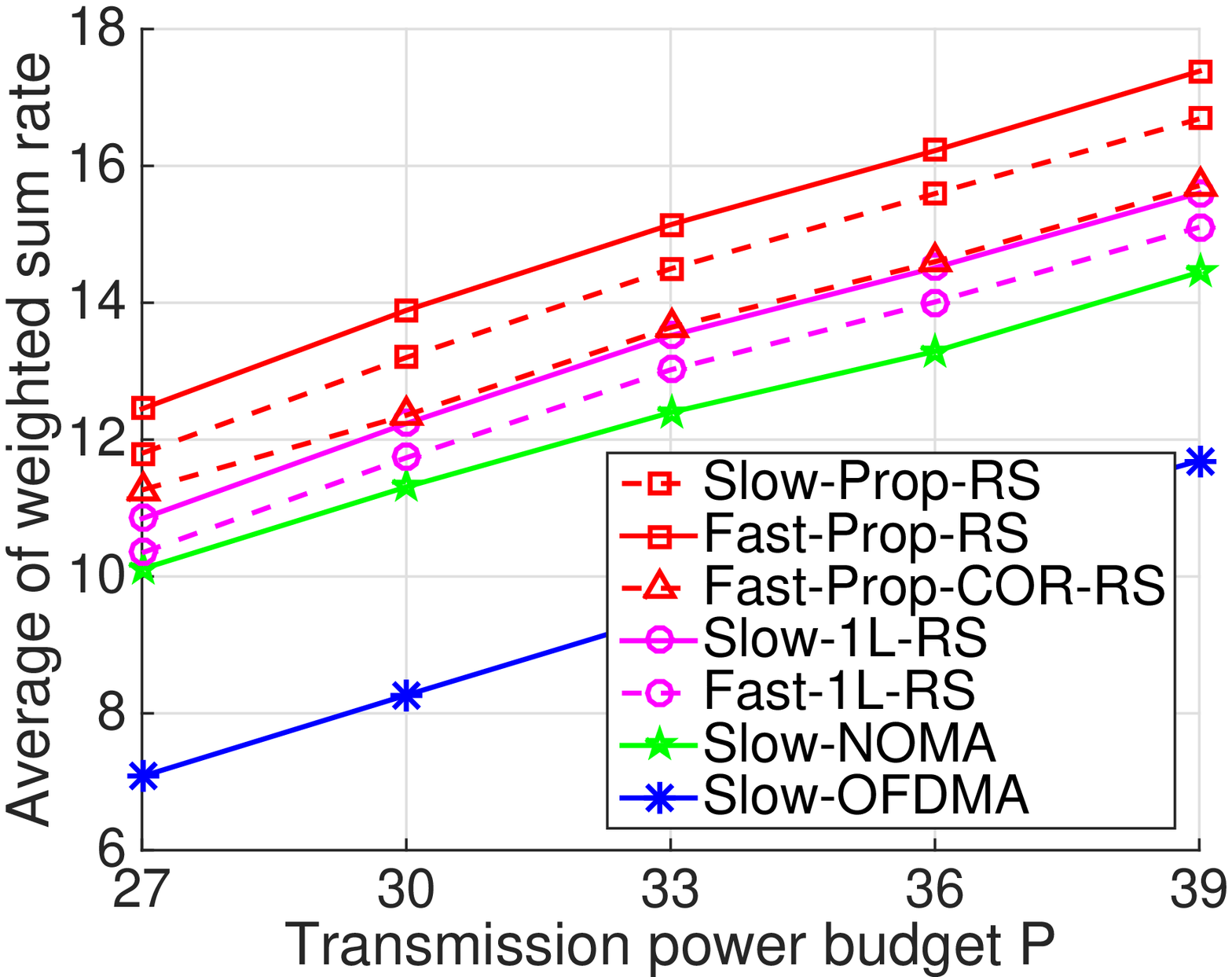}}}
  \subfigure[\small{Weighted sum rate versus $G$.}]
 {\resizebox{4.5cm}{!}{\includegraphics{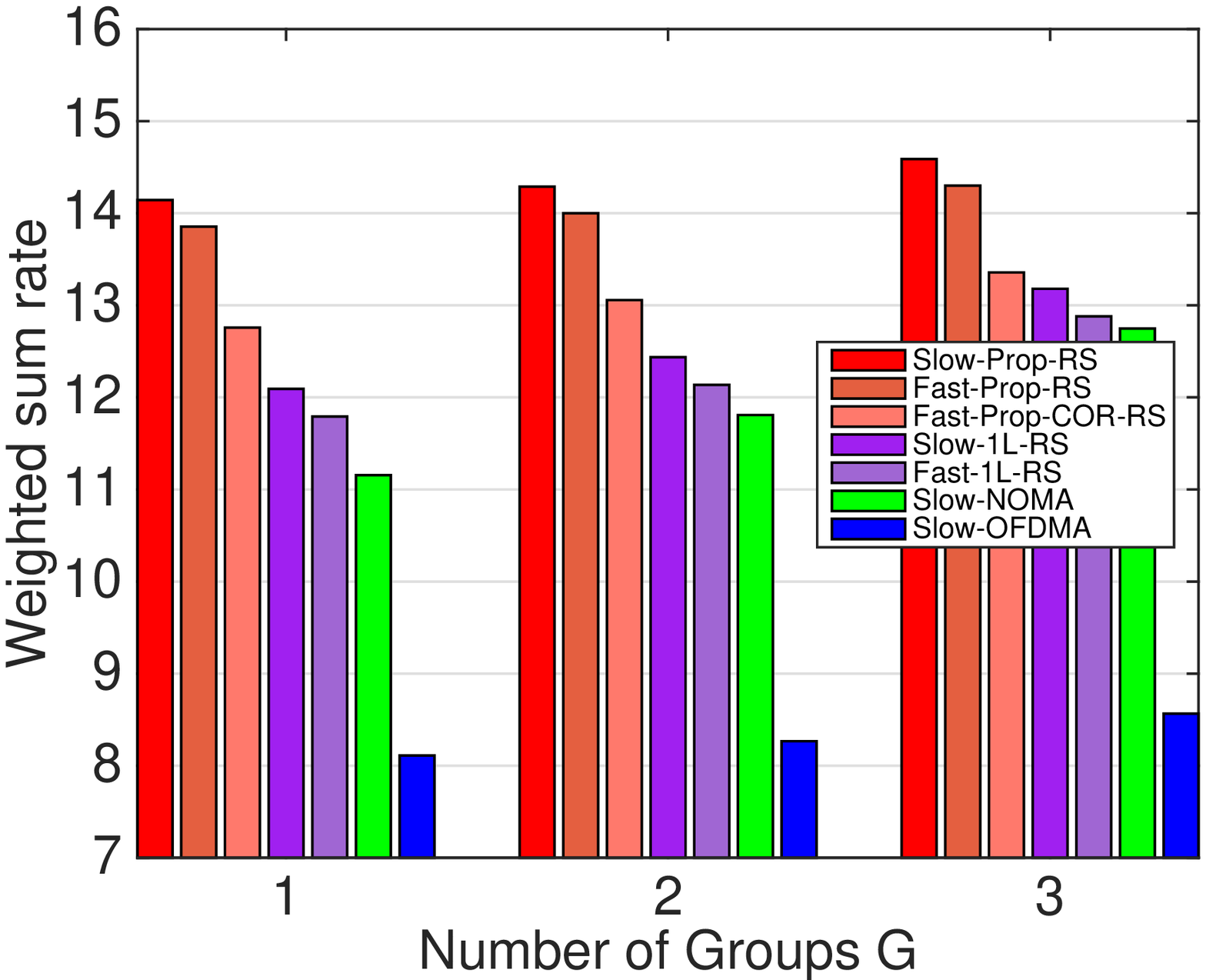}}}
 \end{center}
   \caption{\small{Weighted sum rate in the case of spatially correlated channel.}}
   \label{fig:rateVSPM_OR}
\end{figure*}

\begin{figure*}[t]
\begin{center}
 \subfigure[\small{Rates of transmission units of Slow-Prop-RS.}]
 {\resizebox{4.5cm}{!}{\includegraphics{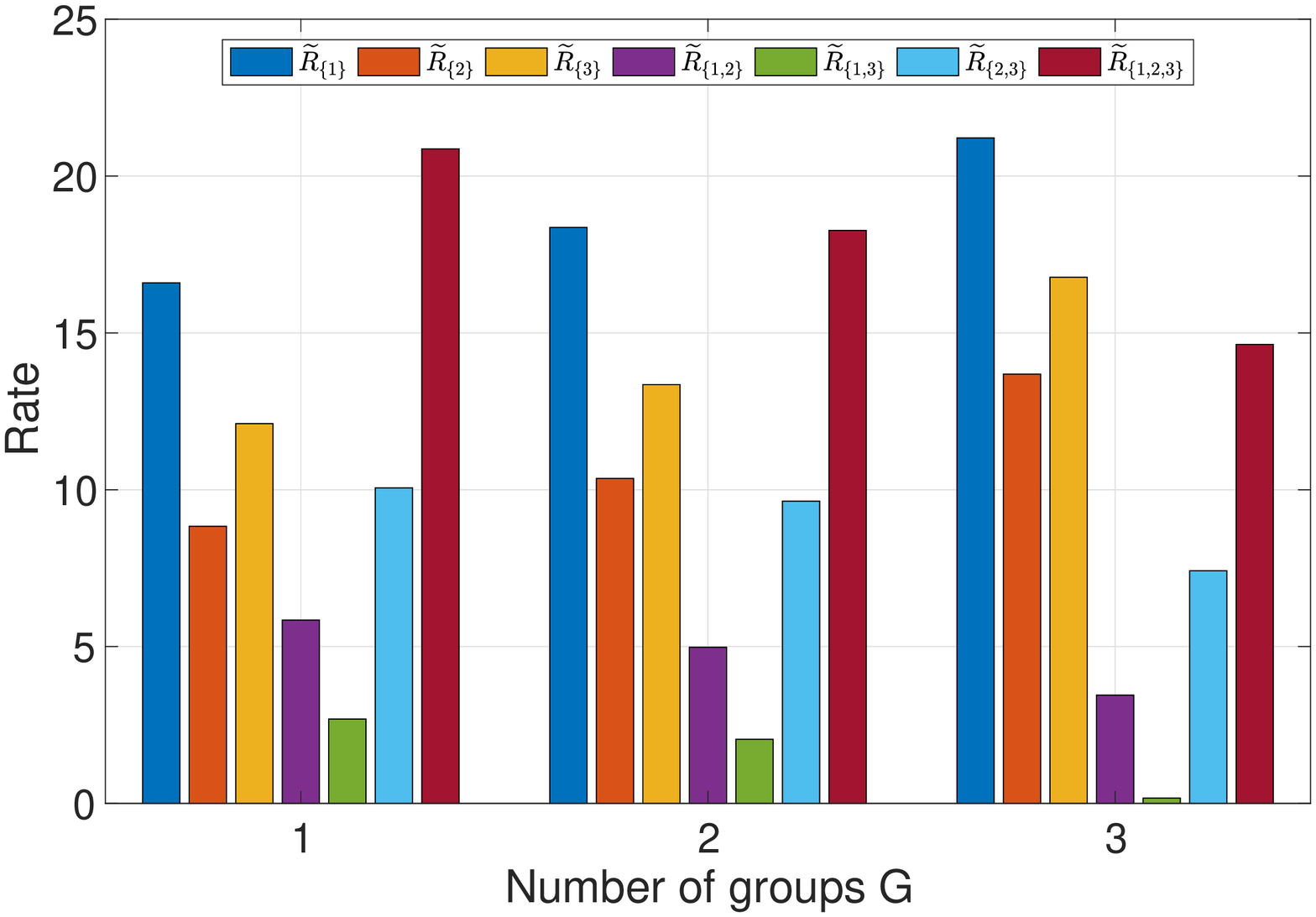}}}
 \subfigure[\small{Rates of transmission units of Fast-Prop-RS.}]
 {\resizebox{4.5cm}{!}{\includegraphics{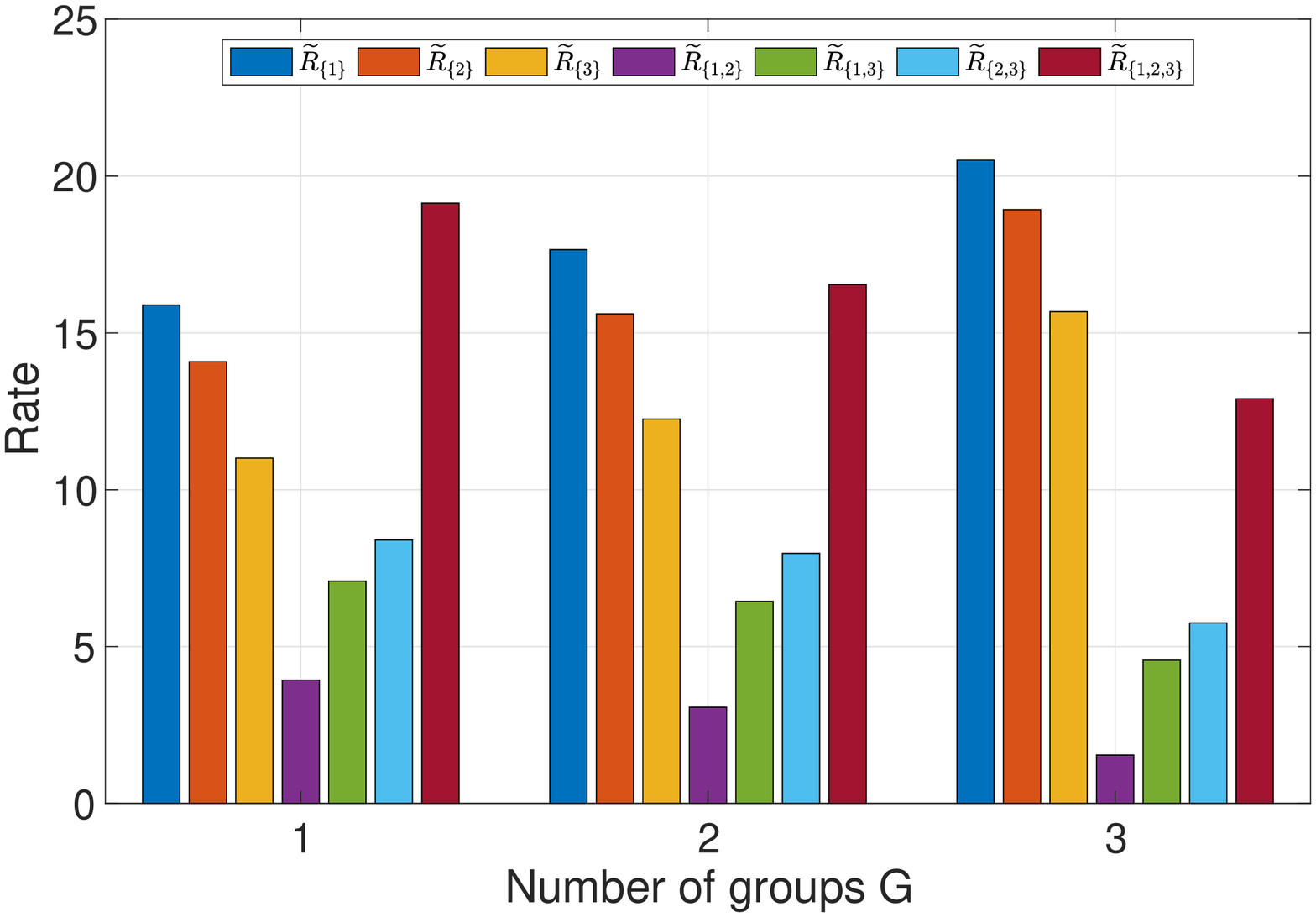}}}
 \subfigure[\small{Rates of transmission units of Fast-Prop-COR-RS.}]
 {\resizebox{4.5cm}{!}{\includegraphics{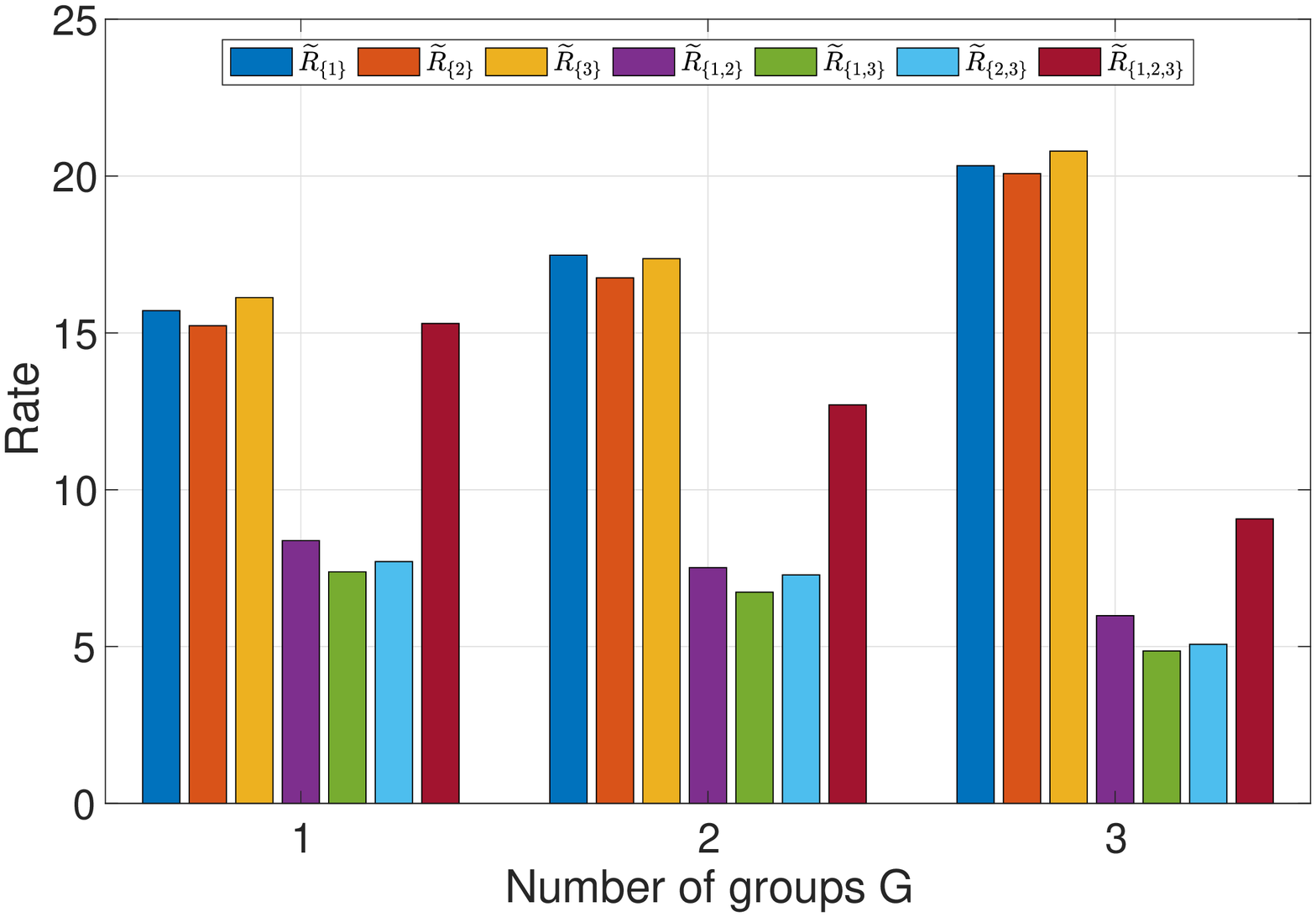}}}
 \end{center}
   \caption{\small{Rates of transmission units of the proposed solutions versus $G$ in the case of spatially correlated channel.}}
   \label{fig:rate_layers}
\end{figure*}

First, we show the weighted sum rate of each scheme in the case of the spatially correlated channel. Fig.~\ref{fig:rateVSPM_OR} illustrates the weighted sum rate versus the number of transmit antennas $M$, the total transmission power budget $P$, and the number of user groups $G$ in the one-ring scattering model, respectively, in the case of the spatially correlated channel. From Fig.~\ref{fig:rateVSPM_OR}, we have the following observations.
Firstly, the weighted sum rate of each scheme increases with $M$, $P$, and $G$. Secondly, in the slow fading and fast fading scenarios, the proposed solutions outperform the corresponding baseline schemes. The gain of Slow-Prop-RS (Fast-Prop-RS or Fast-Prop-COR-RS) over Slow-1L-RS (Fast-1L-RS) is because each proposed solution \textcolor{black}{unleashes the full potential} of the flexibility of rate splitting. The gain of Slow-Prop-RS over \textcolor{black}{Slow-NOMA} arises because the cost for \textcolor{black}{Slow-NOMA} to suppress interference \textcolor{black}{is} high. In contrast, rate splitting together with joint decoding partially decodes interference and partially treats interference as noise. The gain of Slow-Prop-RS over Slow-OFDMA comes from an effective nonorthogonal transmission design. Thirdly, Slow-Prop-RS (Slow-1L-RS) outperforms Fast-Prop-RS (Fast-1L-RS). The gain arises from the fact that Slow-Prop-RS (Slow-1L-RS) optimizes the beamforming vectors at each slot according to the instantaneous \textcolor{black}{system channel state}, whereas Fast-Prop-RS (Fast-1L-RS) \textcolor{black}{adopts the time-invariant} beamforming vectors \textcolor{black}{which are optimized according to the channel statistics}. 
Additionally, Fig.~\ref{fig:rateVSPM_OR} (c) shows that \textcolor{black}{the gain of Slow-Prop-RS (Fast-Prop-RS or Fast-Prop-COR-RS) over Slow-1L-RS (Fast-1L-RS) and the gain of Slow-Prop-RS over \textcolor{black}{Slow-NOMA} increase as $G$ decreases,} demonstrating the advantage of flexibly dealing with interference in the presence of channel correlation among users.
Fig.~\ref{fig:rate_layers} shows the rates of the transmission units in each proposed solution \textcolor{black}{versus the number of user groups $G$ in the case of the spatially correlated channel.} We can see that for each proposed solution, $\widetilde{R}_{\{1\}},\widetilde{R}_{\{2\}},$ and $\widetilde{R}_{\{3\}}$ increase with $G$, whereas $\widetilde{R}_{\{1,2\}},\widetilde{R}_{\{1,3\}},\widetilde{R}_{\{2,3\}},$ and $\widetilde{R}_{\{1,2,3\}}$ decrease with $G$. This is because as channel correlation among the users decreases, it is efficient to decode less interference and treat more interference as noise. 

\begin{figure}
  \centering
  \begin{minipage}{.3\columnwidth}
    \centering
    \includegraphics[width=\textwidth]{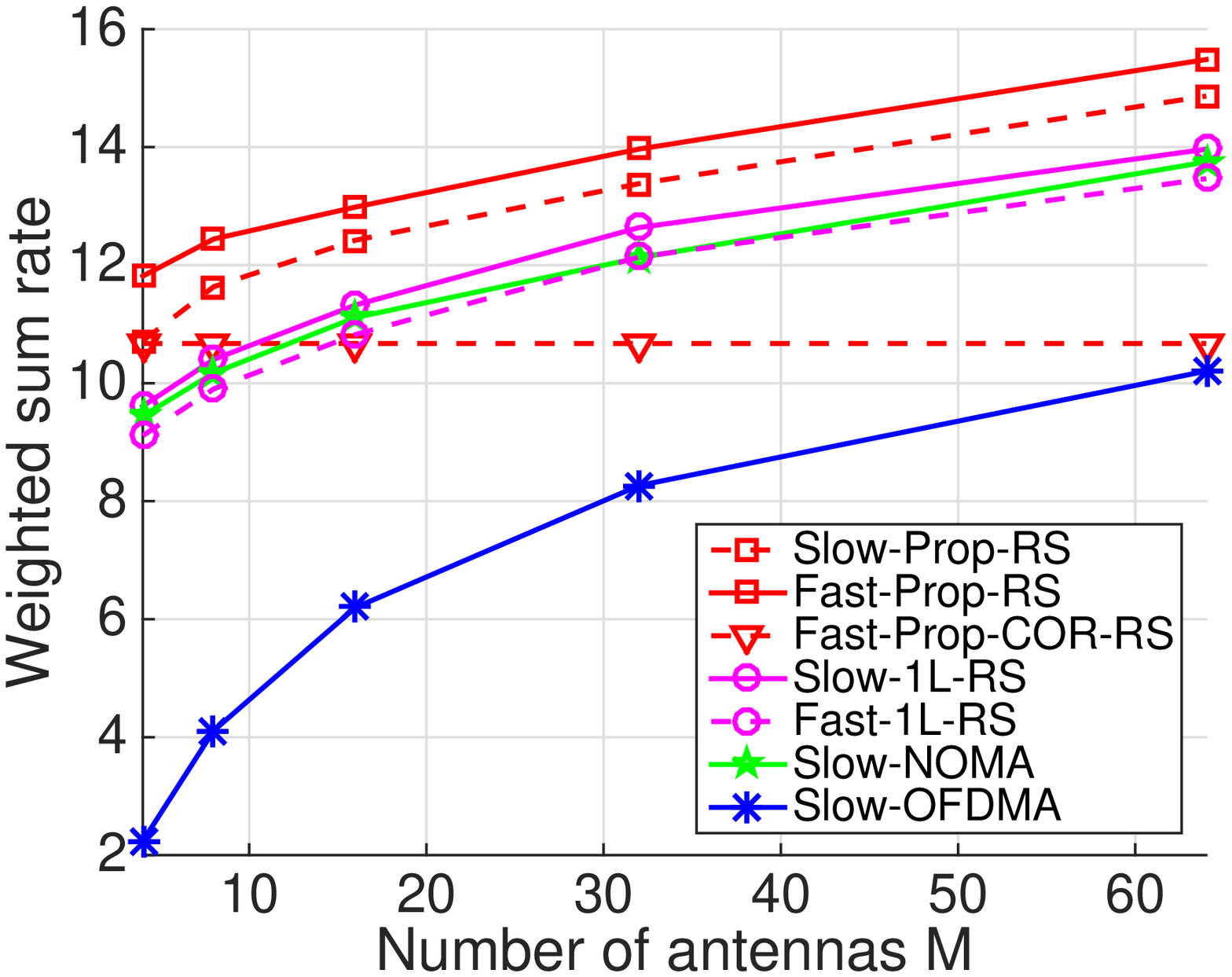}
    \caption{Weighted sum rate versus $M$.}
    \label{fig:rateVSPM_iidM}
  \end{minipage}%
  \begin{minipage}{.3\columnwidth}
    \centering
    \includegraphics[width=\textwidth]{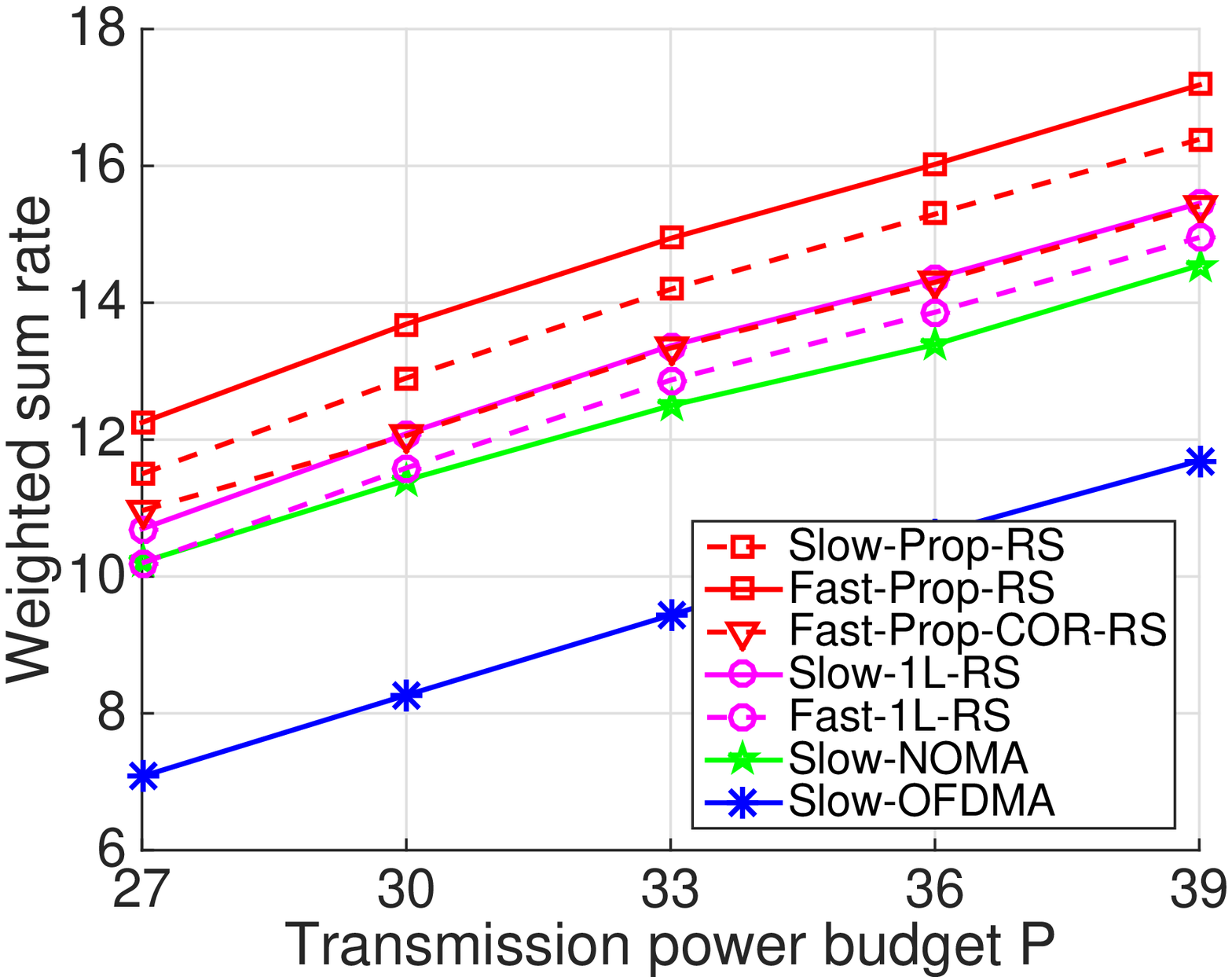}
    \caption{\textcolor{black}{Weighted sum rate versus $P$.}}
    \label{fig:rateVSPM_iidP}
  \end{minipage}
    \begin{minipage}{.3\columnwidth}
    \centering
    \vspace{0.45cm}
    \includegraphics[width=\textwidth]{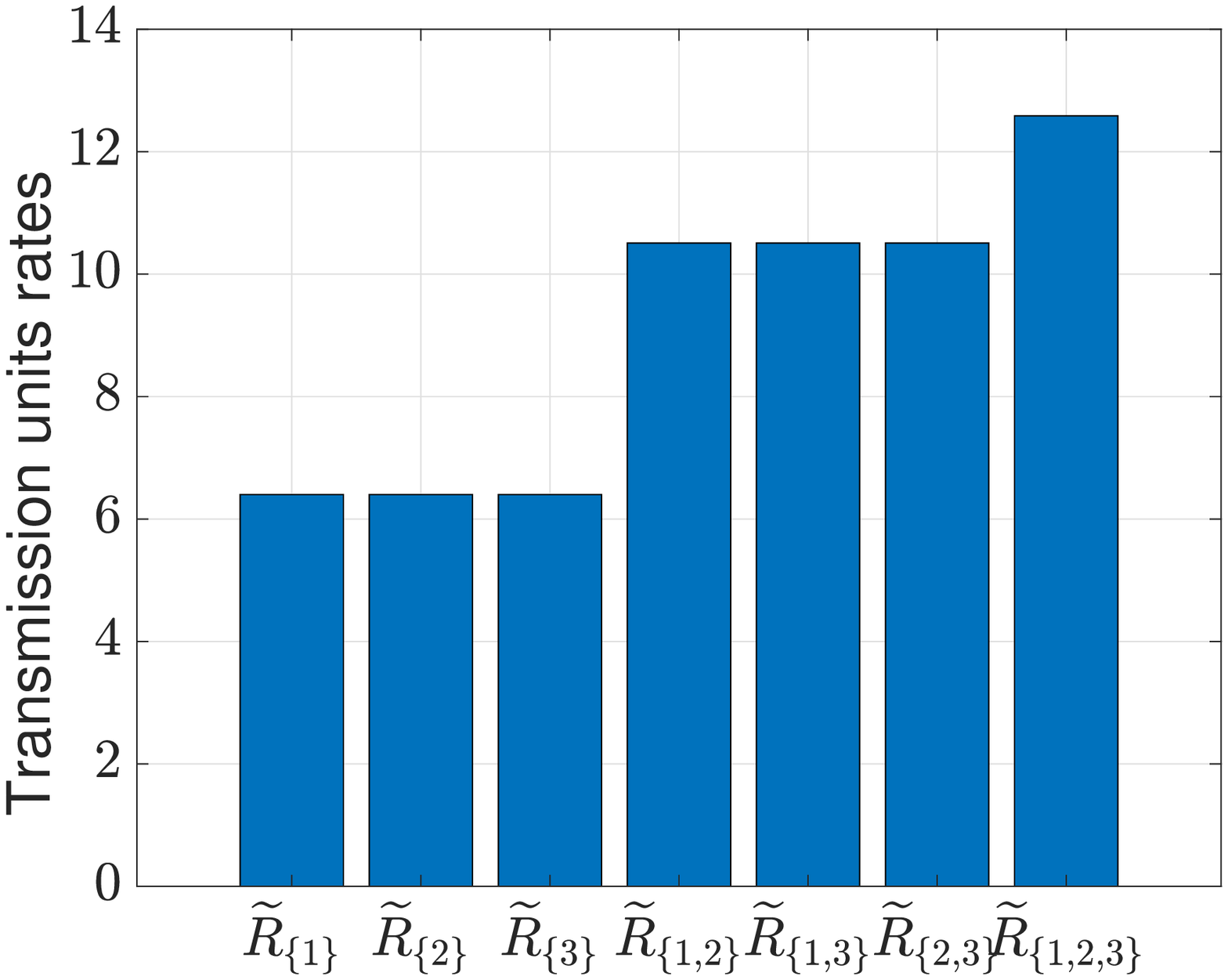}
    \caption{Rates of transmission units of Fast-Prop-IID-RS.}
    \label{fig:layers_iidx}
  \end{minipage}
\end{figure}

Next, we show the weighted sum rate of each scheme in the case of the i.i.d. channel. Fig.~\ref{fig:rateVSPM_iidM} and Fig.~\ref{fig:rateVSPM_iidP} plot the weighted sum rate versus the number of transmit antennas $M$ and the total transmission power budget $P$ \textcolor{black}{in the case of the i.i.d. channel.} \textcolor{black}{We observe from these two figures} that the weighted sum rate of Fast-Prop-IID-RS does not change with $M$. This is because Problem~\ref{prob:UMwT_LMsg_slots_approxi_special_iid} is irrelevant to $M$. Besides, we have the same observations as from Fig.~\ref{fig:rateVSPM_OR} (a) and (b). Fig.~\ref{fig:layers_iidx} shows the rates of the transmission units of \textcolor{black}{Fast-Prop-IID-RS in the case of} the i.i.d. channel. We can see from Fig.~\ref{fig:layers_iidx} that $\widetilde{R}_{\{1\}},\widetilde{R}_{\{2\}}$, and $\widetilde{R}_{\{3\}}$ are identical, and $\widetilde{R}_{\{1,2\}},\widetilde{R}_{\{1,3\}}$, and $\widetilde{R}_{\{2,3\}}$ are identical. \textcolor{black}{The reasons are as follows. The general} multicast setup is symmetric \textcolor{black}{w.r.t.} all users and their requesting messages, i.i.d. channel is considered, and time-invariant beamforming is adopted in the fast fading scenario.

\section{Conclusion}
\textcolor{black}{While applications such as content delivery are responsible for a large and increasing fraction of Internet traffic, general multicast communication will play a central role for future 6G and beyond networks.} This paper investigated the optimization of general rate splitting for general multicast. We optimized the transmission beamforming vectors and rates of sub-message units to maximize the weighted sum average rate and the weighted sum ergodic rate in the slow fading and fast fading scenarios, respectively. We proposed iterative algorithms to obtain KKT points and low-complexity solutions in both scenarios using various optimization techniques. The proposed optimization framework generalizes the existing ones for rate splitting for unicast, \textcolor{black}{unicast with a common message,} single-group multicast, and multi-group multicast. Numerical results demonstrate notable gains of the proposed solutions over existing schemes and reveal the impact of channel correlation among users on the performance of general rate splitting for general multicast.

\textcolor{black}{There are still some key aspects that we leave for future investigations. 
One direction is to go beyond linear approaches and investigate nonlinear precoders such as binning\cite{Entropy,ISIT17}. Another interesting perspective is general multicast with partial channel state information at the transmitter side\cite{TCOM20_new,JSAC21}.}




\end{document}